\documentclass[journal,onecolumn]{IEEEtran}
\usepackage{setspace} 
\usepackage{enumerate}
\usepackage{graphicx}
\usepackage{lipsum}
\usepackage{graphicx}
\usepackage{epstopdf}
\usepackage{csquotes}
\usepackage{subfig}
\usepackage{mathtools}
\usepackage{amssymb, amsmath,amsthm, amsfonts}
\usepackage{ifthen}
\usepackage{tikz}
\usepackage{enumerate}
\usepackage{verbatim}
\usepackage{pifont}
\usepackage{bm}  
\usepackage{cite}
\usepackage{stackengine}
\usepackage{bbm}

\newcommand{\RN}[1]{%
  \textup{\uppercase\expandafter{\romannumeral#1}}%
}

\DeclareMathOperator*{\argmin}{arg\,min}
\DeclareSymbolFont{bbold}{U}{bbold}{m}{n}
\DeclareSymbolFontAlphabet{\mathbbold}{bbold}

\newtheorem{theorem}{Theorem}

\newtheorem{lemma}[theorem]{Lemma}
\newtheorem{prop}[theorem]{Proposition}

\newtheorem{remark}[theorem]{Remark}
\newtheorem{definition}[theorem]{Definition}
\newtheorem{example}{Example}

\newcommand{\ind}{\mathbbm 1}
\newcommand{\Y}{\mathcal{Y}}
\newcommand{\Ucal}{\mathcal{U}}
\newcommand{\W}{\mathcal{W}}
\newcommand{\V}{\mathcal{V}}
\newcommand{\X}{\mathcal{X}}

\newcommand{\Z}{\mathcal{Z}}

\newcommand\blfootnote[1]{%
	\begingroup
	\renewcommand\thefootnote{}\footnote{#1}%
	\addtocounter{footnote}{-1}%
	\endgroup
}
\newcommand{\revised}[1]{{\color{black} #1}}

\def\h2{\tilde h}

\def\hm1{\hat h_{-1}}

\begin{document}
\title{Privacy-aware Distributed Hypothesis Testing}

\author{\IEEEauthorblockN{Sreejith~Sreekumar, Asaf Cohen and 
Deniz G\"und\"uz } }

\maketitle

 \blfootnote{This work has been supported in part by the European Research Council Starting Grant project BEACON (grant agreement number 677854). A part of this work was presented at the IEEE Information theory Workshop (ITW), Guangzhou, 2018 \cite{ITW-18}.   S. Sreekumar was with the Dept. of Electrical and Electronic Engineering, Imperial College London, at the time of this work. He is now with the School of Electrical and Computer Engineering, Cornell University, Ithaca, NY 14850, USA (email: sreejithsreekumar@cornell.edu). A. Cohen is with the Department of Electrical and Computer Engineering, Ben-Gurion University of the Negev, Beer-Sheva 8410501, Israel
(e-mail: coasaf@bgu.ac.il). D. G\"{u}nd\"{u}z is with the Dept. of Electrical and Electronic Engineering, Imperial College London, London SW72AZ, UK (e-mail: d.gunduz@imperial.ac.uk).}

\begin{abstract}
A distributed binary hypothesis testing (HT) problem involving two parties, a remote observer and a detector, is studied. The remote observer has access to a discrete memoryless source, and communicates its observations to the detector via a rate-limited noiseless channel. The detector observes another discrete memoryless source, and performs a binary hypothesis test on the joint distribution of its own observations with those of the observer. While the goal of the observer is to maximize the type II error exponent of the test for a given type I error probability constraint, it also wants to keep a private part of its observations as oblivious to the detector as possible. Considering both equivocation and average distortion under a causal disclosure assumption as possible measures of privacy, the trade-off between the communication rate from the observer to the detector, the type II error exponent, and privacy is studied. For the general HT problem, we establish  single-letter inner bounds on both the rate-error exponent-equivocation and rate-error exponent-distortion trade-offs. Subsequently, single-letter characterizations for both trade-offs are obtained (i) for  testing against conditional independence of the observer's observations from those of the detector, given some additional side-information at the detector; and (ii) when the communication rate constraint over the channel is zero. 
Finally, we show by providing a counterexample that, the strong converse which holds for distributed HT without a privacy constraint, does not hold when a privacy constraint is imposed. This implies that, in general, the rate-error exponent-equivocation and  rate-error exponent-distortion trade-offs are not independent of the type I error probability constraint.
\end{abstract}

\section{Introduction} \label{intro}
Data inference and privacy are often contradicting objectives. In many multi-agent system, each agent/user  reveals information about its data to a remote service,  application or authority, which in turn, provides certain utility to the users based on their data. Many emerging networked systems can be thought of in this context, from social networks to smart grids and communication networks.
While obtaining  the promised utility is the main goal of the users, privacy of data that is shared is becoming increasingly important.
Thus, it is critical that users reveal only the information relevant for obtaining the desired utility, while maximum possible privacy  is retained for their sensitive information.

In distributed learning applications,  typically the goal is to learn the joint probability distribution of data available at different locations. In some cases, there may be prior knowledge about the joint distribution, for example, that it belongs to a certain set of known probability distributions. In such a scenario, the nodes communicate their observations to the detector, which then applies hypothesis testing (HT) on the underlying joint distribution of the data based on its own observations and those received from other nodes. 
However, with the efficient data mining and machine learning algorithms available today, the detector can illegitimately infer some unintended private information from the data provided to it exclusively for HT purposes. Such threats are becoming increasingly imminent as large amounts of seemingly irrelevant yet sensitive data are collected from users, such as in medical research \cite{AJ-2010}, social networks \cite{GA-2005}, online shopping \cite{MF-2010} and smart grids \cite{GD-2016}.
Therefore, there is an inherent trade-off between the utility acquired by sharing data and the associated privacy leakage.

In this paper, we study distributed HT (DHT)  with a privacy constraint, in which, an \textit{observer} communicates its observations to a \textit{detector} over a noiseless rate-limited channel of rate $R$ nats per observed sample. 
Using the data received from the observer, the detector performs binary HT on the joint distribution of its own observations and those of the observer. The performance of the HT is measured by the asymptotic exponential rate of decay of the type II error probability, known as the type II error exponent (or error-exponent henceforth), for a given constraint on the type I error probability (definitions will be given below). While the goal is to maximize the performance of the HT, the observer also wants to maintain a certain level of privacy against the detector for some latent private data that is correlated with its observations. We are interested in characterizing the trade-off between the communication rate from the observer to the detector over the channel, error-exponent achieved by the HT and the amount of information leakage of  private data. A special case of HT known as \textit{testing against conditional independence} (TACI) will be of particular interest. In TACI, the detector tests whether its own observations are independent of those at the observer, conditioned on  additional side information available at the detector. 

\subsection{Background}
Distributed HT without any privacy constraint has been studied extensively from an information theoretic perspective in the past, although many open problems remain. The fundamental results for this problem are first established in \cite{Ahlswede-Csiszar}, which includes a single-letter lower bound on the optimal error-exponent and a  \textit{strong converse} result which states that the optimal error-exponent is independent of the constraint on the type I error probability. Exact single-letter characterization of the optimal error-exponent for the testing against independence (TAI) problem, i.e., TACI with no side information at the detector, is also obtained. The lower bound established in \cite{Ahlswede-Csiszar} is further improved in \cite{Han} and \cite{Shimokawa}. Strong converse is studied in the context of  complete data compression and zero-rate compression in \cite{Han} and \cite{Shalaby-pap},  respectively, where, in the former, the observer communicates to the detector using a message set of size two, while in the latter using a message set whose size grows sub-exponentially with the number of observed samples.  
The TAI problem with multiple observers remains open (similar to several other distributed compression problems when a non-trivial fidelity criterion is involved); however, the optimal error-exponent  is obtained in \cite{Zhao-Lai} when the sources observed at different observers follow a certain Markov relation. The scenario in which, in addition to HT, the detector is also interested in obtaining a reconstruction of the observer's source, is studied in \cite{Katz-estdetjourn}. 
The authors characterize the trade-off between the achievable error-exponent and the average distortion between the observer's observations and the detector's reconstruction. 
The TACI is first studied in \cite{Rahman-Wagner}, where the optimality of a random binning based encoding scheme is shown. The optimal error-exponent for TACI over a noisy communication channel is established in \cite{SG_isit17}. Extension of this work to general HT over a noisy channel is considered in \cite{SD_2020}, where lower bounds on the optimal error-exponent are obtained by using a separation based scheme and also using hybrid coding for the communication between the observer and the detector. 
The TACI with a single observer and multiple detectors is studied in \cite{Sadaf_Wigger_Timo}, where each detector tests for the conditional independence of its own observations from those of the observer.
The general HT version of this problem over a noisy broadcast channel and DHT over a multiple access channel is explored in \cite{Sadaf-Wigger-HTN}.
While all the above works consider the asymmetric objective of maximizing the error-exponent under a constraint on the type I error probability, the trade-off between the exponential rate of decay of both the type I and type II error probabilities are considered in \cite{HK-1989,HK-2017,WK-2017}. 
 
Data privacy has been a hot topic of research in the past decade, spanning across multiple disciplines in computer and computational sciences. Several practical schemes have been proposed that deal with the protection or violation of data privacy in different contexts, e.g., see \cite{Bayardo-Agrawal, Rakesh-Srikant,Bert15,Yael2000,Hay08,Narayanan-Shmatikov}.
 More relevant for our work, HT under mutual information and maximal leakage privacy constraints have been studied in \cite{Liao-Lal-Tan-Cal} and \cite{Jiao-Lal-Cal}, respectively, where the observer uses a \textit{memoryless privacy mechanism} to convey a noisy version of its observed data to the detector. The detector performs HT on the probability distribution of the observer's data, and the optimal privacy mechanism that maximizes the error-exponent while satisfying the privacy constraint is analyzed. Recently, a distributed  version of this problem has been studied in \cite{GSSV-2018}, where the observer applies a  privacy mechanism to its observed data prior to further coding for compression, and the goal at the detector is to perform a HT on the joint distribution of its own observations with those of the observer. In contrast with \cite{Liao-Lal-Tan-Cal}, \cite{Jiao-Lal-Cal} and \cite{GSSV-2018}, we study \textit{DHT with a privacy constraint}, but without considering a separate privacy mechanism at the observer. In Section \ref{Prelims}, we will further discuss the differences between the system model considered here and that of \cite{GSSV-2018}.

It is important to note here that the  data privacy  problem is fundamentally different from that of  data security against an eavesdropper or an adversary. In data security, sensitive data is to be protected against an external malicious agent distinct from the legitimate parties in the system. The techniques for guaranteeing data security usually involve either cryptographic methods in which the legitimate parties are assumed to have additional resources unavailable to the adversary (e.g., a shared private key) or the availability of better communication channel conditions (e.g., using wiretap codes). However, in data privacy problems, the sensitive data is to be protected from the same legitimate party that receives the messages and  provides the utility; and hence, the above mentioned techniques for guaranteeing data security are not applicable. Another model frequently used in the context of information-theoretic security assumes the availability of different side-information at the legitimate receiver and the eavesdropper \cite{GEP-08, GEP-08-ISIT}. A DHT problem with security constraints formulated along these lines is studied in \cite{MP-2015}, where the authors propose an inner bound on the rate-error exponent-equivocation trade-off. While our model is  related to that in \cite{MP-2015} when the side-information at the detector and eavesdropper coincide, there are some important differences which will be highlighted in Section \ref{relprevwork}.   
 
Many different privacy measures have been considered in the literature to quantify the amount of private information leakage, such as k-anonymity \cite{Sweeney-2002}, differential privacy \cite{Dwork-2006}, mutual information leakage \cite{Calmon-Fawaz,PF-2014,Cal-Ali-Muriel}, maximal leakage \cite{Issa-Kamath-Wagner-2016}, and total variation distance \cite{Rassouli-Gunduz-TIFS19} to count a few; see \cite{Wagner-Eckhoff} for a detailed survey.  Among these, mutual information between the private and revealed information (or, equivalently, the  \textit{equivocation} of private information given the revealed information) is perhaps the most commonly used measure in the information theoretic studies of privacy. It is well known that a necessary and sufficient condition to guarantee  statistical independence between two random variables is to have zero mutual information between them.  Furthermore, the average information leakage measured using an arbitrary privacy measure is upper bounded by a constant multiplicative factor of that measured by mutual information \cite{PF-2014}. It is also shown in \cite{Calmon-Fawaz} that a differentially private scheme is not necessarily private when the information leakage is measured by mutual information. This is done by constructing an example that is differentially private, yet the mutual information leakage is arbitrarily high. Mutual information based measures have also been used in cryptographic security studies. For example, the notion of semantic security defined in \cite{GM-1984} is shown to be equivalent to a measure based on mutual information in \cite{BTV-2012}. 

A  rate-distortion approach to privacy is first explored by Yamamoto in  \cite{Yamamoto-1988} for a rate-constrained noiseless channel, where, in addition to a distortion constraint for legitimate data, a minimum distortion requirement is enforced for the private part. Recently, there have been several works that have used distortion as a security or privacy metric in several different contexts, such as side-information privacy in discriminatory lossy source coding \cite{TSP-2013} and rate distortion theory of secrecy systems \cite{Schieler-Cuff-2014}\cite{Gaurav-thesis}.  
 \revised{More specifically, in \cite{Schieler-Cuff-2014}, the distortion based security measure is analyzed under a \textit{causal disclosure assumption}, in which the data samples to be protected are causally revealed to the eavesdropper (excluding the current sample), yet the average distortion over the entire block has to satisfy a desired lower bound. This assumption  ensures that distortion as a secrecy measure is more \textit{robust} (see \cite[Section I-A]{Schieler-Cuff-2014}), and could in practice model scenarios in which the sensitive data to be protected is eventually available to the eavesdropper with some delay, but the protection of the current data sample is important. 
 In this paper, we will consider both equivocation and average distortion under a causal disclosure assumption as measures of privacy.} In \cite{LOG-2019}, error-exponent of a HT adversary is considered as a privacy measure. This can be considered as the opposite setting to ours, in the sense that, while the goal here is to increase the error-exponent under a privacy leakage constraint, the goal in \cite{LOG-2019} is to reduce the error-exponent under a constraint on possible transformations that can be applied on the data.  
 
The amount of private information leakage that can be tolerated depends on the specific application at hand. While it may be possible to tolerate a moderate amount of information leakage in applications like online shopping or social networks, it may no longer be the case in matters related to information sharing among government agencies or corporations. While it is obvious that maximum privacy can be attained by revealing no information, this typically comes at the cost of zero utility. On the other hand,  maximum utility can be achieved by revealing all the  information, but at the cost of minimum privacy. Characterizing the optimal trade-off between the utility and the minimum privacy leakage between these two extremes is a fundamental and challenging research problem. 


\subsection*{Main Contributions}
The main contributions of this work are as follows. 
\begin{enumerate}[(i)]
    \item In Section \ref{ghtpconst}, Theorem \ref{thm:genhtequ} (resp. Theorem \ref{thm:genhtdist}), we establish a single-letter inner bound on the rate-error exponent-equivocation (resp. rate-error exponent-distortion) trade-off for DHT with a privacy constraint. The  distortion and equivocation privacy constraints we consider, that is given in \eqref{distexpstr} and \eqref{strongprivconstequ}, respectively, are slightly stronger than what is usually considered in the literature (stated in \eqref{weakdistconst} and \eqref{weakeqconst}, respectively).
    \item Exact characterizations are obtained for some important special cases in Section \ref{optresultsspcl}. More specifically, a single-letter characterization of the optimal rate-error exponent-equivocation (resp. rate-error exponent-distortion) trade-off is established for:
\begin{enumerate}
    \item TACI with a privacy constraint (for vanishing type I error probability constraint) in Section \ref{TACIspclcase}, Proposition \ref{equivocation-noiseless} (resp. Proposition  \ref{distortion-noiseless}),
     \item  DHT with a privacy constraint for zero-rate compression in Section \ref{zerorateHT}, Proposition  \ref{thm:zerorategenhtequ} (resp. Proposition \ref{thm:zerorategenhtdist}).
\end{enumerate}
Since the  optimal trade-offs in Propositions  \ref{thm:zerorategenhtdist} and  \ref{thm:zerorategenhtequ} are independent of the constraint on the type I error probability, they are strong converse results in the context of HT.
\item Finally, in Section \ref{Strongconverse}, we provide a counterexample showing that for a positive rate $R>0$, the strong converse result does not hold in general for TAI with a privacy constraint.
\end{enumerate}

The organization of the paper is as follows. Basic notations are introduced in Section \ref{Notasection}. The problem formulation and associated definitions are given in Section \ref{Probform}. Main results are presented in Sections \ref{ghtpconst} to \ref{Strongconverse}. The proofs of the results are presented either in the Appendix or immediately after the statement of the result. Finally, Section \ref{sec:conclusion} concludes the paper with some open problems for future research.



\section{Preliminaries} \label{Prelims}
\subsection{Notations} \label{Notasection}
$\mathbb{N}$, $\mathbb{R}$ and $\mathbb{R}_{\geq 0}$  stand for the set of natural numbers, real numbers and non-negative real numbers, respectively.   For $a \in \mathbb{R}_{\geq 0}$, $[a] := \{i \in \mathbb{N}, ~i \leq a \}$ and for $a \in \mathbb{R}$,  $a^+:=\max\{0,a\}$ ($:=$ represents equality by definition). Calligraphic letters, e.g., $\mathcal{A}$, denotes sets, while $|\mathcal{A}|$ and $\mathcal{A}^c$ denotes its cardinality and complement, respectively. 
$\ind(\cdot)$ denotes the indicator function, while $O(\cdot)$, $o(\cdot)$ and $\Omega(\cdot)$ stands for the standard asymptotic notations of Big-O, Little-O and Big-$\Omega$, respectively. For a real sequence $\{a_n\}_{n \in \mathbb{N}}$ and $b \in \mathbb{R}$,  $a_n \xrightarrow{(n)} b$ represents $\lim_{n \rightarrow \infty}a_n=b$. Similar notations apply for asymptotic inequalities, e.g. $a_n \overset{(n)}{\geq} b$, means that $\lim_{n \rightarrow \infty}a_n \geq b$.  Throughout this paper, the base of the logarithms is taken to be $e$, and whenever the range of the summation is not specified, it means summation over the entire support, e.g., $\sum_{u} $ denotes $\sum_{u \in \Ucal} $. 

All the random variables (r.v.'s) considered in this paper are discrete with finite support unless specified otherwise.  We denote r.v.'s, their realizations and support by upper case, lower case and calligraphic letters (e.g., $X$, $x$ and $\X$), respectively. The joint probability distribution of r.v.'s $X$ and $Y$ is denoted by $P_{XY}$, while their marginals are denoted by $P_X$ and $P_Y$. The set of all probability distributions with support $\X$ and $\X \times \Y$ are represented by $\mathcal{P}(\X)$ and $\mathcal{P}(\X \times \Y)$, respectively. For $j,i \in \mathbb{N}$, the random vector $(X_i, \ldots,X_j)$, $j \geq i$, is denoted by $X_i^j$, while $X^j$ stands for $(X_1, \ldots,X_j)$. Similar notation holds for the vector of realizations. $X-Y-Z$ denotes a Markov chain relation between the r.v.'s $X$, $Y$ and $Z$. $\mathbb{P}_P(\mathcal{E})$  denotes the probability of event $\mathcal{E}$   with respect to the probability measure induced by distribution $P$, and $\mathbb{E}_P[\cdot]$ denotes the corresponding expectation. The subscript $P$ is omitted when the distribution involved is clear from the context. For two probability distributions $P$ and $Q$ defined on a common support, $P<<Q$ denotes that $P$ is absolutely continuous with respect to $Q$.  

Following the notation in \cite{Csiszar-Korner}, for $P_X \in \mathcal{P}(\X)$ and $\delta \geq 0$, the $P_X$-typical set is
\begin{align}
   \mathcal{T}_{[P_X]_{\delta}}^n:=\left\lbrace x^n \in \X^n:~\left|P_X(x')-\frac{1}{n}\sum_{i=1}^n\ind(x_i=x')\right| \leq \delta,~\forall~x' \in \X\right\rbrace, \notag
\end{align} 
and  the $P_X$-type class (set of sequences of type or empirical distribution $P_X$) is $\mathcal{T}_{P_X}^n:= \mathcal{T}_{[P_X]_{0}}^n$. The set of all possible types of sequences of length $n$ over an  alphabet $\mathcal{X}^n$ and the set of types in $\mathcal{T}_{[P_X]_{\delta}}^n$ are denoted by $\mathcal{P}^n(\X)$  and $\mathcal{P}_n\left(\mathcal{T}_{[P_X]_{\delta}}^n\right)$, respectively. Similar notations apply for pairs and larger combinations of r.v.'s, e.g., $\mathcal{T}_{[P_{XY}]_{\delta}}^n$, $\mathcal{T}_{P_{XY}}^n$, $\mathcal{P}^n(\X \times \Y)$ and $\mathcal{P}_n\left(\mathcal{T}_{[P_{XY}]_{\delta}}^n\right)$. The conditional $P_{Y|X}$ type class of a sequence $x^n \in \X^n$ is 
\begin{align}
\mathcal{T}_{P_{Y|X}}^n(x^n):=\big\{y^n:(x^n,y^n) \in \mathcal{T}_{P_{XY}}^n\big\}.
\end{align}
The standard information-theoretic quantities like Kullback-Leibler (KL) divergence between distributions $P_X$ and $Q_X$, the entropy of $X$ with distribution $P_X$, the conditional entropy of $X$ given $Y$ and the mutual information  between  $X$ and $Y$ with joint distribution $P_{XY}$, are denoted by
 $D(P_X||Q_X)$,  $H_{P_X}(X)$, $H_{P_{XY}}(X|Y)$ and $I_{P_{XY}}(X;Y)$, respectively. 
 When the distribution of the r.v.'s involved are clear from the context, the last three quantities are denoted simply by $H(X)$, $H(X|Y)$ and $I(X;Y)$, respectively.
Given  realizations $X^n=x^n$ and $Y^n=y^n$, $H_e(x^n|y^n)$ denotes the conditional empirical entropy given by
 \begin{align}
 H_e(y^n|x^n):= H_{P_{\tilde X \tilde Y}}(\tilde Y|\tilde X),
 \end{align}
 where $P_{\tilde X \tilde Y}$ denotes the joint type of $(x^n,y^n)$. Finally, the total variation between probability distributions $P_X$ and $Q_X$ defined on the same support $\X$ is 
\begin{align}
    ||P_X-Q_X||:=\frac 12  \sum_{x \in \X} |P_X(x)-Q_X(x)|. \notag
\end{align}
\begin{figure}[t]
\centering
\includegraphics[trim=0cm 0cm 0cm 0cm, clip, width= 0.6\textwidth]{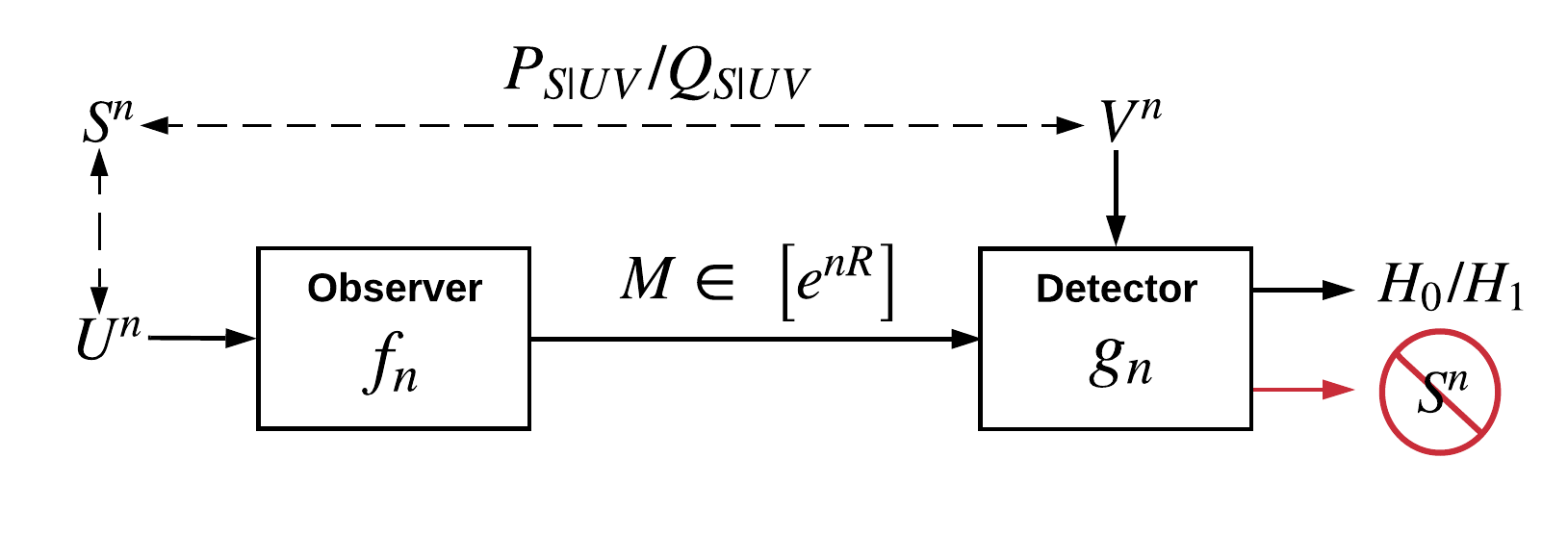}
\caption{DHT with a privacy constraint.} \label{htadversarial}
\end{figure}
\subsection{Problem formulation}\label{Probform}
 Consider the  HT setup illustrated in Fig. \ref{htadversarial}, where $(U^n,V^n,S^n)$ denote $n$ independent and identically distributed (i.i.d.) copies of triplet of r.v.'s $(U,V,S)$. The observer observes $U^n$ and sends  the message index $M$ to the detector over an error-free channel, where $M \sim f_n(\cdot|U^n)$ and $f_n: \Ucal^n  \rightarrow \mathcal{P}(\mathcal{M})$, $\mathcal{M}=[e^{nR}]$. 
Given its own observation $V^n$, the detector performs a HT on the joint distribution 
of $U^n$ and $V^n$ with null hypothesis
\begin{align*}
H_0: (U^n,V^n) \sim  \prod_{i=1}^n P_{UV},
\end{align*} and alternate hypothesis 
\begin{align*}
H_1: (U^n,V^n) \sim  \prod_{i=1}^n Q_{UV}.
\end{align*}
Let  $H$ and  $\hat H$ denote the r.v.'s corresponding to the true hypothesis and the output of the HT, respectively, with support $\mathcal{H}=\hat{\mathcal{H}}=\{0,1\}$, where $0$ denotes the null hypothesis and $1$ the alternate hypothesis. 
Let $g_n: \mathcal{M} \times \V^n \rightarrow  \mathcal{P}(\hat{\mathcal{H}})$ denote the decision rule at the detector, which outputs $\hat H \sim g_n(M,V^n)$. Then, the type I and type  II error probabilities achieved by a $\left(f_n, g_n\right)$ pair are given by
\begin{align}
 \alpha_n \left(f_n, g_n \right) := \mathbb{P}(\hat H=1|H=0)= P_{\hat H}(1), \notag
\end{align}
and 
\begin{align}
  \beta_n \left(f_n, g_n \right):= \mathbb{P}(\hat H=0|H=1)= Q_{\hat H}(0), \notag
\end{align}
 respectively, where
\begin{align}
  P_{\hat H}(1)&= \sum_{u^n,m,v^n} \left[ \prod_{i=1}^n P_{UV}(u_i,v_i) \right] ~f_n(m|u^n) ~g_n(1|m,v^n),  \notag 
 \end{align}
 and
 \begin{align}
 Q_{\hat H}(0)&= \sum_{u^n,m,v^n}\left[ \prod_{i=1}^n Q_{UV}(u_i,v_i)\right]~ f_n(m|u^n)~g_n(0|m,v^n). \notag
\end{align}

Let $P_{U^nV^nS^n M\hat H}$ and $Q_{U^nV^nS^n M\hat H}$ denote the joint distribution of $(U^n,V^n,S^n, M,\hat H)$ under the null and alternate hypotheses, respectively. For a given type I error probability constraint $\epsilon$, define the minimum type II error probability over all possible detectors as 
\begin{align}
&\bar \beta_n \left(f_n, \epsilon \right)  := \inf_{g_n}  \beta_n \left(f_n,~g_n \right), \label{deft2erroradv} \\
&\mbox{ such that } \alpha_n \left( f_n,~g_n \right) \leq  \epsilon. \notag 
\end{align}
The performance of HT is measured by the error-exponent achieved by the test for a given constraint $\epsilon$ on the type I error probability, i.e.,   $\liminf_{n \rightarrow \infty} -\frac{1}{n}\log \left( \bar \beta_n(f_n,\epsilon) \right)$. Although the goal of the detector is to maximize the error-exponent achieved for the  HT, it is also curious about the latent r.v. $S^n$ that is correlated with $U^n$. $S^n$ is referred to as the \textit{private} part of $U^n$, which is distributed i.i.d. according to the joint distribution $P_{SUV}$ and $Q_{SUV}$
under the null and alternate hypothesis, respectively. It is desired to keep the private part as concealed  as possible from the detector.  
We consider two measures of privacy for $S^n$ at the detector. The first is the \textit{equivocation} defined as $ H(S^n|M,V^n)$. The second one is the \textit{average distortion} between $S^n$ and its reconstruction $\hat S^n$ at the detector, measured according to an arbitrary bounded additive distortion metric $d: \mathcal{S} \times \hat{\mathcal{S}} \rightarrow [0, D_m]$ with multi-letter distortion defined as 
\begin{align}
d(s^n, \hat s^n) := \sum_{i=1}^n d(s_i, \hat s_i). \label{multlettdistconst}
\end{align}
\revised{We will assume the \textit{causal disclosure} assumption, i.e., $\hat S_i$ is a function of $S^{i-1}$ in addition to $(M,V^n)$.} The goal is to ensure that the error-exponent for HT is maximized, while satisfying the constraints on the type I error probability $\epsilon$ and the privacy of $S^n$. In the sequel, we study the trade-off between the rate, error-exponent (henceforth also referred to simply as the error exponent) and privacy achieved in the above setting. Before delving into that, a few definitions are in order.
\begin{definition} \label{regdiststr}
For a given type I error probability constraint $\epsilon$, a rate-error exponent-distortion tuple $(R, \kappa, \Delta_0, \Delta_1)$ is \emph{achievable}, if there exists a sequence of encoding and decoding functions $f_n: \Ucal^n  \rightarrow \mathcal{P}(\mathcal{M})$,  and $g_n:\mathcal{M} \times \V^n \rightarrow \mathcal{P}(\hat{\mathcal{H}})$ such that 
\begin{align}  
\liminf_{n \rightarrow \infty} \frac{-\log \left( \bar \beta_n(f_n,\epsilon) \right)}{n} & \geq \kappa, \label{t2eedef} 
\end{align}
and for any $\gamma>0$, there exists an $n_0 \in \mathbb{N}$ such that
\begin{align}  
 \inf_{\big\{g^{(r)}_{i,n}\big\}_{i=1}^n} \mathbb{E}\left[d \left(S^{n}, \hat S^{n} \right) |H=j\right] & \geq  n \Delta_j-\gamma,~ \forall~n \geq n_0,  ~j=0,1, \label{distexpstr}
\end{align}
where  $\hat S_i \sim g^{(r)}_{i,n}(\cdot|M,V^n,S^{i-1})$, and 
$g^{(r)}_{i,n}: [e^{nR}]  \times \V^n \times \mathcal{S}^{i-1}   \rightarrow \mathcal{P}(\hat{\mathcal{S}}_i)$ denotes an arbitrary stochastic reconstruction map at the detector.
The rate-error exponent-distortion region $\mathcal{R}_d(\epsilon)$ is the closure of the set of all such achievable $(R, \kappa, \Delta_0, \Delta_1)$ tuples for a given $\epsilon$.
\end{definition}
\begin{definition}\label{regequstr}
For a given type I error probability constraint $\epsilon$, a rate-error exponent-equivocation\footnote{It is well-known that equivocation as a privacy measure is a special case of average distortion  under the causal disclosure assumption and log-loss distortion metric \cite{Schieler-Cuff-2014}. However, we provide a separate definition of the rate-error exponent-equivocation  region for completeness.} $(R, \kappa, \Lambda_0, \Lambda_1)$ tuple is \emph{achievable}, if there exists a sequence of encoding and decoding functions $f_n: \Ucal^n  \rightarrow \mathcal{P}(\mathcal{M})$ and $g_n:[e^{nR}] \times \V^n \rightarrow \mathcal{P}(\hat{\mathcal{H}})$ such that  \eqref{t2eedef} is satisfied, and for any $\gamma>0$, there exists a $n_0 \in \mathbb{N}$ such that
\begin{align}  
     H(S^{n}|M,V^{n},H=i) \geq n \Lambda_i-\gamma,~\forall~n \geq n_0,~i \in \{0,1\}. \label{strongprivconstequ}
 \end{align}
The rate-error exponent-equivocation region $\mathcal{R}_e(\epsilon)$ is the closure of the set of all such achievable $(R, \kappa, \Lambda_0,\Lambda_1)$ tuples for a given $\epsilon$.
\end{definition}
Note that the privacy measures considered in \eqref{distexpstr} and \eqref{strongprivconstequ} are stronger than 
\begin{align}
 &\liminf_{n \rightarrow \infty} \inf_{\big\{g^{(r)}_{i,n}\big\}_{i=1}^n} \mathbb{E}\left[\frac 1n d \left(S^n, \hat S^n \right) |H=i\right] \geq \Delta_i, ~i=0,1, \label{weakdistconst} \\
 \mbox{and }& \liminf_{n \rightarrow \infty}  \frac 1n H(S^n|M,V^n,H=i) \geq \Lambda_i,~i=0,1, \label{weakeqconst}
\end{align}
respectively. To see this for the equivocation privacy measure, note that if $ H(S^{n}|M,V^{n},H=i) = n \Lambda^*_i-n^a$, $i=0,1$, for some $a \in (0,1)$, then an equivocation pair $(\Lambda^*_0,\Lambda^*_1)$ is achievable under the constraint given in \eqref{weakeqconst}, while it is not achievable under the constraint given in \eqref{strongprivconstequ}. 

\subsection{Relation to Previous Work} \label{relprevwork}
Before stating our results, we briefly highlight the differences between our system model and the ones studied in  \cite{GSSV-2018} and \cite{MP-2015}. In \cite{GSSV-2018}, the observer applies a privacy mechanism to the data before releasing it to the transmitter, which performs further encoding prior to transmission to the detector. More specifically, the observer checks if $U^n \in \mathcal{T}^n_{[P_U]_{\delta}}$ and if successful, sends the output of a memoryless privacy mechanism  applied to $U^n$, to the transmitter. Otherwise, it outputs a  $n$-length zero-sequence. The privacy mechanism plays the role of randomizing the data (or adding noise) in order to achieve the desired privacy. 
Such randomized privacy mechanisms are popular in privacy studies, and   have been used in \cite{Liao-Lal-Tan-Cal}\cite{Jiao-Lal-Cal} \cite{WBI-2017}. In our model, the tasks of coding for privacy and compression are done jointly by utilizing all the available data samples $U^n$. Also, while we consider the equivocation (and average distortion) between the revealed information and the private part as the privacy measure, in \cite{GSSV-2018}, the mutual information between the  observer's observations and the output of the memoryless mechanism is the privacy measure. \revised{As a result of these differences, there exist some points in the rate-error exponent-privacy trade-off that are achievable in our model, but not in \cite{GSSV-2018}. For instance, a perfect privacy condition $\Lambda_0=0$ for testing against independence in \cite[Theorem 2]{GSSV-2018} would imply that the error-exponent is also zero, since the output of the memoryless mechanism has to be independent of the observer's observations (under both hypotheses). However, as we later show in Example \ref{perfprivex}, a positive error-exponent is achievable while guaranteeing perfect privacy in our model.}

\revised{On the other hand, the difference between our model and  \cite{MP-2015} arises from the difference in the privacy constraint as well as the privacy measure. Specifically, the goal in \cite{MP-2015} is to keep $U^n$ private from an illegitimate eavesdropper,  while the objective here is to keep a r.v. $S^n$ that is correlated with $U^n$ private from the detector. Also,  we consider the more general average distortion (under causal disclosure) as a privacy measure, in addition to equivocation in \cite{MP-2015}. Moreover, as already noted, the equivocation privacy constraint in \eqref{strongprivconstequ} is more stringent than \eqref{weakeqconst} that is considered in \cite{MP-2015}. To satisfy the distortion  requirement or the stronger equivocation privacy constraint in \eqref{strongprivconstequ}, we require that the aposteriori probability distribution of $S^n$ given the observations $(M,V^n)$ at the detector is close in some sense to a desired ``target" memoryless distribution. To achieve this, we use a stochastic encoding scheme to induce the necessary randomness for $S^n$ at the detector, which, to the best of our knowledge, has not been considered previously in the context of DHT. Consequently, the analysis of the type I and type II error-probabilities and privacy achieved are novel.  Another subtle yet important difference is that the marginal distributions of $U^n$ and the side-information at the eavesdropper are assumed to be the same under the null and alternate hypotheses in \cite{MP-2015}, which is not the case here. This necessitates separate analysis for the privacy achieved under the two hypotheses. }

Next, we state some supporting results that will be useful later for proving the main results.
\subsection{Supporting Results}
Let  
\begin{align}
g_{\mathcal{A}_{n}}^{(d)}( m,v^n)= \ind \left( (m,v^n) \in \mathcal{A}_{n}^c \right) \label{determdet}
\end{align}
 denote a deterministic detector with acceptance region  $\mathcal{A}_{n} \subseteq [e^{nR}]\times \mathcal{V}^n$   for $H_0$ and $\mathcal{A}_{n}^c$ for  $H_1$. Then, the type I and type II error probabilities are given by
\begin{align}
     \alpha_n \left(f_n, g_n \right) &:= P_{M V^{n}}(\mathcal{A}_{n}^c)= \mathbb{E}_P \big[ \ind \left((M, V^n) \in \mathcal{A}_{n}^c \right)\big], \label{type1det} \\
     \beta_n \left(f_n, g_n \right) &:= Q_{M V^n} (\mathcal{A}_{n})= \mathbb{E}_Q \big[ \ind \left((M, V^n) \in \mathcal{A}_{n} \right)\big]. \label{type2det}
\end{align}
\begin{lemma} \label{lem:determdet}
Any error-exponent that is achievable is also achievable by a deterministic detector
of the form given in \eqref{determdet}
 for some $\mathcal{A}_{n} \subseteq [e^{nR}]\times \mathcal{V}^n$, where $\mathcal{A}_{n} $ and $\mathcal{A}_{n}^c$ denote the acceptance regions for $H_0$ and $H_1$, respectively.
\end{lemma}
The proof of Lemma \ref{lem:determdet} is given in Appendix \ref{prooflemmdetrmdet} for completeness. Due to Lemma \ref{lem:determdet},  henceforth we restrict our attention to a deterministic $g_n$ as given in \eqref{determdet}. 

The next result shows that without  loss of generality (w.l.o.g), it is also sufficient to consider $g^{(r)}_{i,n}$ (in Definition \ref{regdiststr}) to be a deterministic function of the form 
\begin{align}
g^{(r)}_{i,n}=\{\bar{\phi}_{i,n}(\cdot,\cdot,\cdot)\}_{i=1}^n \label{deterdecadv}
\end{align}
for the minimization in \eqref{distexpstr}, where  $\bar{\phi}_{i,n}: \mathcal{M} \times \V^n \times \mathcal{S}^{i-1} \rightarrow \hat {\mathcal{S}}$, $i \in [n]$, denotes an arbitrary deterministic function.

\begin{lemma}\label{determfnatadv}
The infimum in  \eqref{distexpstr} is achieved by a deterministic function $g^{(r)}_{i,n}$ as given in \eqref{deterdecadv}, and hence it is sufficient to restrict to such deterministic $g^{(r)}_{i,n}$  in \eqref{distexpstr}.
\end{lemma}
The proof of Lemma \ref{determfnatadv} is given in Appendix \ref{determadvfnproof}. Next, we state some lemmas that will be handy for upper bounding the amount of privacy leakage in the proofs of the main results stated below. The following one is a well known result proved in \cite{Csiszar-Korner} that upper bounds the difference in entropy of two r.v.'s (with a common support) in terms of the  total variation distance between their probability distributions. 
\begin{lemma} \cite[Lemma 2.7]{Csiszar-Korner} \label{contentrtv}
Let $P_X$ and $Q_X$ be distributions defined on a common support $\X$ and let $\rho := ||P_X-Q_X||$.  Then, for $\rho \leq \frac{1}{4}$
\begin{align}
  |H_{P_X}-H_{Q_X}| \leq -2\rho \log\left(\frac{2 \rho}{|\X|}\right).  \notag
\end{align}
\end{lemma}
The next lemma will be handy in proving Theorem \ref{thm:genhtequ}, Theorem \ref{thm:genhtdist}, Proposition \ref{thm:zerorategenhtdist} and the counter-example for strong converse presented in Section \ref{Strongconverse}.  
\begin{lemma} \label{approxdistmequzero}
Let $(X^n,Y^n)$ denote $n$ i.i.d. copies of r.v.'s $(X,Y)$, and 
$P_{X^nY^n}=\prod_{i=1}^n P_{XY}$ and $Q_{X^nY^n}=\prod_{i=1}^n Q_{XY}$ denote two joint probability distributions on $(X^n,Y^n)$. For $\delta>0$, define 
\begin{align}
 \Pi(x^n, \delta,P_X) := \ind \left(x^n \notin \mathcal{T}_{[P_X]_{\delta}}^n \right). \label{nottypical}
\end{align}
 If $P_X \neq Q_X$, then for $\delta>0$ sufficiently small, there exists $\bar \delta>0$ and $n_0(\delta,|\X|,|\Y|) \in \mathbb{N}$ such that for all $n \geq n_0(\delta,|\X|,|\Y|)$,
\begin{align}
\Vert Q_{Y^n}(\cdot)-Q_{Y^n|\Pi(X^n, \delta,P_X)}(\cdot|1) \Vert  \leq e^{-n \bar \delta}. \label{althypapproxdist} 
\end{align}
If $P_X = Q_X$, then for any $\delta>0$, there exists $\bar \delta>0$ and $n_0(\delta,|\X|,|\Y|) \in \mathbb{N}$ such that for all $n \geq n_0(\delta,|\X|,|\Y|)$,
\begin{align}
\Vert Q_{Y^n}(\cdot)-Q_{Y^n|\Pi(X^n, \delta,P_X)}(\cdot|0) \Vert & \leq  e^{-n \bar \delta}, \label{althypapproxdist2} 
\end{align}
Also, for any $\delta>0$, there exists $\bar \delta>0$ and $n_0(\delta,|\X|,|\Y|) \in \mathbb{N}$ such that for all $n \geq n_0(\delta,|\X|,|\Y|)$,
\begin{align}
  \Vert P_{Y^n}(\cdot)-P_{Y^n|\Pi(X^n, \delta,P_X)} (\cdot|0)\Vert &  \leq e^{-n \bar \delta}. \label{nullhypapproxdist}
\end{align}
\end{lemma}
\begin{IEEEproof}
The proof is presented in Appendix \ref{supplem}.
\end{IEEEproof}

In the next section, we  establish an inner bound on $\mathcal{R}_e(\epsilon)$ and $\mathcal{R}_d(\epsilon)$.
\section{MAIN RESULTS} \label{ghtpconst}
The following two theorems are the main results of this paper providing inner bounds for $\mathcal{R}_e(\epsilon)$ and $\mathcal{R}_d(\epsilon)$, respectively.
\begin{theorem} \label{thm:genhtequ}
For $\epsilon \in (0,1)$, $(R,\kappa, \Lambda_0, \Lambda_1) \in \mathcal{R}_e(\epsilon)$ if there exists an auxiliary r.v. $W$, such that $(V,S)-U-W$, and 
\begin{align}
R &\geq I_{P}(W;U|V), \label{rtconsteqgen} \\
 \kappa &\leq \kappa^*(P_{W|U},R), \label{expconsteqgen} \\
  \Lambda_0 &\leq H_{P}(S|W,V), \label{distconsteq0gen} \\
  \Lambda_1 &\leq \ind \left( P_U=Q_U \right) ~ H_{Q}(S|W,V)+\ind \left( P_U \neq Q_U \right)~ H_{Q}(S|V), \label{distconsteq1gen}
\end{align}
where
\begin{align}
\kappa^*(P_{W|U},R) & := \min \left(E_1(P_{W|U}),~ E_2(R, P_{W|U}) \right), \notag \\
E_1(P_{W|U})&:= \min_{P_{\tilde U \tilde V  \tilde W } \in \mathcal{L}_1(P_{UW}, P_{VW}) } D(P_{\tilde U \tilde V  \tilde W} || Q_{UV}P_{W|U}), \label{expterm1} \\
E_2(R, P_{W|U}) &:= 
\begin{cases}  \label{expterm2}
\substack{ \min \\ P_{\tilde U \tilde V    \tilde W} \in \mathcal{L}_2(P_{UW}, P_{V}) }~ D(P_{\tilde U \tilde V   \tilde W} || Q_{UV}P_{W|U})+(R-I_{P}(U;W|V)), & \mbox{if } I_{P}(U;W)>R, \\   \infty, & \mbox{otherwise,} 
 \end{cases}\\
\mathcal{L}_1(P_{UW}, P_{VW})&:= \{P_{\tilde U \tilde V   \tilde W} \in \mathcal{P}(\Ucal \times \V \times\W) : P_{\tilde U \tilde W}=P_{UW},~ P_{\tilde V    \tilde W}=P_{VW} \}, \notag \\
\mathcal{L}_2(P_{UW}, P_{V}) &:= \{P_{\tilde U \tilde V   \tilde W} \in \mathcal{P}(\Ucal \times \V \times\W) : P_{\tilde U \tilde W}=P_{UW},~ P_{\tilde V   }=P_{V}, ~  H_{P}(W|V) \leq H(\tilde W| \tilde V) \}, \notag \\
P_{SUVW}&:= P_{SUV} P_{W|U}, ~\mbox{ and } Q_{SUVW}:= Q_{SUV} P_{W|U}. \notag
\end{align}
\end{theorem}
\begin{theorem} \label{thm:genhtdist}
For a given bounded additive distortion measure $d(\cdot,\cdot)$ and  $\epsilon \in (0,1)$, $(R,\kappa, \Delta_0, \Delta_1) \in \mathcal{R}_d(\epsilon)$ if there exist an auxiliary r.v. $W$ and deterministic functions $\phi: \W \times \V \rightarrow \hat S$ and $\phi': \V \rightarrow \hat S$, such that $(V,S)-U-W$ and \eqref{rtconsteqgen}, \eqref{expconsteqgen}, 
\begin{align}
 \Delta_0 &\leq  \min_{\phi(\cdot, \cdot)} \mathbb{E}_P\left[d \left(S, \phi(W,V)\right)\right], \label{distconst0} \\
 \mbox{and } \Delta_1 &\leq \ind \left( P_U=Q_U \right)~ \min_{\phi(\cdot, \cdot)}\mathbb{E}_{Q}\left[d \left(S, \phi(W,V)\right)\right]+\ind \left( P_U \neq Q_U \right)~\min_{\phi'(\cdot)} \mathbb{E}_Q \left[d \left(S, \phi'(V)\right)\right], \label{distconst1}
\end{align}
are satisfied, where  $P_{SUVW}$ and $Q_{SUVW}$ are as defined in Theorem \ref{thm:genhtequ}.
\end{theorem}
The proof of Theorem \ref{thm:genhtequ} and Theorem \ref{thm:genhtdist} is given in Apppendix \ref{genHTdist}.
While the rate-error exponent trade-off in Theorem \ref{thm:genhtequ} and Theorem \ref{thm:genhtdist} is the same as that achieved by the Shimokawa-Han-Amari (SHA) scheme \cite{Shimokawa}, the coding strategy achieving it is different due to the requirement of the privacy constraint.
As mentioned above, in order to obtain a single-letter lower bound for the achievable distortion (and achievable equivocation) of the private part at the detector, it is required that the aposteriori probability distribution of $S^n$ given the observations $(M,V^n)$ at the detector is close in some sense to a desired ``target" memoryless distribution. For this purpose, we use the so-called likelihood encoder \cite{Cuff-2013,Song-Cuff-Poor-2016} (at the observer) in our achievability scheme. The likelihood encoder is a stochastic encoder that induces the necessary randomness for $S^n$ at the detector, and  to the best of our knowledge has not been used before in the context of DHT.  The analysis of the type I and type II error probabilities and the privacy achieved by our scheme is novel and 
involves the application of the well-known \textit{channel resolvability} or \textit{soft-covering lemma} \cite{Wyner-1975-comminfo, Han-Verdu-1993,Cuff-2013}. Properties of the total variation distance between probability distributions mentioned in \cite{Schieler-Cuff-2014}  play a key role in this analysis. The analysis also reveals the interesting fact that the coding schemes in Theorem \ref{thm:genhtequ} and Theorem \ref{thm:genhtdist}, although quite different from the SHA scheme, achieves the same lower bound on the error-exponent. 

Theorems \ref{thm:genhtequ} and \ref{thm:genhtdist} provide single-letter inner bounds on $\mathcal{R}_d(\epsilon)$ and $\mathcal{R}_e(\epsilon)$, respectively. A complete computable characterization of these regions would require a matching converse. This is a hard problem, since such a characterization is not available even for the DHT problem  without a privacy constraint, in general (see \cite{Ahlswede-Csiszar}). However, it is known that a single-letter characterization of the rate-error exponent region exists for the special case of TACI \cite{Rahman-Wagner}. In the next section, we show that TACI with a privacy constraint also admits a single-letter characterization, in addition to other optimality results. 

\section{Optimality Results for Special cases} \label{optresultsspcl}
\subsection{ TACI  with a Privacy Constraint} \label{TACIspclcase}
Assume that the detector observes two discrete memoryless sources $Y^n$ and $Z^n$, i.e.,  $V^n=(Y^n,Z^n)$.
In TACI, the detector tests for the conditional independence of $U$ and $Y$, given $Z$.
Thus, the joint distribution of the r.v.'s under the null and alternate hypothesis are given by
\begin{subequations} \label{tacinullandalt}
\begin{equation} \label{tacinullhypdis}
    H_0:~ P_{SUYZ}:= P_{S|UYZ} P_{U|Z} P_{Y|UZ} P_{Z}, 
\end{equation}
\mbox{and}
\begin{equation} \label{tacialthypdis}
   H_1:~Q_{SUYZ} := Q_{S|UYZ} P_{U|Z}P_{Y|Z} P_Z, 
\end{equation}
\end{subequations}
respectively.

Let $ \mathcal{R}_e$ and $ \mathcal{R}_d$ denote the rate-error exponent-equivocation and rate-error exponent-distortion regions, respectively, for the case of vanishing type I error probability constraint, i.e.,
 \begin{align}
 \mathcal{R}_e &:= \lim_{\epsilon \rightarrow 0}\mathcal{R}_e(\epsilon) \mbox{ and }\mathcal{R}_d := \lim_{\epsilon \rightarrow 0}\mathcal{R}_d(\epsilon). \notag 
  \end{align}
 Assume that the privacy constraint under the alternate hypothesis is inactive.  
 Thus, we are interested in characterizing the set of all tuples $(R, \kappa, \Lambda_0, \Lambda_1) \in \mathcal{R}_e $ and $(R, \kappa, \Delta_0, \Delta_1)\in \mathcal{R}_d$, where
 \begin{align}
    \Lambda_1 &\leq \Lambda_{min}:=H_{Q}(S|U,Y,Z), \notag \\
    \mbox{ and }\Delta_1 &\leq \Delta_{min}:= \min_{\phi(u, y,z)} \mathbb{E}_{Q}\left[d \left(S, \phi(U,Y,Z)\right)\right].
 \end{align}
  Note that $\Lambda_{min}$ and $\Delta_{min}$ correspond to the equivocation and average distortion of $S^n$ at the detector, respectively, when $U^n$ is  available directly at the detector under the alternate hypothesis.
 The above assumption is motivated by scenarios, in which, the observer is more eager to protect $S^n$ when there is a correlation between its own observation and that of the detector. Consider the following example of user privacy in the context of online shopping, in which the observer and detector correspond to a consumer and an online shopping portal, respectively. A consumer would like to share some information about his/her shopping behaviour, e.g., shopping history and preferences, with the shopping portal in order to get better deals and recommendations on relevant products. The shopping portal would like to determine whether the consumer belongs to its target age group (e.g., below 30 years old) before sending special offers to this customer. Assuming that the shopping patterns of the users within and outside the target age groups are independent,  the shopping portal  performs an independence test to check if the consumer's shared data is correlated with the data of its own customers.  If the consumer is indeed within the target age group, the shopping portal would like to gather more information about this potential customer, particular interests, more accurate age estimation, etc.; while the user is reluctant to provide any further information. In this example, $U^n$, $S^n$ and $Y^n$ corresponds to shopping behaviour, more information about the customer, and customers data available to the shopping portal, respectively. 
 
 For the above mentioned case, we have the following results.
\begin{prop}\label{equivocation-noiseless}
For the HT given in \eqref{tacinullandalt}, $(R,\kappa, \Lambda_0, \Lambda_{min}) \in \mathcal{R}_e$ if and only if there exists an auxiliary r.v. $W$, such that  $(Z,Y,S)-U-W$, and 
\begin{align}
\kappa &\leq I_P(W;Y|Z), \label{expconsteq}\\
 R &\geq I_P(W;U|Z), \label{rtconsteq} \\
 \Lambda_0 &\leq H_{P}(S|W,Z,Y),
 \label{distconsteq0}
\end{align}
for some joint distribution of the form $P_{SUYZW}:= P_{SUYZ} P_{W|U} $. 
\end{prop}
\begin{IEEEproof}
For TACI, the inner bound in Theorem \ref{thm:genhtequ} yields that for $\epsilon \in (0,1)$, $(R,\kappa, \Lambda_0, \Lambda_1) \in \mathcal{R}_e(\epsilon)$ if there exists an auxiliary r.v. $W$, such that  $(Y,Z,S)-U-W$, and 
\begin{align}
R &\geq I_{P}(W;U|Y,Z), \label{rtconsteqgensp} \\
 \kappa &\leq \kappa^*(P_{W|U},R), \label{expconsteqgensp} \\
 \Lambda_0 &\leq H_{P}(S|W,Y,Z), \label{distconsteq0gensp} \\
  \Lambda_1 &\leq  H_{Q}(S|W,Y,Z), \label{distconsteq1gensp}
\end{align}
where
\begin{align}
\kappa^*(P_{W|U},R) & := \min \left(E_1(P_{W|U}),~ E_2(R, P_{W|U}) \right), \notag\\
E_1(P_{W|U})&:= \min_{P_{\tilde U \tilde Y  \tilde Z \tilde W } \in \mathcal{L}_1(P_{UW}, P_{YZW}) } D(P_{\tilde U \tilde Y  \tilde Z  \tilde W} || Q_{UYZ}P_{W|U}), \label{expterm1sp}  \\
E_2(R, P_{W|U}) &:= 
\begin{cases}  \label{expterm2sp}
\substack{ \min \\ P_{\tilde U \tilde Y  \tilde Z \tilde W} \in \mathcal{L}_2(P_{UW}, P_{YZ}) }~ D(P_{\tilde U \tilde Y  \tilde Z   \tilde W} || Q_{UYZ}P_{W|U})+(R-I_{P}(U;W|Y,Z)), & \mbox{if } I_{P}(U;W)>R, \\   \infty, & \mbox{ otherwise}, 
 \end{cases}\\
\mathcal{L}_1(P_{UW}, P_{YZW})&:= \{P_{\tilde U \tilde Y  \tilde Z  \tilde W} \in \mathcal{P}(\Ucal \times \Y \times \Z \times\W) : P_{\tilde U \tilde W}=P_{UW},~ P_{\tilde Y \tilde Z    \tilde W}=P_{YZW} \}, \notag \\
\mathcal{L}_2(P_{UW}, P_{YZ}) &:= \{P_{\tilde U \tilde Y  \tilde Z   \tilde W} \in \mathcal{P}(\Ucal \times \Y  \times \Z \times\W) : P_{\tilde U \tilde W}=P_{UW},~ P_{\tilde Y  \tilde Z   }=P_{YZ}, ~  H_{P}(W|Y,Z) \leq H(\tilde W| \tilde Y  \tilde Z) \}, \notag \\
P_{SUYZW}&:= P_{SUYZ} P_{W|U}, ~ Q_{SUYZW}:= Q_{S|YZ}P_{U|Z} P_{Y|Z} P_Z P_{W|U}. \notag
\end{align}
Note that since $(Y,Z,S)-U-W$, we have
\begin{align}
  I_{P}(W;U) \geq I_{P}(W;U|Y,Z). \label{binnnotnec}
\end{align}
Let  $\mathcal{B}'  := \{P_{W|U}: I_{P}(U;W|Z) \leq R\}$.
Then, for $P_{W|U} \in \mathcal{B}' $, we have,
\begin{align}
E_1(R,P_{W|U})&=  \min_{P_{\tilde U \tilde Y  \tilde Z \tilde W} \in \mathcal{L}_1(P_{UW}, P_{YZW}) } D(P_{\tilde  U \tilde Y  \tilde Z \tilde W } || Q_{UYZ}P_{W|U}) = I_{P}(Y;W|Z), \notag  \\
E_2(R,P_{W|U})& \geq I_{P}(U;W|Z)-I_{P}(U;W|Y,Z)= I_{P}(Y;W|Z). \notag 
\end{align}
Hence,
\begin{align}
    \kappa^*(P_{W|U},R) \geq I_{P}(Y;W|Z). \label{expexampel}
\end{align}
By noting that $\Lambda_{min} \leq H_{Q}(S|W,Y,Z) $ (by the data processing inequality), we have shown that for $\Lambda_1 \leq \Lambda_{min}$, $(R, \kappa,\Lambda_0, \Lambda_1) \in \mathcal{R}_e$ if \eqref{expconsteq}-\eqref{distconsteq0} are satisfied. This completes the proof of achievability.

\textit{Converse:}
Let $(R, \kappa, \Lambda_0,\Lambda_1) \in \mathcal{R}_e$. Let $T$ be a r.v. uniformly distributed over $[n]$ and independent of all the other r.v.'s $(U^n,Y^n,Z^n,S^n,M)$. Define an auxiliary r.v. $W := (W_{T},T)$, where \revised{$W_{i} := ( M,Y^{i-1},S^{i-1},Z^{i-1},Z_{i+1}^n)$}, $i \in [n]$. Then,  we have for sufficiently large $n$ that
\begin{flalign}
nR &\geq  H_P(M) \geq H_P(M|Z^n) \geq  I_P(M;U^n|Z^n) \notag \\
&=\sum\nolimits_{i=1}^n I_P(M;U_{i}|U^{i-1},Z^n) \notag \\
&= \sum\nolimits_{i=1}^n I_P(M,U^{i-1},Z^{i-1},Z_{i+1}^n;U_i|Z_i) \label{memorylessprop} \\
&\revised{= \sum\nolimits_{i=1}^n I_P(M,U^{i-1},Z^{i-1},Z_{i+1}^n,Y^{i-1},S^{i-1};U_i|Z_i)} \label{markprophyp} \\
& \revised{\geq  \sum\nolimits_{i=1}^n I_P(M,Z^{i-1},Z_{i+1}^n,Y^{i-1},S^{i-1};U_i|Z_i)} \notag \\
& =  \sum\nolimits_{i=1}^n I_P(W_{i};U_i|Z_i) = nI_P(W_T;U_T|Z_T,T) \notag  \\
&=n I_P(W_T,T;U_T|Z_T) \label{auxtimeshar} &&
\end{flalign}
\begin{flalign}
&= n I_P(W;U|Z). \label{rtchnbnd} &&
\end{flalign}
Here, \eqref{memorylessprop} follows since the sequences $(U^n,Z^n)$ are memoryless; 
  \eqref{markprophyp} follows since \revised{$(Y^{i-1},S^{i-1})-(M,U^{i-1},Z^n)-U_i$} form a Markov chain; and, \eqref{auxtimeshar} follows from the fact that $T$ is independent of all the other r.v.'s.
  
The equivocation of $S^n$ under the null hypothesis can be bounded as follows.
 \begin{flalign}
 H(S^n|M, Y^n,Z^n, H=0) &=  \sum\nolimits_{i=1}^n H(S_i|M,S^{i-1},Y^n,Z^n, H=0) \notag \\
 & \revised{\leq  \sum\nolimits_{i=1}^n H(S_i|M,Y^{i-1},S^{i-1},  Z^{i-1},Z_{i+1}^n,Y_i,Z_i, H=0)} \label{condredentr}  \\
 & = \sum\nolimits_{i=1}^n H(S_i|W_i, Y_i, Z_i, H=0)\notag \\
 &= n H(S_T|W_T,Y_T,Z_T,T, H=0) \notag \\
 &=n H_{P}(S|W,Y,Z), \label{eqvocbnd} &&
\end{flalign}  
where $ P_{SUYZW}=P_{SUYZ}P_{W|U}$ for some  conditional distribution $P_{W|U}$. In \eqref{condredentr}, we used the fact that conditioning reduces entropy.

Finally, we prove the upper bound on $\kappa$. For any encoding function $f_n$ and decision region $\mathcal{A}_n \subseteq \mathcal{M} \times \Y^n \times \Z^n$ for $H_0$ such that $\epsilon_n \rightarrow 0$, we have,
\begin{flalign}
D\left(P_{MY^nZ^n}||Q_{MY^nZ^n}\right) &\geq P_{MY^nZ^n}(\mathcal{A}_n) \log\left(\frac{P_{MY^nZ^n}(\mathcal{A}_n)}{Q_{MY^nZ^n}(\mathcal{A}_n)} \right) + P_{MY^nZ^n}(\mathcal{A}_n^c) \log\left(\frac{P_{MY^nZ^n}(\mathcal{A}_n^c)}{Q_{MY^nZ^n}(\mathcal{A}_n^c)} \right) \label{dpikldiverg}\\
& \geq -H(\epsilon_n)-(1-\epsilon_n)\log \left(\bar \beta_n\left(f_n,\epsilon_n\right)\right). \notag&&
\end{flalign}
Here, \eqref{dpikldiverg} follows from the log-sum inequality \cite{Csiszar-Korner}. Thus,
\begin{flalign}
\limsup_{n \rightarrow \infty}\frac{-\log\left(\bar \beta_n\left(f_n,\epsilon_n\right)\right)}{n}&  \leq \limsup_{n \rightarrow \infty} \frac 1n D\left(P_{MY^nZ^n}|| Q_{MY^nZ^n}\right) \notag \\
& = \limsup_{n \rightarrow \infty} ~ \frac 1n I_P(M;Y^n|Z^n) \label{kldivtomutinf}\\
&=H_P(Y|Z)-\liminf_{n \rightarrow \infty} \frac 1n H_P(Y^n|M,Z^n), \label{finalbndexpsing}&&
\end{flalign}
where \eqref{kldivtomutinf} follows since $Q_{MY^nZ^n}= P_{MZ^n}P_{Y^n|Z^n}$.
The last term can be single-letterized as follows:
\begin{flalign}
 H_P(Y^n|M,Z^n) &=\sum\nolimits_{i=1}^n H_P(Y_i|Y^{i-1}, M,Z^n) \notag \\
 &\revised{ \geq \sum\nolimits_{i=1}^n H_P(Y_i|Y^{i-1},S^{i-1}, M,Z^n)} \notag \\
 &=\sum\nolimits_{i=1}^n  H_P(Y_i|Z_i,W_{i})  \notag \\
 &= n H_P(Y_T|Z_T,W_{T},T) \notag \\
 &=n H_P(Y|Z,W). \label{rtbndsrc}&&
\end{flalign}
Substituting \eqref{rtbndsrc} in \eqref{finalbndexpsing}, we obtain
\begin{align}
    \kappa \leq I_P(Y;W|Z).
\end{align}
 Also, note that $(Z,Y)-U-W$ holds. To see this, note that  $(U_i,Y_i,Z_i,S_i)$ are i.i.d across  $i \in [n]$. Hence, any information in $W_i$ on $(Y_i,Z_i,S_i)$ is only through $M$ as a function of $U_i$, and so given $U_i$, $W_i$ is independent of $(Y_i,Z_i,S_i)$. The above Markov chain then follows from the fact that $T$ is independent of $(U^n,Y^n,Z^n,S^n,M)$. This completes the proof of the converse and the theorem.
\end{IEEEproof}

Next, we state the result for TACI with a distortion privacy constraint, where  the distortion is measured using an arbitrary distortion measure $d(\cdot, \cdot)$. Let $\Delta_{min}:= \min_{\phi(u, y,z)} \mathbb{E}_{Q}\left[d \left(S, \phi(U,Y,Z)\right)\right]$.

\begin{prop} \label{distortion-noiseless}
For the HT given in \eqref{tacinullandalt}, $(R,\kappa, \Delta_0, \Delta_{min}) \in \mathcal{R}_d$ if and only if there exist an auxiliary  r.v. $W$ and a deterministic function $\phi: \W \times \Y \times \Z \rightarrow \hat{\mathcal{S}}$ such that 
\begin{align}
R & \geq  I_P(W;U|Z), \label{eq:ratedist} \\
 \kappa  &\leq I_P(W;Y|Z), \label{eq:expdist} \\
\Delta_0 &\leq  \min_{\phi(\cdot,\cdot,\cdot)} \mathbb{E}_{P}\left[d \left(S, \phi(W,Y,Z)\right)\right],\label{eq:distconstnull}
 \end{align}
for some $P_{SUYZW}$ 
 as defined in Proposition \ref{equivocation-noiseless}.
\end{prop}



\begin{IEEEproof}
The proof of achievability follows from Theorem \ref{thm:genhtdist}, similarly to the way Proposition \ref{equivocation-noiseless} is obtained from Theorem \ref{thm:genhtequ}. Hence, only differences will be highlighted.  Similar to the inequality $\Lambda_{min} \leq H_{Q}(S|U,Y,Z)$ in the proof of Proposition \ref{equivocation-noiseless}, we need to prove the inequality $ \Delta_{min} \leq  \mathbb{E}_{Q}\left[d \left(S, \phi(W,Y,Z)\right)\right]$, where $Q_{SUYZW}:=Q_{SUYZ}P_{W|U}$ for some conditional distribution $P_{W|U}$. This  can be shown as follows:
\begin{flalign}
  & \min_{\phi(\cdot, \cdot,\cdot)}  \mathbb{E}_{Q}\left[d \left(S, \phi(W,Y,Z)\right)\right] \notag \\
  &= \sum_{u,y,z}Q_{UYZ}(u,y,z)~\sum_{w}P_{W|U}(w|u)~\min_{\phi(w,y,z)}~\sum_{s}Q_{S|UYZ}(s|u,y,z)~d(s,\phi(w,y,z)) \notag \\
   & \geq \sum_{u,y,z}Q_{UYZ}(u,y,z)~\sum_{w,s}P_{W|U}(w|u)~Q_{S|UYZ}(s|u,y,z)~d(s,\phi^*(u,y,z)) \label{distlessurev} \\
   & \geq \sum_{u,y,z}Q_{UYZ}(u,y,z)~\min_{\phi(u,y,z)}~\sum_{w,s}P_{W|U}(w|u)~Q_{S|UYZ}(s|u,y,z)~d(s,\phi(u,y,z)) \notag \\
   &=\sum_{u,y,z}Q_{UYZ}(u,y,z)~\min_{\phi(u,y,z)}~\sum_{s}Q_{S|UYZ}(s|u,y,z)~d(s,\phi(u,y,z)) \notag \\
   &=  \min_{\phi(\cdot, \cdot,\cdot)}  \mathbb{E}_{Q}\left[d \left(S, \phi(U,Y,Z)\right)\right]:= \Delta_{min}, \notag &&
\end{flalign}
where, in \eqref{distlessurev}, $\phi^*(u,y,z)$ is chosen such that
\begin{align}
    \phi^*(u,y,z):=\argmin_{\phi(w,y,z),w \in \W} \sum_{s}Q_{S|UYZ}(s|u,y,z)~d(s,\phi(w,y,z)). \notag 
\end{align}

 \vspace{5 pt}
\textit{Converse:}
Let  $W=(W_T,T)$ denote the auxiliary r.v. defined in the converse of Proposition \ref{equivocation-noiseless}. Inequalities \eqref{eq:ratedist} and \eqref{eq:expdist} follow similarly as obtained in Proposition \ref{equivocation-noiseless}. We prove \eqref{eq:distconstnull}. Defining $\revised{\tilde{\phi}_n(M,Y^n,Z^n,S^{i-1},i):= \bar{\phi}_{i,n}(M,Y^n,Z^n,S^{i-1})} $, we have
\begin{flalign}
\min_{g^{(r)}_{i,n}}\mathbb{E}\left[d \left(S^n, \hat S^n\right) \Big|H=0 \right]&=\revised{ \min_{\{\tilde{\phi}_n(m,y^n,z^n,s^{i-1},i)\}_{i=1}^n}\mathbb{E}\left[ \sum_{i=1}^n d \left(S_i, \tilde{\phi}_n(M,Y^n,Z^n,S^{i-1},i) \right) \Big| H=0\right]} \label{determfun} \\
 &= \min_{\{\tilde{\phi}_n(\cdot,\cdot,\cdot,\cdot,\cdot)\}_{i=1}^n}\mathbb{E}\left[ \sum_{i=1}^n d \left(S_i, \tilde{\phi}_n(W_i,Z_i,Y_i,Y_{i+1}^n,i) \right) \Big|H=0\right] \notag \\
  & \leq  \min_{\{\phi(w_i,z_i,y_i,i)\}}\mathbb{E}\left[ \sum_{i=1}^n d \left(S_i, \phi(W_i,Z_i,Y_i,i) \right) \Big|H=0\right] \notag \\
  & = n\min_{\{\phi(\cdot,\cdot,\cdot,\cdot)\}}\mathbb{E}\left[ \mathbb{E} \left[ d \left(S_T, \phi(W_T,Z_T,Y_T,T) \right) \big|T\right] ~\Big|H=0\right] \notag \\
& =n \min_{\{\phi(\cdot,\cdot,\cdot,\cdot)\}}\mathbb{E} \left[ d \left(S_T, \phi(W_T,Z_T,Y_T,T) \right)\Big|H=0\right] \notag \\
&=n\min_{\{\phi(w,z,y)\}}\mathbb{E} \left[ d \left(S, \phi(W,Z,Y) \right)\Big|H=0\right], \notag &&
\end{flalign}
where  \eqref{determfun} is due to \eqref{determfunach} (in Appendix \ref{determadvfnproof}).
Hence,  any $\Delta_0$ satisfying \eqref{distexpstr} satisfies
\begin{align}
\Delta_0 &\leq \min_{\{\phi(w,z,y)\}}\mathbb{E}_P \left[ d \left(S, \phi(W,Z,Y) \right)\right]. \notag   
\end{align}
This completes the proof of the converse and the theorem.
\end{IEEEproof}

A more general version of Proposition \ref{equivocation-noiseless} and Proposition  \ref{distortion-noiseless} is claimed in \cite{ITW-18} as Theorem 7 and Theorem 8, respectively, in which a  privacy constraint under the alternate hypothesis is also imposed. However, we have identified a mistake in the converse proof; and hence, a single-letter characterization for this general problem remains open.

To complete the single-letter characterization in Proposition \ref{equivocation-noiseless} and Proposition  \ref{distortion-noiseless}, we bound the alphabet size of the auxiliary r.v. $W$ in the following lemma, whose proof is given in Appendix \ref{cardbndauxproof}.
\begin{lemma} \label{cardauxrv}
In Proposition \ref{equivocation-noiseless} and \ref{distortion-noiseless}, it suffices to consider auxiliary r.v.'s $W$ such that  $|\W| \leq |\Ucal|+2$. 
\end{lemma}
The proof of Lemma \ref{cardauxrv} uses standard arguments based on the Fenchel-Eggleston-Carath\'eodory's theorem and is given in Appendix \ref{cardbndauxproof}.
\begin{remark}
 When $Q_{S|UYZ}=Q_{S|YZ}$, a tight single-letter characterization of $\mathcal{R}_e$ and $\mathcal{R}_d$ exists even if the privacy constraint is active under the alternate hypothesis. This is due to the fact that  given $Y^n$ and $Z^n$, $M$ is independent of $S^n$ under the alternate hypothesis. In this case, $(R,\kappa, \Lambda_0, \Lambda_1) \in \mathcal{R}_e$ if and only if there exists an auxiliary r.v. $W$, such that  $(Z,Y,S)-U-W$, and
\begin{align}
\kappa &\leq I_P(W;Y|Z), \label{expconsteqsp} \\
 R &\geq I_P(W;U|Z), \label{rtconsteqsp}  \\
 \Lambda_0 &\leq H_{P}(S|W,Z,Y),
 \label{distconsteq0sp}\\
 \Lambda_1 &\leq H_{Q}(S|Z,Y),  \label{distconsteq1sp}
\end{align}
for some $P_{SUYZW}$ as in Proposition \ref{equivocation-noiseless}. 
Similarly, we have that 
$(R,\kappa, \Delta_0, \Delta_1) \in \mathcal{R}_d$ if and only if there exist an auxiliary  r.v. $W$ and a deterministic function $\phi: \W \times \Y \times \Z \rightarrow \hat{\mathcal{S}}$ such that \eqref{expconsteqsp}, \eqref{rtconsteqsp},
\begin{align}
\Delta_0 &\leq  \min_{\phi(\cdot,\cdot,\cdot)} \mathbb{E}_{P}\left[d \left(S, \phi(W,Y,Z)\right)\right],\label{eq:distconstnullsp}\\
\Delta_1 &\leq  \min_{\phi(\cdot, \cdot,\cdot)} \mathbb{E}_{Q}\left[d \left(S, \phi(Y,Z)\right)\right],\label{eq:distconstaltsp}
 \end{align}
are satisfied for some $P_{SUYZW}$ as in Proposition \ref{equivocation-noiseless}.

\end{remark}

The computation of the trade-off given in Proposition  \ref{equivocation-noiseless} is challenging in spite of the cardinality bound on the auxiliary r.v. $W$ provided by Lemma \ref{cardauxrv}, as closed form solutions do not exist in general. Below, we provide an example  where such a solution does exist. 
\begin{example} \label{rep-tradeoffeg}
Let $\V=\Ucal=\mathcal{S}=\{0,1\}$, $V=Y$, $Z=$ constant, $V-S-U$, $P_U(0)=Q_U(0)=0.5$, $P_{S|U}(0|0)=P_{S|U}(1|1)=Q_{S|U}(0|0)=Q_{S|U}(1|1)=1-q$, $P_{V|S}(0|0)=P_{V|S}(1|1)=1-p$ and $Q_{V|S}(0|0)=Q_{V|S}(1|1)=0.5$. Then, $(R,\kappa,\Lambda_0,0) \in \mathcal{R}_e$ if  there exists $r \in [0,0.5]$ such that 
\begin{align}
    R &\geq 1-h_b(r), \label{ratecex} \\
    \kappa &\leq 1-h_b((r*q)*p), \label{expcex} \\
    \Lambda_0 &\leq h_b(p)+h_b(q*r)-h_b(p*(q*r)), \label{equcex}
\end{align}
where, for $a,b \in \mathbb{R}$, $a*b:=(1-a)\cdot b+(1-b) \cdot a$, and  $h_b:[0,1] \mapsto [0,1]$ is the binary entropy function given by
\begin{align}
    h_b(t)=-(1-t) \log(1-t)-t \log(t). \notag
\end{align}
The above characterization\footnote{Numerical computation shows that the characterization given  in \eqref{ratecex}-\eqref{equcex} is exact even when $q \in (0,1)$.} is exact for $q=0$, i.e.,  $(R,\kappa,\Lambda_0,0) \in \mathcal{R}_e$ only if there exists $r \in [0,0.5]$ such that \eqref{ratecex}-\eqref{equcex} are satisfied.
\end{example}
\begin{IEEEproof}
Taking $\mathcal{W}=\{0,1\}$, and $P_{W|U}(0|0)=P_{W|U}(1|1)=1-r$, the constraints defining the trade-off given in Proposition \ref{equivocation-noiseless} simplifies to 
\begin{align}
I(U;W)&= 1-h_b(r), \notag \\
    I(V;W) &= 1-h_b((r*q)*p) \notag \\
   H(S|V,W) &= H(S|W)-I(S;V|W)=H(S|W)+H(V|S)-H(V|W)=h_b(r*q)+h_b(p)-h_b(p*(q*r)). \notag
\end{align}
On the other hand, if $q=0$, note that $S=U$. Hence, the same constraints can be bounded as follows:
\begin{align}
I(U;W)&= 1-H(U|W), \notag \\
    I(V;W) &= 1-H(V|W) \leq 1-h_b\left(h_b^{-1}(H(U|W))*p \right) \label{applymgl1} \\
   H(U|V,W)&=H(U|W)+H(V|U)-H(V|W)  \leq  h_b(p)+H(U|W)-h_b\left(h_b^{-1}(H(U|W))*p \right) \label{applymgl2},
\end{align}
where $h_b^{-1}:[0,1] \mapsto [0,0.5]$ is the inverse of the binary entropy function. Here, the inequality in \eqref{applymgl1} and \eqref{applymgl2} follows by an application of Mrs. Gerber's lemma \cite{Elgamalkim}, since  $V=U\oplus N_p$ under the null hypothesis and $N_p \sim Ber(p)$ is independent of $U$ and $W$. Also, $\Lambda_{min}=0$ since $S=U$. Noting that $H(U|W) \in [0,1]$, and defining $r:=h_b^{-1}(H(U|W)) \in [0,0.5]$, the result follows.
\end{IEEEproof}
\begin{figure}[t]
\centering
\includegraphics[trim=0cm 0cm 0cm 0cm, clip, width= 0.7\textwidth]{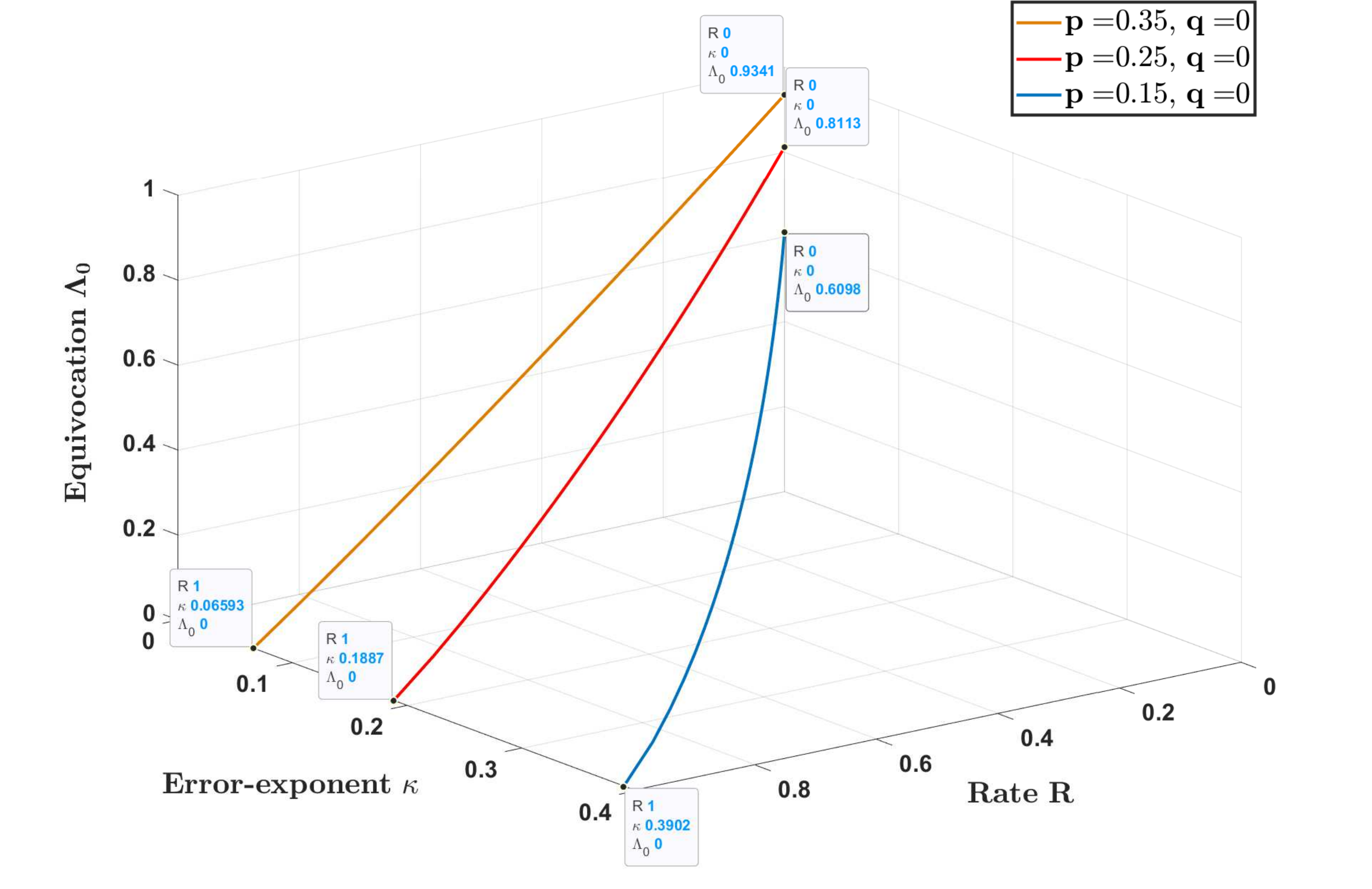}
\centering \caption{$(R,\kappa,\Lambda_0)$ trade-off at the boundary of $\mathcal{R}_e$ in Example \ref{rep-tradeoffeg} (Axes units are in bits)} \label{figREPtradeoff}
\end{figure}
\begin{figure} 
  \subfloat[ $R-\kappa$ trade-off (Axes units are in bits)]{%
      \label{figREtradeoff} \includegraphics[width=0.5\textwidth]{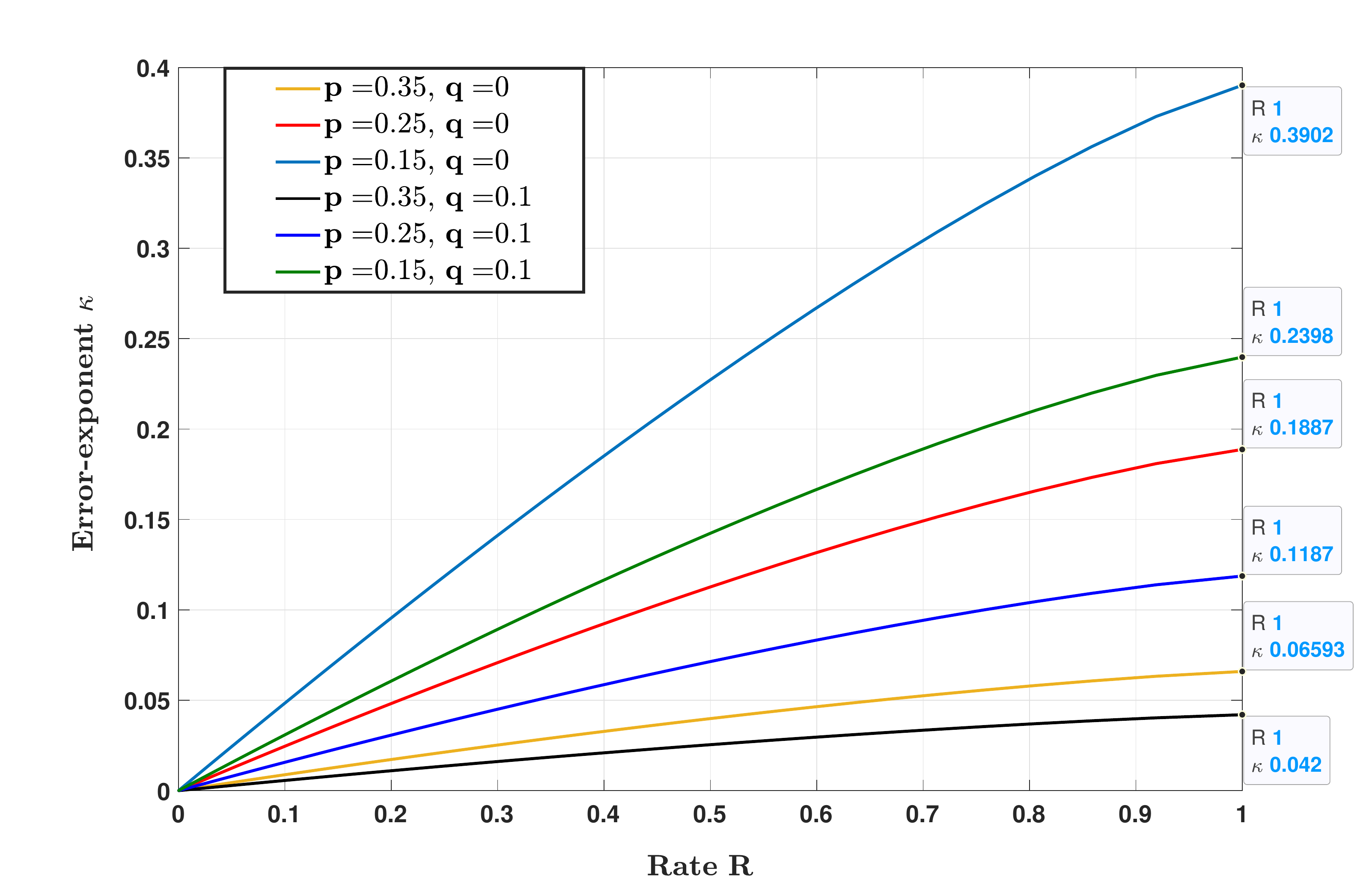}}  
          \subfloat[ $\kappa-\Lambda_0$ trade-off (Axes units are in bits) ]{\label{figEPtradeoff}
        \includegraphics[width=0.5\textwidth]{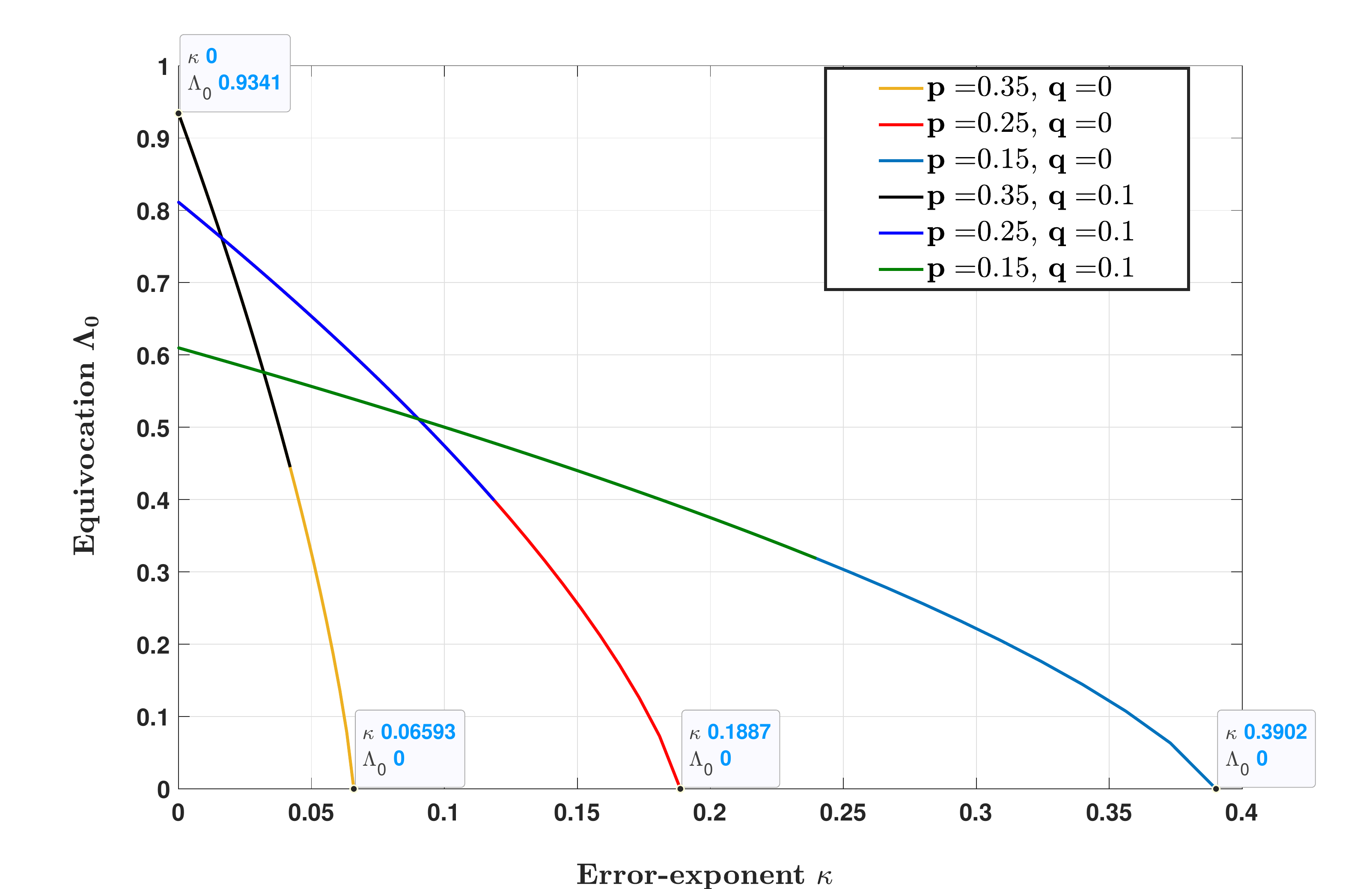}} 
           \caption{Projections of Fig. \ref{figREPtradeoff} in the $R-\kappa$ plane and  $\kappa-\Lambda_0$ plane} 
\end{figure}



Fig. \ref{figREPtradeoff} depicts the curve $\left(1-h_b(r),1-h_b(p*(q*r)),h_b(p)+h_b(r*q)-h_b(p*(r*q))\right)$ for $q=0$ and $p \in \{0.15,0.25,0.35\}$, as $r$ is varied in the range  $[0,0.5]$. The projection of this curve on the  $R-\kappa$ and $\kappa-\Lambda_0$  plane is shown  in  Fig. \ref{figREtradeoff} and Fig. \ref{figEPtradeoff}, respectively, for $q \in \{0,0.1 \}$ and the same values of $p$. As expected, the error-exponent $\kappa$ increases with rate $R$ while the equivocation $\Lambda_0$ decreases with $\kappa$ at the boundary of $\mathcal{R}_e$.

Proposition \ref{equivocation-noiseless} (resp. Proposition  \ref{distortion-noiseless}) provide a characterization of $\mathcal{R}_e$ (resp. $\mathcal{R}_d$)  under the condition of vanishing type I error probability constraint. Consequently, the converse part of these results are known as \textit{weak converse} results in the context of HT. In the next subsection, we establish the optimal error exponent-privacy trade-off for the special case of zero-rate compression. This trade-off is independent of the type I error probability constraint $\epsilon \in (0,1)$, and hence known as a \textit{strong converse} result.
\subsection{Zero-rate compression} \label{zerorateHT}
 Assume the following zero-rate constraint on the communication between the observer and the detector,
\begin{align}
    \lim_{n \rightarrow \infty}\frac{\log(|\mathcal{M}|)}{n}=0. \label{zeroratemsg}
\end{align}
 Note that \eqref{zeroratemsg} does not imply that $|\mathcal{M}|=0$, i.e., nothing can be transmitted, but that the message set cardinality can grow at most  sub-exponentially in $n$.  Such a scenario is motivated practically by low power or low bandwidth constrained applications in which communication is costly. 
Proposition \ref{thm:zerorategenhtdist} and Proposition \ref{thm:zerorategenhtequ} stated below provide an optimal single-letter characterization of $\mathcal{R}_d(\epsilon)$ and $\mathcal{R}_e(\epsilon)$ in this case.  While the coding schemes in the achievability part of these results are inspired from that in \cite{Han}, the analysis of privacy achieved at the detector is new.
Lemma \ref{approxdistmequzero} serves as a crucial tool for this purpose. We next state the results. Let 
  \begin{subequations}  \label{maxdistposs}
  \begin{align} 
     \Delta_0^{max}&:=\min_{\phi'(\cdot)} \mathbb{E}_P\left[d \left(S, \phi'(V)\right)\right],  \\
     \mbox{and }   \Delta_1^{max}&:=\min_{\phi'(\cdot)} \mathbb{E}_Q\left[d \left(S, \phi'(V)\right)\right].  
       \end{align}
  \end{subequations}

\begin{prop} \label{thm:zerorategenhtdist} 
For $\epsilon \in (0,1)$, $(0,\kappa, \Delta_0, \Delta_1) \in \mathcal{R}_d(\epsilon)$ if and only if it satisfies,
\begin{align}
\kappa &\leq \min_{P_{\tilde U  \tilde V} \in \mathcal{L}'(P_{U}, P_{V})} D(P_{\tilde U  \tilde V}||Q_{UV}), \label{expconsteqgenzr} \\
 \Delta_0 &\leq  \Delta_0^{max}, \label{distconsteq0genzr} \\
  \Delta_1 &\leq   \Delta_1^{max}, \label{distconsteq1genzr}
\end{align}
where $\phi': \V \rightarrow \hat S$ is a deterministic function and
\begin{align}
\mathcal{L}'(P_{U}, P_{V})&= \{P_{\tilde U \tilde V  } \in \mathcal{P}(\Ucal \times \V ) : P_{\tilde U}=P_{U},~ P_{\tilde V  }=P_{V} \}. \notag 
\end{align}
\end{prop}
\begin{IEEEproof}
First, we prove that $(0, \kappa, \Delta_0, \Delta_1)$ satisfying \eqref{expconsteqgenzr}-\eqref{distconsteq1genzr} is achievable. While the encoding and decoding scheme is the same as that in \cite{Han}, we mention  it for the sake of completeness. \\
\textbf{\underline{Encoding:}} The observer sends the message $M=1$ if $U^n \in \mathcal{T}_{[P_U]_{\delta}}^n$, $\delta>0$, and $M=0$ otherwise.  \\
\textbf{\underline{Decoding:}} The detector declares $\hat H=0$ if $M=1$ and $V^n \in \mathcal{T}_{[P_{V}]_{\delta}}^n$, $\delta>0$. Otherwise, $\hat H=1$ is declared. 

We analyze the type I and type II error probabilities for the above scheme. Note that for any $\delta>0$, the weak law of large numbers implies that
\begin{align}
 \mathbb{P}\left(U^n \in \mathcal{T}_{[P_U]_{\delta}}^n \cap V^n \in \mathcal{T}_{[P_{V}]_{\delta}}^n)\big|H=0 \right) = \mathbb{P}\left(M=1 \cap V^n \in \mathcal{T}_{[P_{V}]_{\delta}}^n)\big|H=0 \right)\xrightarrow{(n)} 1.   \notag
\end{align}
 Hence, the type I error probability tends to zero, asymptotically. The type II error probability can be written as follows:
 \begin{flalign}
 \beta_n(f_n, g_n)&= \mathbb{P}(U^n \in \mathcal{T}_{[P_U]_{\delta}}^n \cap V^n \in \mathcal{T}_{[P_{V}]_{\delta}}^n)|H=1 ) \notag \\
     &=  \sum_{\substack{u^n \in \mathcal{T}_{[P_U]_{\delta}}^n, \\ v^n \in \mathcal{T}_{[P_{V}]_{\delta} }}  } Q_{U^n V^n }(u^n,v^n) \leq (n+1)^{|\Ucal||\V|}  e^{-n(\kappa^*-O(\delta))}= e^{-n \left(\kappa^*-\frac{|\Ucal||\V|\log(n+1)}{n}-O(\delta) \right)}, \notag &&
 \end{flalign}
 where
 \begin{align}
     \kappa^*=\min_{P_{\tilde U  \tilde V} \in \mathcal{L}'(P_{U}, P_{V})} D(P_{\tilde U  \tilde V}||Q_{UV}). \notag
 \end{align}
Next, we lower bound the average distortion for $S^n$ achieved by this scheme at the detector. 
Defining
\begin{align}
  \Pi(U^n,\delta,P_U)&:=\ind \left(U^n \notin \mathcal{T}_{[P_U]_{\delta}}^n \right), \label{defiudelta}\\
     \rho_n^{(0)}(\delta)&:=\left\| P_{S^nV^n}(\cdot)-P_{S^nV^n|\Pi(U^n,\delta,P_U)}(\cdot|0)\right\|, \label{totvarprbdist1},\\
       \rho_n^{(1)}(\delta)&:=\left\| Q_{S^nV^n}(\cdot)-Q_{S^nV^n|\Pi(U^n,\delta,P_U)}(\cdot|1)\right\|, \label{totvarprbdist2} \\
       \phi'_n(v^n)&:=(\phi'(v_1),\cdots,\phi'(v_n)), \notag
 \end{align}
we can write\revised{
 \begin{flalign}
   &\Big| \min_{\{\bar \phi_i(m,v^n,s^{i-1})\}_{i=1}^n}\mathbb{E}\left[d\left(S^n, \hat S^n\right) \big|H=0 \right]-n\min_{\phi'(v)}\mathbb{E}_P\left[d \left(S, \phi'(V)\right)  \right] \Big| \notag\\
&=\Big| \min_{\{\bar \phi_i(m,v^n,s^{i-1})\}_{i=1}^n}\mathbb{E}\left[d\left(S^n, \hat S^n\right) \big|H=0 \right]-\min_{ \phi'_n(v^n)}\mathbb{E}\left[d \left(S^n,  \phi'_n(V^n)\right) \big|H=0 \right] \Big| \notag \\
& \leq \Big| \min_{\{\bar \phi_i(m,v^n,s^{i-1})\}_{i=1}^n}\mathbb{E}\left[d \left(S^n, \hat S^n\right) \big|H=0 \right]-\mathbb{P}\left(M=1|H=0\right) \min_{ \phi'_n(v^n)}\mathbb{E}\left[d \left(S^n, \phi'_n(V^n)\right) \big|M=1,H=0 \right] \Big| \notag \\
& \qquad + \mathbb{P}\left(M=0|H=0\right) \min_{ \phi'_n(v^n)}\mathbb{E}\left[d \left(S^n,  \phi'_n(V^n)\right) \big|M=0,H=0 \right] \notag \\
& \leq \Big| \min_{\{\bar \phi_i(m,v^n,s^{i-1})\}_{i=1}^n}\mathbb{E}\left[d \left(S^n, \hat S^n\right) \big|H=0 \right]- \min_{ \phi'_n(v^n)}\mathbb{E}\left[d \left(S^n, \phi'_n(V^n)\right) \big|M=1,H=0 \right] \Big| \notag \\
& \qquad +\mathbb{P}\left(M=0|H=0\right) \left[\min_{ \phi'_n(v^n)}\mathbb{E}\left[d \left(S^n, \phi'_n(V^n)\right) \big|M=1,H=0 \right]+ \min_{ \phi'_n(v^n)}\mathbb{E}\left[d \left(S^n, \phi'_n(V^n)\right) \big|M=0,H=0 \right]\right]  \notag \\ 
& =\Big| \min_{\{\bar \phi_i(m,v^n,s^{i-1})\}_{i=1}^n}\mathbb{E}\left[d \left(S^n, \hat S^n\right) \big|H=0 \right]-\min_{ \phi'_n(v^n)}\mathbb{E}\left[d \left(S^n, \phi'_n(V^n)\right) |\Pi(U^n,\delta,P_U)=0,H=0 \right] \Big| \notag \\
&~ +\mathbb{P}\left(\Pi(U^n,\delta,P_U)=1\big|H=0\right) \bigg[\min_{ \phi'_n(v^n)}\mathbb{E}\left[d \left(S^n, \phi'_n(V^n)\right) \big|M=1,H=0 \right] \notag \\
&\qquad \qquad \qquad \qquad  \qquad \qquad \qquad \qquad + \min_{ \phi'_n(v^n)}\mathbb{E}\left[d \left(S^n, \phi'_n(V^n)\right) \big|M=0,H=0 \right]\bigg]  \label{equmsgzero} \\
& \leq nD_m   \rho_n^{(0)}(\delta)+ 2~e^{-n \Omega(\delta)} nD_m \label{bnddistexpdecr}\\
& \xrightarrow{(n)}0, \label{expdecnullhypdist}&&
 \end{flalign}}
 where \eqref{equmsgzero} is since  $\Pi(U^n,\delta,P_U)=1-M$  with probability one by the encoding scheme; \eqref{bnddistexpdecr} follows from   
 \begin{align}
     \mathbb{P}\left(\Pi(U^n,\delta,P_U)=1|H=0\right)=\mathbb{P}\left(U^n \notin \mathcal{T}_{[P_U]_{\delta}}^n|H=0\right) \leq e^{-n \Omega(\delta)}
 \end{align}
and \cite[Property 2(b)]{Schieler-Cuff-2014}; and, \eqref{expdecnullhypdist} is due to \eqref{nullhypapproxdist}. Similarly,  it can be shown using  \eqref{althypapproxdist2} that if $Q_U = P_U$, then
 \begin{flalign}
     \bigg| \min_{\{\bar{\phi}_{i,n}(m,v^n,s^{i-1})\}_{i=1}^n}\mathbb{E}\left[d \left(S^n, \hat S^n\right) |H=1 \right]-n\min_{\phi'(v)}\mathbb{E}_Q\left[d \left(S, \phi'(V)\right) \right] \bigg| \xrightarrow{(n)}0. 
 \end{flalign}
 On the other hand, if $Q_U \neq P_U$ and $\delta $ is small enough, we have
\begin{align}
    \mathbb{P}\left(M=0 |H=1\right)= \mathbb{P}\left(\Pi(U^n,\delta,P_U)=1 |H=1\right) \geq  1-e^{-n(D(P_U||Q_U)-O(\delta))} \xrightarrow{(n)} 1. \label{probtendexp1zr}
\end{align} 
Hence, we can write for $\delta$ small enough,\revised{
  \begin{flalign}
  &\Big| \min_{\{\bar \phi_i(m,v^n,s^{i-1})\}_{i=1}^n}\mathbb{E}\left[d\left(S^n, \hat S^n\right) |H=1 \right]-n\min_{\phi'(v)}\mathbb{E}_Q\left[d \left(S, \phi'(V)\right)  \right] \Big| \notag \\
&=\Big| \min_{\{\bar \phi_i(m,v^n,s^{i-1})\}_{i=1}^n}\mathbb{E}\left[d\left(S^n, \hat S^n\right) |H=1 \right]-\min_{ \phi'_n(v^n)}\mathbb{E}\left[d \left(S^n, \phi'_n(V^n)\right) |H=1 \right] \Big| \notag \\
& \leq \Big| \min_{\{\bar \phi_i(m,v^n,s^{i-1})\}_{i=1}^n}\mathbb{E}\left[d \left(S^n, \hat S^n\right) |H=1 \right]-\mathbb{P}\left(M=0|H=0\right) \min_{ \phi'_n(v^n)}\mathbb{E}\left[d \left(S^n, \phi'_n(V^n)\right) |M=0,H=1 \right] \Big| \notag \\
& \qquad + \mathbb{P}\left(M=1|H=1\right) \min_{ \phi'_n(v^n)}\mathbb{E}\left[d \left(S^n, \phi'_n(V^n)\right) |M=1,H=1 \right] \notag \\
& \leq \Big| \min_{\{\bar \phi_i(m,v^n,s^{i-1})\}_{i=1}^n}\mathbb{E}\left[d \left(S^n, \hat S^n\right) |H=1 \right]- \min_{ \phi'_n(v^n)}\mathbb{E}\left[d \left(S^n, \phi'_n(V^n)\right) |M=0,H=1 \right] \Big| \notag \\
& \qquad +\mathbb{P}\left(M=1|H=1\right) \left[\min_{ \phi'_n(v^n)}\mathbb{E}\left[d \left(S^n, \phi'_n(V^n)\right) |M=1,H=1 \right]+ \min_{ \phi'_n(v^n)}\mathbb{E}\left[d \left(S^n, \phi'_n(V^n)\right) |M=0,H=1 \right]\right]  \notag \\
& =\Big| \min_{\{\bar \phi_i(m,v^n,s^{i-1})\}_{i=1}^n}\mathbb{E}\left[d \left(S^n, \hat S^n\right) |H=1 \right]-\min_{ \phi'_n(v^n)}\mathbb{E}\left[d \left(S^n, \phi'_n(V^n)\right) |\Pi(U^n,\delta,P_U)=1,H=1 \right] \Big| \notag \\
&~ +\mathbb{P}\left(\Pi(U^n,\delta,P_U)=0|H=1\right) \bigg[\min_{ \phi'_n(v^n)}\mathbb{E}\left[d \left(S^n, \phi'_n(V^n)\right) |M=1,H=1 \right] \notag \\
&\qquad \qquad \qquad \qquad  \qquad \qquad \qquad \qquad+\min_{ \phi'_n(v^n)}\mathbb{E}\left[d \left(S^n, \phi'_n(V^n)\right) |M=0,H=1 \right]\bigg]  \label{msgzeroind2} \\
& \leq nD_m   \rho_n^{(1)}(\delta)+ 2~e^{-n (D(P_U||Q_U)-O(\delta))} nD_m \label{bnddistexpdecrnew}\\
& \xrightarrow{(n)}0, \label{expdecnullhypdistnew}&&
 \end{flalign}}
 where \eqref{msgzeroind2} is since  $\Pi(U^n,\delta,P_U)=1-M$  with probability one;  \eqref{bnddistexpdecrnew} is due to \eqref{probtendexp1zr} and \cite[Property 2(b)]{Schieler-Cuff-2014}; and, \eqref{expdecnullhypdistnew} follows from \eqref{althypapproxdist}.
 This completes the proof of the achievability.

We next prove the converse. Note that by the strong converse result in \cite{Shalaby-pap}, the right hand side (R.H.S) of \eqref{expconsteqgenzr} is an upper bound on the achievable error-exponent for all $\epsilon \in (0,1)$ even without a privacy constraint (hence, also with a privacy constraint). Also,
\begin{flalign}
\min_{g^{(r)}_{i,n}}\mathbb{E}\left[d \left(S^n, \hat S^n\right) |H=0 \right] &\leq  \min_{\{\phi'(v_i)\}_{i=1}^n}  \sum_{i=1}^n \mathbb{E}_{P_{S_iV_i}} \left[ d \left(S_i, \phi'(V_i) \right) \right] \label{zeroratebnd1conv} \\
 & =n \min_{\{\phi'( v)\}}\mathbb{E}_{P}   \left[  d(S,\phi'(V)) \right]. \notag &&
\end{flalign} 
Here, \eqref{zeroratebnd1conv} follows from the fact that the detector can always reconstruct $\hat S_i$ as a function of $V_i$ for $i \in [n]$. 
Similarly,
\begin{align}
&\min_{g^{(r)}_{i,n}}\mathbb{E}\left[d \left(S^n, \hat S^n\right) |H=1 \right] \leq n \min_{\{\phi'( v)\}}\mathbb{E}_{Q}   \left[  d(S,\phi'(V)) \right]. \notag 
\end{align} 
Hence, any achievable $\Lambda_0$ and $\Lambda_1$ has to satisfy \eqref{distconsteq0genzr} and \eqref{distconsteq1genzr}, respectively. This completes the proof.
\end{IEEEproof}

The following Proposition is the analogous result to Proposition \ref{thm:zerorategenhtdist} when the  privacy measure is equivocation.

\begin{prop} \label{thm:zerorategenhtequ}
For $\epsilon \in (0,1)$, $(0,\kappa, \Lambda_0, \Lambda_1) \in \mathcal{R}_e(\epsilon)$ if and only if it satisfies \eqref{expconsteqgenzr} and 
\begin{align}
 \Lambda_0 &\leq H_{P}(S|V), \label{equconsteq0genzr} \\
  \Lambda_1 &\leq  H_{Q}(S|V). \label{equconsteq1genzr}
\end{align}
\end{prop}
\begin{IEEEproof}
For proving the achievablity part, the encoding and decoding scheme is the same as in Proposition \ref{thm:zerorategenhtdist}. Hence, the analysis of the error-exponent given in Proposition \ref{thm:zerorategenhtdist} holds. To lower bound the equivocation of $S^n$ at the detector, defining $\Pi(U^n,\delta,P_U)$, $\rho_n^{(0)}(\delta)$ and $\rho_n^{(1)}(\delta)$
as in \eqref{defiudelta}-\eqref{totvarprbdist2}, we can write 
 \begin{flalign}
      &\big|nH_P(S|V)-H(S^n|M,V^n,H=0)\big| \notag \\
     &=\big|H(S^n|V^n,H=0)-H(S^n|M,V^n,H=0)\big| \notag \\
     &\leq  \big|H(S^n,V^n|H=0)-H(S^n,V^n|M,H=0)\big| \notag \\
      & \leq \big|H(S^n,V^n|H=0)-\mathbb{P}\left(M=1|H=0\right)~H(S^n,V^n|M=1,H=0)\big|+\mathbb{P}\left(M=0|H=0\right)H(S^n,V^n|M=0,H=0) \notag \\
      &\leq \big|H(S^n,V^n|H=0)-H(S^n,V^n|M=1,H=0)\big|\notag \\
      & \qquad +\mathbb{P}\left(M=0|H=0\right) \big(H(S^n,V^n|M=1,H=0)+H(S^n,V^n|M=0,H=0)  \big)\notag \\
     & \leq \big|H(S^n,V^n|H=0)-H(S^n,V^n|\Pi(U^n,\delta,P_U)=0,H=0)\big| \notag \\
     & \qquad +\mathbb{P}\left(\Pi(U^n,\delta,P_U)=1|H=0\right)   \big(H(S^n,V^n|M=1,H=0)+H(S^n,V^n|M=0,H=0)  \big) \notag \\
     & \overset{(n)}{\leq} -2 \rho_n^{(0)}(\delta) \log \left(\frac{\rho_n^{(0)}(\delta) }{|\mathcal{S}|^n|\V|^n} \right)+ 2~e^{-n \Omega(\delta)} \log \left(|\mathcal{S}|^n|\V|^n \right) \label{applylembndent}\\
     & \xrightarrow{(n)} 0, \label{tend0exp} &&
 \end{flalign}
 where \eqref{applylembndent} follows due to Lemma \ref{contentrtv}, \cite[Lemma 2.12]{Csiszar-Korner} and the fact that entropy of a r.v. is bounded by the logarithm of cardinality of its support; and, \eqref{tend0exp}  follows from \eqref{nullhypapproxdist} in Lemma \ref{approxdistmequzero} since $\delta>0$. In a similar way, it can be shown using  \eqref{althypapproxdist2} that if $Q_U = P_U$, then
 \begin{align}
   &\big|H(S^n|V^n,H=1)-H(S^n|M,V^n,H=1)\big| \xrightarrow{(n)}0.
 \end{align}
 On the other hand, if $Q_U \neq P_U$ and $\delta $ is small enough, we can write,
 \begin{flalign}
      &|nH_Q(S|V)-H(S^n|M,V^n,H=1)| \notag \\
     &=|H(S^n|V^n,H=1)-H(S^n|M,V^n,H=1)| \notag  \\
     &\leq  |H(S^n,V^n|H=1)-H(S^n,V^n|M,H=1)| \notag \\
     & \leq |H(S^n,V^n|H=1)-H(S^n,V^n|M=0,H=1)| \notag \\
     & \qquad +\mathbb{P}\big(\Pi(U^n,\delta,P_U)=0|H=1\big) \left(H(S^n,V^n|M=0,H=1)+H(S^n,V^n|M=1,H=1) \right) \notag &&
\end{flalign}
\begin{flalign}
     & \leq -2 \rho_n^{(1)}(\delta) \log \left(\frac{\rho_n^{(1)}(\delta) }{|\mathcal{S}|^n|\V|^n} \right)+ 2~e^{-n  (D(P_U||Q_U)-O(\delta))}\log \left(|\mathcal{S}|^n|\V|^n \right) \label{tendstozeroalt}, &&
 \end{flalign}
 where \eqref{tendstozeroalt} follows from Lemma \ref{contentrtv} and \eqref{probtendexp1zr}.
It  follows from \eqref{althypapproxdist} in Lemma \ref{approxdistmequzero} that for $\delta>0$ sufficiently small, $\rho_n^{(1)}(\delta)  \leq e^{-n \bar \delta}$ for some $\bar \delta>0$, thus implying that the R.H.S. of \eqref{tendstozeroalt} tends to zero. This completes the proof of achievability. 

The converse follows from the results in \cite{Han} and \cite{Shalaby-pap} that 
the R.H.S of 
\eqref{expconsteqgenzr} is the optimal error-exponent achievable for all values of $\epsilon \in (0,1)$ even when there is no privacy constraint, and the following inequality
\begin{align}
 H(S^n|M,V^n, H=j)   \leq  H(S^n|V^n, H=j),~j=0,1. \label{zerortineq}
 \end{align}
This concludes the proof of the Proposition.
\end{IEEEproof}

In Section \ref{Probform}, we mentioned that it is possible to achieve a positive error-exponent with perfect privacy in our model. Here, we provide an example of TAI with an equivocation privacy constraint under both hypothesis, and show that  perfect privacy is possible. Recall that TAI is a special case of TACI, in which, $Z=$ constant, and hence, the null and alternate hypothesis are given by
\begin{align}
   & H_0: (U^n,Y^n) \sim \prod_{i=1}^n P_{UY}, \notag\\
    \mbox{and }& H_1: (U^n,Y^n) \sim \prod_{i=1}^n P_UP_Y. \notag
\end{align}
\begin{example} \label{perfprivex}
  Let
 $\mathcal{S}=\Ucal=\{0,1,2,3\}$, $\Y=\{0,1\}$,  
 \[     
    P_{SU}= 0.125 \cdot 
\begin{bmatrix}
   1      &  1 & 0 & 0\\
    1 & 1  & 0 & 0 \\ 0 & 0 &1 &1 \\ 0 & 0 & 1 & 1 \end{bmatrix}
    , ~~
    P_{Y|U}=
 \begin{bmatrix}
  1 & 0 \\
    0 & 1 \\
    1 & 0 \\
    0 & 1
    \end{bmatrix},
\]
$P_{SUY}:= P_{SU}P_{Y|U}$ and $Q_{SUY}:= P_{SU} P_Y$, where  $P_Y= \sum_{u \in \Ucal}P_{U}(u) P_{Y|U}(y|u)$. Then, we have  $H_Q(S|Y)=H_P(S)=H_P(U)=2$ bits. Also, noting that under the null hypothesis, $Y=U~ mod~ 2$, $H_P(S|Y)=2$ bits. It follows from  the inner bound given by equations \eqref{rtconsteqgensp}-\eqref{distconsteq1gensp}, and, \eqref{binnnotnec} and \eqref{expexampel} that $(R, \kappa, \Lambda_0, \Lambda_1) \in \mathcal{R}_e(\epsilon)$, $\epsilon \in (0,1)$ if 
\begin{align*}
   R &\geq I_{P}(W;U), \\
 \kappa &\leq I_P(W;Y),  \\
 \Lambda_0 &\leq H_{P}(S|W,Y),  \\
  \Lambda_1 &\leq  H_{Q}(S|W,Y)=H_Q(S|W),
\end{align*}
where $P_{SUYW}:=P_{SUY}P_{W|U}$ and $Q_{SUYW}:=Q_{SUY}P_{W|U}$ for some conditional distribution $P_{W|U}$.
If we set $W:= U~ mod~ 2$,  then we have  $I_P(U;W)=1$ bit, $I_P(Y;W)=H_P(Y)=1$ bit, $H_P(S|W,Y)=H_P(S|Y)=2$ bits, and $H_Q(S|W)=H_P(S|Y)=2$ bits. Thus, by revealing only $W$ to the detector, it is possible to achieve a  positive error-exponent while ensuring maximum privacy under both the null and alternate hypothesis, i.e., the tuple $(1,1,2,2) \in \mathcal{R}_e(\epsilon)$, $\forall ~ \epsilon \in (0,1)$. 
\end{example}

\section{A Counterexample to the Strong Converse }\label{Strongconverse}
Ahlswede and Csisz\'{a}r  obtained a strong converse result for the DHT problem without a privacy constraint in \cite{Ahlswede-Csiszar}, where they showed that for any positive rate $R$, the optimal achievable error-exponent is independent of the type I error probability constraint $\epsilon$. Here, we explore whether a similar result holds in our model, in which, an additional privacy constraint is imposed. We will show through a counterexample that this is not the case in general. The basic idea used in the counterexample is a \enquote{time-sharing} argument which is used to construct from a given coding scheme that achieves the optimal rate-error-exponent-equivocation trade-off under a vanishing type I error probability constraint, a new coding scheme that satisfies the given type I error probability constraint $\epsilon^*$ and the same error-exponent as before, yet achieves a higher equivocation for $S^n$ at the detector. This concept has been used previously in other contexts, e.g., in the characterization of the first-order maximal channel coding rate of additive white gaussian noise (AWGN) channel in the finite block-length regime \cite{YP-Thesis}, and subsequently in the characterization of the second order maximal coding rate in the same setting \cite{YCDP-2015}. However, we will provide a self-contained proof of the counterexample by utilizing Lemma \ref{approxdistmequzero} for this purpose. 

Assume that the joint distribution $P_{SUV}$ is such that $H_P(S|U,V)< H_P(S|V)$.
Proving the strong converse amounts to showing that any  $(R, \kappa, \Lambda_0,\Lambda_1) \in \mathcal{R}_e(\epsilon)$ for some $\epsilon \in (0,1)$ also belongs to $\mathcal{R}_e$. Consider TAI problem with an equivocation privacy constraint, in which, $R \geq H_P(U)$ and $\Lambda_1 \leq \Lambda_{min}$. Then, from the optimal single-letter characterization of $\mathcal{R}_e $ given in Proposition \ref{equivocation-noiseless}, it follows by taking $W=U$ that $(H_P(U), I_P(V;U), H_P(S|V,U), \Lambda_{min}) \in \mathcal{R}_e$. Note that $I_P(V;U)$ is the maximum error-exponent achievable for any type I error probability constraint $\epsilon \in (0,1)$, even when $U^n$ is observed directly at the detector. Thus, for vanishing type I error probability constraint $\epsilon \rightarrow 0$ and $\kappa=I_P(V;U)$, the term $H_P(S|V,U)$ denotes the maximum achievable equivocation for $S^n$ under the null  hypothesis. From the proof of Proposition \ref{equivocation-noiseless}, the coding scheme achieving this tuple is as follows:
\begin{enumerate}
    \item Quantize $u^n$ to codewords  in $\mathcal{B}_n= \{u^n(j) \in  \mathcal{T}_{[P_U]_{\delta}}^n, ~j \in [e^{n(H_P(U)+\eta)}]\}$ and send the index of quantization to the detector, i.e.,  if $u^n \in \mathcal{T}_{[P_U]_{\delta}}^n$, send $M=j$, where $j$ is the index of $u^n$ in $\mathcal{B}_n$. Else, send $M=0$.
    \item At the detector, if $M=0$, declare $\hat H=1$. Else, declare $\hat H=0$ if $(u^n(M),v^n) \in \mathcal{T}_{[P_{UV}]_{\delta'}}^n$ for some $\delta'> \delta$, and $\hat H=1$ otherwise.
\end{enumerate}
The type I error probability of the above scheme tends to zero asymptotically with $n$. 
Now, for a fixed $\epsilon^*>0$, consider a modification of this coding scheme as follows:
\begin{enumerate}
    \item If $u^n \in \mathcal{T}_{[P_U]_{\delta}}^n$, send $M=j$ with  probability $1-\epsilon^*$, where $j$ is the index of $u^n$ in $\mathcal{B}_n$, and with probability $\epsilon^*$, send $M=0$. If $u^n \notin \mathcal{T}_{[P_U]_{\delta}}^n$, send $M=0$.
     \item At the detector, if $M=0$, declare $\hat H=1$. Else, declare $\hat H=0$ if $u^n(M),v^n) \in \mathcal{T}_{[P_{UV}]_{\delta'}}^n$ for some $\delta'> \delta$, and $\hat H=1$ otherwise.
\end{enumerate}
 It is easy to see that for this modified coding scheme, the type I error probability is asymptotically equal to $\epsilon^*$, while the error-exponent remains the same as $I(V;U)$ since the probability of declaring $\hat H=0$ is decreased. Recalling that $ \Pi(u^n,\delta,P_U):=\ind \left( u^n \notin \mathcal{T}_{[P_{U}]_{\delta}}^n\right)$, we also have\\
\begin{flalign}
   & \frac{1}{n}H\big(S^n|M,V^n, H=0\big)\notag \\
   &= (1-\gamma_n)(1-\epsilon^*) \frac{1}{n } H\big(S^n|U^n,V^n,\Pi(U^n, \delta,P_U)=0,H=0 \big) 
    +(1-\gamma_n)~\epsilon^*~ \frac 1n H\big(S^n|M=0,V^n,\Pi(U^n, \delta,P_U)=0,H=0\big) \notag \\
    & \qquad +\gamma_n~ \frac{1}{n}H\big(S^n|M=0,V^n,\Pi(U^n, \delta,P_U)=1,H=0\big) \notag  \\
      & \geq (1-\gamma_n)(1-\epsilon^*)~ \big(H_P\left(S|U,V\right)-\gamma_n''\big) +(1-\gamma_n)~\epsilon^*~ \frac 1n H\big(S^n|M=0,V^n,\Pi(U^n, \delta,P_U)=0,H=0\big) \notag \\
       & \qquad +\gamma_n~ \frac{1}{n}H\big(S^n|M=0,V^n,\Pi(U^n, \delta,P_U)=1,H=0\big) \label{typicalbnd}\\
       &> (1-\gamma_n)(1-\epsilon^*) ~\big(H_P\left(S|U,V\right)-\gamma_n''\big) 
    +(1-\gamma_n)~\epsilon^*~ \left(H_P(S|U,V)-\frac{\gamma_n'}{n}\right) \notag \\
       & \qquad +\gamma_n~ \frac{1}{n}H\big(S^n|M=0,V^n,H=0,\Pi(U^n, \delta,P_U)=1\big) \label{dataprocequivoc} \\
    & = (1-\gamma_n)(1-\epsilon^*)~ \big(H_P\left(S|U,V\right)-\gamma_n''\big) +(1-\gamma_n)~\epsilon^* \left(H_P\left(S|U,V\right)-\frac{\gamma_n'}{n}\right)+\gamma_n''' \label{singlletterequ} \\
    &= (1-\gamma_n)H_P\left(S|U,V\right)-\bar \gamma_n, \label{finbndstrngconv} &&
\end{flalign}
where $\{\gamma_n''\}_{n \in \mathbb{N}}$  denotes some sequence of positive numbers such that $\gamma_n'' \xrightarrow{(n)} 0$, and
\begin{align}
  &  \gamma_n:= \mathbb{P}\left(U^n \notin \mathcal{T}_{[P_U]_{\delta}}^n\big|H=0 \right) \leq e^{-n \Omega(\delta)} \xrightarrow{(n)} 0, \label{expdecaytyp} \\
  &\gamma_n':=-2 \rho_n^* \log\left(\frac{2   \rho_n^*}{|\mathcal{S}|^n} \right),   \notag\\
  &\rho_n^*:= \left\| P_{S^nV^n|\Pi(U^n, \delta,P_U), M}(\cdot|0,0)-P_{S^nV^n}(\cdot) \right\|=\left\| P_{S^nV^n|\Pi(U^n, \delta,P_U)}(\cdot|0)-P_{S^nV^n}(\cdot) \right\|, \label{markovchnequmIu} \\
     & \gamma_n''':= \frac{\gamma_n}{n}H(S^n|M=0,V^n,,H=0,\Pi(U^n, \delta,P_U)=1)  \xrightarrow{(n)} 0, \label{decaytozero1}\\
& \bar \gamma_n:=(1-\gamma_n)(1-\epsilon^*)\gamma_n''+(1-\gamma_n)~\epsilon^* \frac{\gamma_n'}{n}-\gamma_n'''. \notag
\end{align}
   Equation \eqref{typicalbnd} follows similarly to the proof of Theorem 1 in \cite{Villard-Pablo-journal}. Equation  \eqref{dataprocequivoc} is obtained as follows:
  \begin{flalign}
      & \frac 1n H\left(S^n|M=0,V^n,I_{U}(U^n, \delta)=0, H=0\right)\notag \\
      &\geq \frac 1n H\left(S^n|V^n,H=0\right)-\frac{\gamma_n'}{n} \label{eqtotvarapply} \\
        &>H_{P}(S|U,V)-\frac{\gamma_n'}{n}. \label{strictlygreater}&&
  \end{flalign}
Here, \eqref{eqtotvarapply} is obtained by an application of Lemma \ref{contentrtv}; and \eqref{strictlygreater} is due to the assumption that $H_P(S|U,V)< H_P(S|V)$.

It follows from Lemma \ref{approxdistmequzero} that $\rho_n^* \xrightarrow{(n)} 0$, which in turn implies that 
  \begin{align}
\frac{\gamma_n'}{n}   \xrightarrow{(n)} 0. \label{tendstozerofin}
  \end{align}
From  \eqref{expdecaytyp}, \eqref{decaytozero1} and \eqref{tendstozerofin}, we have that $\bar \gamma_n \xrightarrow{(n)} 0$. 
Hence, equation \eqref{finbndstrngconv} 
implies that $(H_P(U),I_P(V;U),\Lambda_0^*, \Lambda_{min}) \in \mathcal{R}_e(\epsilon^*)$ for some $\Lambda_0^*> H_P(S|U,V)$. Since $(H_P(U),I_P(V;U),\Lambda_0^*,\Lambda_{min}) \notin \mathcal{R}_e$, this implies that in general, the strong converse does not hold for HT with an equivocation privacy constraint. The same counterexample can be used in a similar manner to show that the strong converse does not hold  for HT with an average distortion privacy constraint either.   
\section{Conclusions} \label{sec:conclusion}
We have studied the DHT problem with a privacy constraint, with equivocation and average distortion under a causal disclosure assumption as the measures of privacy. We have established a single-letter inner bound on the rate-error exponent-equivocation and rate-error exponent-distortion trade-offs. We have also obtained the optimal rate-error exponent-equivocation and rate-error exponent-distortion trade-offs for two special cases, when the communication rate over the channel is zero, and for TACI under a privacy constraint. 
It is interesting to note that the strong converse for DHT does not hold when there is an additional privacy constraint in the system. 
Extending these results to the case when the communication between the observer and detector takes place over a noisy communication channel is an interesting avenue for future research.

\begin{appendices} 
\section{Proof of Lemma \ref{lem:determdet}} \label{prooflemmdetrmdet}
Note that for a stochastic detector, the type I and type II error probabilities are linear functions of $P_{\hat H|M,V^n}$. As a result, for each fixed $n$ and $f_n$, $ \alpha_n \left(f_n, g_n \right)$ and  $ \beta_n \left(f_n, g_n \right)$ for a stochastic detector $g_n$ can be thought of as the type I and type II errors achieved by  \enquote{time sharing} among a finite number of deterministic detectors. To see this, consider some ordering on the elements of the set $\mathcal{M} \times \V^n$ and let $\nu_i:=P_{\hat H|M,V^n}(0|i)$, $i \in [1:N] $, where $i$ denotes the $i^{th}$ element of $\mathcal{M} \times \V^n$ and $N=|\mathcal{M} \times \V^n|$. Then, we can write 
  \[     
    P_{\hat H|M,V^n}=  
\begin{bmatrix}
   \nu_1      &  1-\nu_1\\
    \nu_2 & 1-\nu_2 \\ \vdots &\vdots\\\nu_N & 1-\nu_N  \end{bmatrix}.
\]
Then, it is easy to see that $ P_{\hat H|M,V^n}= \sum_{i=1}^N \nu_i I_i$, where $I_i:=[e_i ~ 1-e_i ]$ and $e_i$ is an $N$ length vector with 1 at the $i^{th}$ component and 0 elsewhere.
Now, suppose $(\alpha_n^{(1)}, \beta_n^{(1)})$ and $(\alpha_n^{(2)},  \beta_n^{(2)})$ denote the pair of type I and type II error probabilities achieved by deterministic detectors $g_{n}^{(1)}$ and $g_{n}^{(2)}$, respectively. Let $\mathcal{A}_{1,n}$ and $\mathcal{A}_{2,n}$ denote their corresponding acceptance regions for $H_0$.  Let $g_{n}^{(\theta)}$ denote the stochastic detector formed by using  $g_{n}^{(1)}$ and $g_{n}^{(2)}$ with probabilities $\theta$ and $1-\theta$, respectively. From the above mentioned linearity property, it follows that $g_{n}^{(\theta)}$ achieves  type I and type II error probabilities of   $\alpha_n \left(f_n,g_{n}^{(\theta)}\right)=\theta  \alpha_n^{(1)}+(1-\theta) \alpha_n^{(2)}$ and  $\beta_n\left(f_n,g_{n}^{(\theta)}\right)=\theta \beta_n^{(1)}+(1-\theta) \beta_n^{(2)}$, respectively. Let $r(\theta)=\min(\theta,1-\theta)$. Then, for $\theta \in (0,1)$, 
\begin{align}
 - \frac 1n \log\left( \beta_n\left(f_n,g_{n}^{(\theta)}\right)\right) \leq   \min \left( -\frac 1n \log \beta_n^{(1)}, -\frac 1n \log \beta_n^{(2)} \right)-\frac{1}{n} \log(r(\theta)). \notag
\end{align}
Hence, either
\begin{align}
 \alpha_n^{(1)} \leq \alpha_n \left(f_n,g_{n}^{(\theta)}\right) \mbox{ and }  -\frac 1n \log\beta_n^{(1)} \geq  -\frac 1n \log\left( \beta_n\left(f_n,g_{n}^{(\theta)}\right)\right)+\frac{1}{n} \log(r(\theta)), \notag
\end{align}
or
\begin{align}
 \alpha_n^{(2)} \leq  \alpha_n \left(f_n,g_{n}^{(\theta)}\right) \mbox{ and }  -\frac 1n \log\beta_n^{(2)} \geq  -\frac 1n \log\left( \beta_n\left(f_n,g_{n}^{(\theta)}\right)\right)+\frac{1}{n} \log(r(\theta)). \notag
\end{align}
Thus, since $\frac{1}{n} \log(r(\theta)) \xrightarrow{(n)}0$, a stochastic detector does not offer any advantage over deterministic detectors in the trade-off between the error-exponent and the type I error probability.

\section{Proof of Lemma \ref{determfnatadv}} \label{determadvfnproof}
\revised{
Let $\tilde P^{(\mathcal{C}_n,0)}_{S^nU^nV^nM\hat S^n}= P_{S^nU^nV^nM}~\prod_{i=1}^n\tilde P_{\hat S_i|M,V^n,S^{i-1}}$ and $\tilde P^{(\mathcal{C}_n,1)}_{S^nU^nV^nM\hat S^n}= Q_{S^nU^nV^nM}~\prod_{i=1}^n\tilde P_{\hat S_i|M,V^n,S^{i-1}}$ denote the joint distribution of the r.v.'s $(S^n,U^n,V^n,M,\hat S^n)$ under hypothesis $H_0$ and $H_1$, respectively, where  $\tilde P_{\hat S_i|M,V^n,S^{i-1}}$ denotes $g^{(r)}_{i,n}$ for $i \in [n]$. Then, we have 
\begin{flalign}
&\min_{g^{(r)}_{i,n}}\mathbb{E}\left[d \left(S^n, \hat S^n\right) |H=j \right] \notag \\
&=\min_{\big\{\tilde P_{\hat S_i|M,V^n,S^{i-1}}\big\}_{i=1}^n}\mathbb{E}_{\tilde P^{(j)}} \left[d \left(S^n, \hat S^n \right) \right] \notag\\
&  =\min_{\big\{\tilde P_{\hat S_i|M,V^n,S^{i-1}}\big\}_{i=1}^n} \frac 1n \sum_{i=1}^n \mathbb{E}_{\tilde P^{(j)}} \left[ d \left(S_i, \hat S_i\right)\right]  \notag \\
&  =\frac 1n \sum_{i=1}^n \sum_{(m,v^n,s^{i-1})} \tilde P^{(j)}_{MV^nS^{i-1}}(m,v^n,s^{i-1})~\min_{\tilde P_{\hat S_i|M,V^n,S^{i-1}}(\cdot|m,v^n,s^{i-1}) }~\sum_{\hat s_i } \tilde P_{\hat S_i|M,V^n,S^{i-1}}(\hat s_i|m,v^n,s^{i-1}) ~\notag \\
& \qquad \qquad  \qquad  \qquad \qquad \qquad  \qquad \qquad \qquad  \qquad  \qquad \qquad \qquad  \qquad  \qquad  \mathbb{E}_{\tilde P^{(j)}_{S_i|M,V^n,S^{i-1}}(\cdot|m,v^n,s^{i-1})} \left[ d \left(S_i, \hat s_i\right)\right] \notag \\
& = \frac 1n \sum_{i=1}^n \sum_{m,v^n,s^{i-1}} \tilde P^{(j)}_{MV^nS^{i-1}}(m,v^n,s^{i-1})~\mathbb{E}_{\tilde P^{(j)}_{S_i|M,V^n,S^{i-1}}(\cdot|m,v^n,s^{i-1})} \left[ d \left(S_i, \phi_{ij}(m,v^n,s^{i-1}) \right)\right],  \notag &&
\end{flalign}
where 
\begin{align}
\phi_{ij}(m,v^n,s^{i-1})= \argmin_{\hat s \in \hat{\mathcal{S}}} \mathbb{E}_{ \tilde P^{(j)}_{S_i|M,V^n,S^{i-1}}(\cdot|m,v^n,s^{i-1})}\left[d(S_i,\hat s)\right]. \notag
\end{align}
Continuing, we have
\begin{flalign}
&\min_{g^{(r)}_{i,n}}\mathbb{E}\left[d \left(S^n, \hat S^n\right) |H=j \right] \notag \\
&= \frac 1n \sum_{i=1}^n \sum_{m,v^n,s^{i-1}} \tilde P^{(j)}_{MV^nS^{i-1}}(m,v^n,s^{i-1})~ \min_{\phi_i(m,v^n,s^{i-1})}\mathbb{E}_{\tilde P^{(j)}_{S_i|M,V^n,S^{i-1}}(\cdot|m,v^n,s^{i-1})} \left[ d \left(S_i, \phi_{i}(m,v^n,s^{i-1}) \right)\right]  \notag \\
&= \min_{\{\phi_i(m,v^n,s^{i-1})\}_{i=1}^n}\frac 1n \sum_{i=1}^n \mathbb{E}_{\tilde P^{(j)}} \left[ d \left(S_i, \phi_i(M,V^n,S^{i-1}) \right) \right]. \label{determfunach} &&
\end{flalign} This completes the proof.}
\section{Proof of Lemma \ref{approxdistmequzero}} \label{supplem}
We will first prove \eqref{althypapproxdist}. Fix $\delta>0$. For $\gamma>0$, define the following sets:
\begin{align}
\mathcal{B}_{0,\gamma}^{ \delta}& := \big\{y^n  \in \mathcal{T}_{[P_{Y}]_{\gamma}}^n:  P_{Y^n}(y^n) \geq  P_{Y^n|\Pi(X^n, \delta,P_X)}(y^n|0)\big\}, \notag
\end{align}
\begin{align}
\mathcal{C}_{0,\gamma}^{ \delta}&:= \big\{y^n  \in \mathcal{T}_{[P_{Y}]_{\gamma}}^n:  P_{Y^n}(y^n) <  P_{Y^n|\Pi(X^n, \delta,P_X)}(y^n|0)\big\},\\
\mathcal{B}_{1,\gamma}^{ \delta}& := \big\{y^n  \in \mathcal{T}_{[Q_{Y}]_{\gamma}}^n:  Q_{Y^n}(y^n) \geq  Q_{Y^n|\Pi(X^n, \delta,P_X)}(y^n|0)\big\}, \notag \\
\mathcal{C}_{1,\gamma}^{ \delta}&:= \big\{y^n  \in \mathcal{T}_{[Q_{Y}]_{\gamma}}^n:  Q_{Y^n}(y^n) <  Q_{Y^n|\Pi(X^n, \delta,P_X)}(y^n|0)\big\}, \notag \\
\mathcal{B}_{2,\gamma}^{ \delta}&:= \big\{y^n  \in \mathcal{T}_{[Q_{Y}]_{\gamma}}^n:  Q_{Y^n}(y^n) \geq  Q_{Y^n|\Pi(X^n, \delta,P_X)}(y^n|1)\big\}, \notag \\
\mathcal{C}_{2,\gamma}^{ \delta}&:= \big\{y^n  \in \mathcal{T}_{[Q_{Y}]_{\gamma}}^n:  Q_{Y^n}(y^n) <  Q_{Y^n|\Pi(X^n, \delta,P_X)}(y^n|1)\big\}. \notag
\end{align}
Then, we can write
\begin{flalign}
   & \big\| Q_{Y^n}(\cdot)-Q_{Y^n|\Pi(X^n, \delta,P_X)}(\cdot|1) \big\| \notag \\
   &= \frac 12 \sum_{y^n} \big|Q_{Y^n}(y^n)-Q_{Y^n|\Pi(X^n, \delta,P_X)}(y^n|1)\big| \notag \\
   &=\frac 12  \sum_{y^n \notin \mathcal{T}_{[Q_{Y}]_{\gamma}}^n} \big|Q_{Y^n}(y^n)-Q_{Y^n|\Pi(X^n, \delta,P_X)}(y^n|1)\big| +\frac 12  \sum_{y^n \in \mathcal{T}_{[Q_{Y}]_{\gamma}}^n} \big|Q_{Y^n}(y^n)-Q_{Y^n|\Pi(X^n, \delta,P_X)}(y^n|1)\big|\notag \\
   & \leq \frac 12  \sum_{y^n \notin \mathcal{T}_{[Q_{Y}]_{\gamma}}^n} Q_{Y^n}(y^n)+Q_{Y^n|\Pi(X^n, \delta,P_X)}(y^n|1) +\frac 12  \sum_{y^n \in \mathcal{T}_{[Q_{Y}]_{\gamma}}^n} \big|Q_{Y^n}(y^n)-Q_{Y^n|\Pi(X^n, \delta,P_X)}(y^n|1)\big|. \label{approxdist1} &&
      \end{flalign}
      Next, note that
      \begin{align}
       Q_{Y^n|\Pi(X^n, \delta,P_X)}(y^n|1)&=    Q_{Y^n}(y^n)~\frac{Q_{\Pi(X^n, \delta,P_X)|Y^n}(1|y^n)}{Q_{\Pi(X^n, \delta,P_X)}(1)}  \leq \frac{Q_{Y^n}(y^n)}{Q_{\Pi(X^n, \delta,P_X)}(1)} \leq 2 Q_{Y^n}(y^n), \label{bndtwicealthyp}
      \end{align}
      for sufficiently large $n$ (depending on $|\X|$), since $Q_{\Pi(X^n, \delta,P_X)}(1) \xrightarrow{(n)} 1$. Thus, for $n$ large enough,
      \begin{align}
           &\sum_{y^n \notin \mathcal{T}_{[Q_{Y}]_{\gamma}}^n} Q_{Y^n}(y^n)+Q_{Y^n|\Pi(X^n, \delta,P_X)}(y^n|1) \leq 3  \sum_{y^n \notin \mathcal{T}_{[Q_{Y}]_{\gamma}}^n} Q_{Y^n}(y^n) \leq e^{- n\Omega(\gamma)}.\label{bndterm1dist}
      \end{align}
We can bound the last term in \eqref{approxdist1} as follows:
       \begin{flalign}
   &\sum_{y^n \in \mathcal{T}_{[Q_{Y}]_{\gamma}}^n} \big|Q_{Y^n}(y^n)-Q_{Y^n|\Pi(X^n, \delta,P_X)}(y^n|1)\big| \notag \\
   & =\sum_{y^n \in \mathcal{B}_{2,\gamma}^{ \delta}} Q_{Y^n}(y^n)-Q_{Y^n|\Pi(X^n, \delta,P_X)}(y^n|1)+ \sum_{y^n \in \mathcal{C}_{2,\gamma}^{ \delta}} Q_{Y^n|\Pi(X^n, \delta,P_X)}(y^n|1)-Q_{Y^n}(y^n) \notag \\
   &=\sum_{y^n \in \mathcal{B}_{2,\gamma}^{ \delta}} Q_{Y^n}(y^n)-Q_{Y^n|\Pi(X^n, \delta,P_X)}(y^n|1)+ \sum_{y^n \in \mathcal{C}_{2,\gamma}^{ \delta}} Q_{Y^n|\Pi(X^n, \delta,P_X)}(y^n|1)-Q_{Y^n}(y^n) \notag \\
   &=\sum_{y^n \in \mathcal{B}_{2,\gamma}^{ \delta}} Q_{Y^n}(y^n)\left(1-\frac{Q_{Y^n|\Pi(X^n, \delta,P_X)}(y^n|1)}{Q_{Y^n}(y^n)}\right)+ \sum_{y^n \in \mathcal{C}_{2,\gamma}^{ \delta}}Q_{Y^n}(y^n)\left( \frac{Q_{Y^n|\Pi(X^n, \delta,P_X)}(y^n|1)}{Q_{Y^n}(y^n)}-1 \right) \notag \\
     &=\sum_{y^n \in \mathcal{B}_{2,\gamma}^{ \delta}} Q_{Y^n}(y^n)\left(1-\frac{Q_{\Pi(X^n, \delta,P_X)|Y^n}(1| y^n)}{Q_{\Pi(X^n, \delta,P_X)}(1)}\right)+ \sum_{y^n \in \mathcal{C}_{2,\gamma}^{ \delta}}Q_{Y^n}(y^n)\left( \frac{Q_{\Pi(X^n, \delta,P_X)|Y^n}(1| y^n)}{Q_{\Pi(X^n, \delta,P_X)}(1)}-1 \right) \notag   \\
   &\leq \sum_{y^n \in \mathcal{B}_{2,\gamma}^{ \delta}} Q_{Y^n}(y^n)\left(1-Q_{\Pi(X^n, \delta,P_X)|Y^n}(1| y^n)\right)+ \sum_{y^n \in \mathcal{C}_{2,\gamma}^{ \delta}}Q_{Y^n}(y^n)\left( \frac{1}{Q_{\Pi(X^n, \delta,P_X)}(1)}-1 \right). \label{bndterm1dist2}  &&
   \end{flalign}
Let $P_{\tilde Y}$ denote the type of $y^n$ and define \begin{align}
    E_n(\delta, \gamma) := \min _{P_{\tilde Y } \in \mathcal{P}_n\left(\mathcal{T}_{[Q_{Y}]_{\gamma}}^n\right) }\min_{P_{\tilde X} \in \mathcal{P}_n\left(\mathcal{T}_{[P_X]_{\delta}}^n\right) } D(P_{\tilde X|\tilde Y }||Q_{X|Y}|P_{\tilde Y}). \notag 
\end{align}
Then, for $y^n\in \mathcal{T}_{[Q_{Y}]_{\gamma}}^n$, arbitrary $\tilde \gamma>0$ and $n$ sufficiently large (depending on $|\X|,|\Y|, \delta,\gamma$), it follows from \cite[Lemma 2.6]{Csiszar-Korner} that
\begin{align}
&Q_{\Pi(X^n, \delta,P_X)|Y^n}(1| y^n) \geq 1-e^{-n\left( E_n(\delta, \gamma)-\tilde \gamma \right)},\label{typequfin1} \\
 \mbox{and } &Q_{\Pi(X^n, \delta,P_X)}(1) \geq 1-e^{-n(D(P_{X}||Q_X)-\tilde \gamma)}. \label{typequfin2}
\end{align}
From \eqref{approxdist1}, \eqref{bndterm1dist}, \eqref{bndterm1dist2}, \eqref{typequfin1} and \eqref{typequfin2}, it follows that
\begin{align}
    \Vert Q_{Y^n}(\cdot)-Q_{Y^n|\Pi(X^n, \delta,P_X)}(\cdot|1) \Vert \leq e^{- n\Omega(\gamma)}+e^{-n\left( E_n(\delta, \gamma)-\tilde \gamma\right)}+ e^{-n(D(P_{X}||Q_X)-\tilde \gamma)}. \label{sumthreetrmszero}
\end{align}
We next show that $  E_n(\delta, \gamma) >0$ for sufficiently small $\delta>0$ and $\gamma>0$. This would imply that the R.H.S of \eqref{sumthreetrmszero} converges exponentially to zero (for $\tilde \gamma$ small enough) with exponent $\bar \delta:= \min \left(\Omega(\gamma), E_n(\delta, \gamma)-\tilde \gamma,D(P_{X}||Q_X)-\tilde \gamma \right)$, thus proving \eqref{althypapproxdist}. We can write,
\begin{align}
    E_n(\delta, \gamma)  &\geq  \min _{P_{ \tilde Y } \in \mathcal{T}_{[Q_{Y}]_{\gamma}}^n }\min_{P_{\tilde X} \in \mathcal{T}_{[P_X]_{\delta}}^n } D(P_{\tilde X}||\hat Q_{X}) \label{KLdivconvex} \\
   &\geq 2 \left[  \min _{P_{\tilde Y } \in \mathcal{T}_{[Q_{Y}]_{\gamma}}^n }\min_{P_{\tilde X} \in \mathcal{T}_{[P_X]_{\delta}}^n }\Vert P_{\tilde X} - \hat Q_{X}\Vert^2 \right], \label{pinskerineq}
\end{align}
where 
\begin{align}
 \hat Q_{X}(x) := \sum_{y} P_{\tilde Y }(y) Q_{X|Y}(x|y). \notag 
\end{align}
Here, \eqref{KLdivconvex} follows due to the convexity of KL divergence \eqref{pinskerineq} is due to Pinsker's inequality \cite{Csiszar-Korner}.
We also have from the triangle inequality satisfied by total variation  that,
\begin{align}
    \big\| P_{\tilde X}- \hat Q_{X}\big\| \geq  \big\| P_{X}- Q_{X}\big\| - \big\| P_{\tilde X}- P_{X}\big\| - \big\| \hat Q_{X}-Q_{X}\big\|. \notag 
\end{align}
For  $y^n \in \mathcal{T}_{[Q_{Y}]_{\gamma}}^n $, 
\begin{align}
    \big\| \hat Q_{X}-Q_{X}\big\|  \leq  \big\| Q_{X|Y}P_{\tilde Y }- Q_{XY} \big\| \leq  \big\| P_{\tilde Y}- Q_{Y} \big\| \leq O(\gamma). \notag
\end{align}
Also, for $P_{\tilde X} \in \mathcal{T}_{[P_X]_{\delta}}^n $,
\begin{align}
    \big\| P_{\tilde X}- P_X  \big\| \leq O(\delta). \notag 
\end{align}
 Hence,
\begin{align}
     E_n(\delta, \gamma) &\geq 2\left(\big\| P_{X}- Q_{X}\big\|-O(\gamma)-O(\delta)\right)^2. \notag
\end{align}
Since $P_X \neq Q_X$, $ E_n(\delta, \gamma)>0$  for sufficiently small $\gamma>0$ and $\delta>0$. 
This completes the proof of \eqref{althypapproxdist}.

We next prove \eqref{nullhypapproxdist}. Similar to \eqref{approxdist1} and \eqref{bndtwicealthyp}, we have,
\begin{flalign}
  &  \big\| P_{Y^n}(\cdot)-P_{Y^n|\Pi(X^n, \delta,P_X)} (\cdot|0)\big\| \notag \\
  & \leq \frac 12 \sum_{y^n \notin \mathcal{T}_{[P_{Y}]_{\gamma}}^n} \left[ P_{Y^n}(y^n)+P_{Y^n|\Pi(X^n, \delta,P_X)}(y^n|0) \right]+ \frac 12 \sum_{y^n \in \mathcal{T}_{[P_{Y}]_{\gamma}}^n} \big|P_{Y^n}(y^n)-P_{Y^n|\Pi(X^n, \delta,P_X)}(y^n|0)\big|, \label{approxdist2}&&
\end{flalign}
and 
\begin{align}
    P_{Y^n|\Pi(X^n, \delta,P_X)}(y^n|0) \leq 2 P_{Y^n}(y^n), \label{bndtwicejntdist}
\end{align}
since $P_{\Pi(X^n, \delta,P_X)}(0) \xrightarrow{(n)} 1$.

Also, for $\gamma< \frac{\delta}{|\Y|}$ and sufficiently large $n$ (depending on $\delta, \gamma,|\X|,|\Y|$), we have
\begin{flalign}
 &\sum_{y^n \in \mathcal{T}_{[P_{Y}]_{\gamma}}^n} \big|P_{Y^n}(y^n)-P_{Y^n|\Pi(X^n, \delta,P_X)}(y^n|0)\big| \notag \\
   & =\sum_{y^n \in \mathcal{B}_{0, \gamma}^{ \delta}} P_{Y^n}(y^n)-P_{Y^n|\Pi(X^n, \delta,P_X)}(y^n|0)+ \sum_{y^n \in \mathcal{C}_{0,\gamma}^{ \delta}} P_{Y^n|\Pi(X^n, \delta,P_X)}(y^n|0)-P_{Y^n}(y^n) \notag \\
    &\leq \sum_{y^n \in \mathcal{B}_{0,\gamma}^{ \delta}} P_{Y^n}(y^n)\left(1-P_{\Pi(X^n, \delta,P_X)|Y^n}(0| y^n)\right)+ \sum_{y^n \in \mathcal{C}_{0,\gamma}^{ \delta}}P_{Y^n}(y^n)\left( \frac{1}{P_{ \Pi(X^n, \delta,P_X)}(0)}-1 \right)  \notag \\
 & \leq   \sum_{y^n \in \mathcal{B}_{0,\gamma}^{ \delta}} P_{Y^n}(y^n)e^{-n \Omega(\delta-\gamma|\Y|)}+ \sum_{y^n \in \mathcal{C}_{0,\gamma}^{ \delta}}P_{Y^n}(y^n) e^{-n\Omega(\delta)}  \label{tendszero1} \\
   & \leq e^{-n \Omega(\delta-\gamma|\Y|)}, \label{expprobdec}  &&
\end{flalign}
where to obtain \eqref{tendszero1}, we used
\begin{align}
 &   P_{ \Pi(X^n, \delta,P_X)}(0) \geq 1-e^{-n \Omega(\delta)},   \label{typicconcen1} \\
 \mbox{and }&   P_{\Pi(X^n, \delta,P_X)|Y^n}(0| y^n) \geq 1-e^{-n \Omega(\delta-\gamma|\Y|)}, \mbox{ for } y^n \in \mathcal{B}_{0,\gamma}^{ \delta} \mbox{ and } \gamma< \frac{\delta}{|\Y|} . \label{typicconcen2}
\end{align}
 Here, \eqref{typicconcen1} follows from \cite[Lemma 2.12]{Csiszar-Korner}, and \eqref{typicconcen2} follows from \cite[Lemma 2.10 and Lemma 2.12]{Csiszar-Korner}, respectively. Thus, from \eqref{approxdist2}, \eqref{bndtwicejntdist} and \eqref{expprobdec}, we can write that,
\begin{align}
 \big\| P_{Y^n}(\cdot)-P_{Y^n|\Pi(X^n, \delta,P_X)} (\cdot|0)\big\|
  \leq  e^{-n \Omega(\gamma)}+ e^{-n \Omega(\delta-\gamma|\Y|)} \xrightarrow{(n)} 0.\notag
\end{align}
This completes the proof of \eqref{nullhypapproxdist}. The proof of \eqref{althypapproxdist2} is exactly the same as \eqref{nullhypapproxdist}, with the only difference that the sets $\mathcal{B}_{1,\gamma}^{ \delta}$ and $\mathcal{C}_{1,\gamma}^{ \delta}$ are used in place of  $\mathcal{B}_{0,\gamma}^{ \delta}$ and $\mathcal{C}_{0,\gamma}^{ \delta}$, respectively.

\section{Proof of Theorem \ref{thm:genhtequ} and Theorem \ref{thm:genhtdist}} \label{genHTdist}
We  describe the encoding and decoding operations which are the same for both Theorem \ref{thm:genhtequ} and Theorem \ref{thm:genhtdist}. Fix positive numbers (small) $\eta,\delta>0$, and let $\delta':=\frac{\delta}{2},~\hat\delta := |\Ucal|\delta, ~\tilde \delta:=2 \delta$ and  $\bar{\delta}:= \frac{\delta'}{|\V|}$.\\
 \underline{\textbf{Codebook Generation}}:
 Fix  a finite alphabet $\mathcal{W}$ and a  conditional distribution $P_{W|U}$. 
 Let  $\mathbb{B}_n=\big\{W^n(j),~j \in [M_n']\big\}$,  $M_n':= e^{n(I_P(U:W)+\eta)}$, denote a random codebook such that each $W^n(j)$ is randomly and independently generated according to distribution $\prod_{i=1}^n P_W(w_i)$, where  
\begin{align}
P_{W}(w)= \sum_{u \in \Ucal}P_U(u) P_{W|U}(w|u). \notag
\end{align}
Denote a realization of $\mathbb{B}_n$ by $\mathcal{B}_n$ and the support of $\mathbb{B}_n$ by $\mathfrak{B}_n$. 
\\
 \underline{\textbf{Encoding}}: For a given codebook $\mathcal{B}_n$, let
 \begin{align}
&P_{\mathsf{E}_{\mathsf{u}}}^{(\mathcal{B}_n)}(j|u^n)  := \frac{\prod_{i=1}^n  P_{U|W}(u_i|w_{i}(j))}{\sum_{j'}\prod_{i=1}^n  P_{U|W}(u_i|w_{i}(j')) )}, \label{stochasticenc}
\end{align}
denote the likelihood encoding function. If $I_P(U;W)+\eta+|\Ucal||\W|\frac{\log(n+1)}{n}> R$, the observer performs uniform random binning on the indices in $\left[M_n' \right]$, i.e.,  for each $j \in \left[M_n' \right]$, it selects an index  uniformly at random from the set  $\tilde{\mathcal{M}}_n:=\left[e^{n\left(R-|\Ucal||\W|\frac{\log(n+1)}{n}\right)}\right]$. Denote the random binning function by $f_{\mathbb{B}}$ and  a realization of it by $f_{\mathsf{B}}$. If $I_P(U;W)+\eta+|\Ucal||\W|\frac{\log(n+1)}{n}\leq R$, set $f_{\mathbb{B}}$ as the identity function with probability one, i.e., $f_{\mathbb{B}}(j)=j$. If $u^n \in \mathcal{T}_{[P_{U}]_{\delta'}}^n$,  then the observer outputs the message $m=(t,f_{\mathsf{B}}(j))$ if $I_P(U;W)+\eta+|\Ucal||\W|\frac{\log(n+1)}{n} > R$ or $m=(t,j)$  otherwise, where $j \in [M_n']$  is chosen randomly with probability $P_{\mathsf{E}_{\mathsf{u}}}^{(\mathcal{B}_n)}(j|u^n)$ and $t$ denotes the index of the joint type of $(u^n,w^n(j))$ in the set of types $ \mathcal{P}^n(\Ucal \times \W)$. If $u^n \notin \mathcal{T}_{[P_{U}]_{\delta'}}^n$, the observer outputs the \textit{error} message $M=0$. Note that $|\mathcal{M}| \leq e^{nR}$ since the total number of types in $\mathcal{P}^n(\Ucal \times \W)$ is upper bounded by $(n+1)^{|\Ucal||\W|}$ \cite[Lemma 2.2]{Csiszar-Korner}. Let $\mathbb{C}_n:=(\mathbb{B}_n,f_{\mathbb{B}})$, and let  $\mathcal{C}_n=(\mathcal{B}_n,f_{\mathsf{B}})$ and $\mu_n(\cdot)$ denote its realization and probability distribution, respectively. For a given $\mathcal{C}_n$, let $f_n^{(\mathcal{C}_n)}$ represent the  encoder induced by the above operations, where $f_n^{(\mathcal{C}_n)}: \Ucal^n \rightarrow \mathcal{P}(\mathcal{M})$ and $\mathcal{M}:=[e^{nR}]$. \\
 \underline{\textbf{Decoding}}: 
If $M=0$ or $t \notin \mathcal{T}_{[P_{UW}]_{\delta}}^n$, $\hat H=1$ is declared. Else, given $m=(t,f_{\mathsf{B}}(j))$ and $V^n=v^n$, the detector decodes for a codeword $ \hat w^n:=w^n(\hat j) \in \mathcal{T}_{[P_{W}]_{\hat\delta}}^n$  in the codebook $\mathcal{B}_n$  such that 
\begin{align}
  \hat j&= \argmin_{ \substack{l:~ f_{\mathsf{B}}(l)=f_{\mathsf{B}}(j), \\  w^n(l)\in \mathcal{T}_{[P_{W}]_{\hat\delta}}^n}} H_e(w^n(l)|v^n),   \mbox{ if }I_P(U;W)+\eta+\frac 1n |\Ucal||\W|\log(n+1)> R, \notag \\
    \hat j&=j, \mbox{ otherwise}. \notag 
\end{align}
Denote the above decoding rule by $P_{\mathsf{ED}}^{(\mathcal{C}_n)}$, where $P_{\mathsf{ED}}^{(\mathcal{C}_n)}: \mathcal{M} \times \V^n \rightarrow \mathcal{J}$.
 The detector declares $\hat H=0$ if $(\hat w^n,v^n) \in \mathcal{T}_{[P_{WV}]_{\tilde \delta}}^n$ and $\hat H=1$ otherwise. Let $g_n^{(\mathcal{C}_n)}:\mathcal{M} \times \V^n \rightarrow \hat{\mathcal{H}}$ stand for the decision rule induced by the above operations. \\
\underline{\textbf{System induced distributions and auxiliary distributions}}:\\
 The system induced probability distribution when $H=0$
is given by 
\begin{flalign}
 & \tilde P^{(\mathcal{C}_n,0)}(s^n,u^n,v^n,j,w^n,m, \hat j,\hat w^n) \notag\\
&=\left[\prod_{i=1}^n P_{SUV}(s_i,u_i,v_i,z_i)\right] P_{\mathsf{E}_{\mathsf{u}}}^{(\mathcal{B}_n)}(j|u^n)  \ind(w^n(j)= w^n)\ind(f_{\mathsf{B}}(j)=m) ~\ind\left(\hat j=P_{\mathsf{ED}}^{(\mathcal{C}_n)}(m,v^n)\right) \notag \\& \qquad \ind(w^n(\hat j)= \hat w^n), \qquad \qquad \qquad \qquad \qquad \qquad \qquad \qquad \qquad \qquad \qquad \qquad \qquad \qquad \qquad \qquad \mbox{if } u^n \in \mathcal{T}_{[P_U]_{\delta'}}^n, \label{sysinddistgen}&&
 \end{flalign}
and 
\begin{flalign}
 \tilde P^{(\mathcal{C}_n,0)}(s^n,u^n,v^n,m)=
\left[\prod_{i=1}^n P_{SUV}(s_i,u_i,v_i)\right] \ind(m=0)   ,~\mbox{if } u^n \notin \mathcal{T}_{[P_U]_{\delta'}}^n. \label{sysinddistgen2} &&
 \end{flalign}
Consider two auxiliary distribution $\tilde{\Psi}$ and  $\Psi$ given by
\begin{flalign}
&\tilde \Psi^{(\mathcal{C}_n,0)}(s^n,u^n,v^n,j,w^n,m, \hat j,\hat w^n) \notag \\
& := \left[\prod_{i=1}^n P_{SUV}(s_i,u_i,v_i)\right] P_{\mathsf{E}_{\mathsf{u}}}^{(\mathcal{B}_n)}(j|u^n)  \ind(w^n(j)= w^n)\ind(f_{\mathsf{B}}(j)=m) ~\ind \left(\hat j=P_{\mathsf{ED}}^{(\mathcal{C}_n)}(m,v^n)\right) \ind(w^n(\hat j)= \hat w^n), \label{auxdist1gen}&&
 \end{flalign}
 and 
 \begin{flalign}
&  \Psi^{(\mathcal{C}_n,0)}(s^n,u^n,v^n,j,w^n,m, \hat j,\hat w^n) \notag \\
& := \frac{1}{M_n'}~  \ind(w^n(j)=w^n)~\left[\prod_{i=1}^n P_{U|W}(u_i|w_i)\right] \left[ \prod_{i=1}^n P_{VS|U}(v_i,s_i|u_i)\right]  ~\ind(f_{\mathsf{B}}(j)=m)~\ind \left(\hat j=P_{\mathsf{ED}}^{(\mathcal{C}_n)}(m,v^n)\right) \notag \\
&\qquad  \ind(w^n(\hat j)= \hat w^n). \label{auxdist2gen}&&
 \end{flalign}
Let $\tilde P^{(\mathcal{C}_n,1)}$ and  $\tilde \Psi^{(\mathcal{C}_n,1)}$ denote probability distributions under $H=1$ defined by the R.H.S. of \eqref{sysinddistgen}, \eqref{sysinddistgen2} and \eqref{auxdist1gen} with $P_{SUV}$ replaced by $Q_{SUV}$, and let $\Psi^{(\mathcal{C}_n,1)}$  denote the R.H.S. of  \eqref{auxdist2gen} with $P_{VS|U}$ replaced by $Q_{VS|U}$. Note that the encoder $f_n^{(\mathcal{C}_n)}$ is such that $P_{\mathsf{E}_{\mathsf{u}}}^{(\mathcal{B}_n)}(j|u^n)=   \Psi^{(\mathcal{C}_n,0)}(j|u^n)$ and hence, the only difference between the joint distribution  $ \Psi^{(\mathcal{C}_n,0)}$ and $\tilde \Psi^{(\mathcal{C}_n,0)}$ is the marginal distribution of $U^n$. By the soft-covering lemma \cite{Wyner-1975-comminfo} \cite{Cuff-2013}, it follows that for some $\gamma_1>0$,
\begin{align}
\mathbb{E}_{\mu_n} \left[\left\|   \Psi^{(\mathbb{C}_n,0)}_{U^n}-\tilde \Psi^{(\mathbb{C}_n,0)}_{U^n} \right\| \right] \leq e^{-n\gamma_1 } \xrightarrow{(n)} 0. \label{auxmatchmarggen}
 \end{align}
Hence, from \cite[Property 2(d)]{Schieler-Cuff-2014}, it follows that
\begin{align}
\mathbb{E}_{\mu_n} \left[\left\|  \Psi^{(\mathbb{C}_n,0)}-\tilde \Psi^{(\mathbb{C}_n,0)} \right\|  \right] \leq e^{-n\gamma_1}. \label{auxmatchgen}
 \end{align}
Also, note that the only difference between the distributions $\tilde P^{(\mathcal{C}_n,0)}$ and $\tilde \Psi^{(\mathcal{C}_n,0)}$ is $P_{\mathsf{E}_{\mathsf{u}}}^{(\mathcal{B}_n)}$ when $U^n \notin \mathcal{T}_{[P_U]_{\delta'}}^n$. Since 
 \begin{align}
  \mathbb{P}\left(U^n \notin \mathcal{T}_{[P_U]_{\delta'}}^n|H=0\right) \leq  e^{-n\Omega(\delta')}, \label{msgzeroprob}   
 \end{align}
 it follows that
 \begin{align}
\mathbb{E}_{\mu_n} \left[\left\| \tilde P^{(\mathbb{C}_n,0)} -\tilde \Psi^{(\mathbb{C}_n,0)} \right\|  \right] \leq e^{-n\Omega(\delta')}. \label{sysauxmatchgen2}
 \end{align}
 Equations \eqref{auxmatchgen} and \eqref{sysauxmatchgen2} together imply via  \cite[Property 2(c)]{Schieler-Cuff-2014} that
  \begin{align}
\mathbb{E}_{\mu_n} \left[\left\| \tilde P^{(\mathbb{C}_n,0)} -  \Psi^{(\mathbb{C}_n,0)} \right\| \right] \leq   e^{-n\Omega(\delta')}+e^{-n\gamma_1} \xrightarrow{(n)} 0. \label{sysauxmatchgen}
 \end{align}
 Note that for  $l \in \{0,1\}$, the joint distribution $\Psi^{(\mathcal{C}_n,l)}$ satisfies 
 \revised{
 \begin{align}
  S_i-(w_i(J),V_i)-(M,w^n(J),V^n,S^{i-1}),~i \in [n].\label{Markcondauxdist}
  \end{align}}
  Also, since $I_P(U;W)+ \eta>0$, by the application of soft-covering lemma,
\begin{align}
\mathbb{E}_{\mu_n} \left[\sum_{i=1}^n \left\| P_{W}-  \Psi^{(\mathbb{C}_n,l)}_{W_{i}(J)} \right\| \bigg|H=l \right]  &\leq e^{-\gamma_{3} n} \xrightarrow{(n)}0, ~ l=0,1, \label{auxlettermatchwgen}
\end{align}
for some $\gamma_{3}>0$.

If $Q_U=P_U$, then it again follows from the soft-covering lemma that
\begin{align}
\mathbb{E}_{\mu_n} \left[\left\| \Psi^{(\mathbb{C}_n,1)}_{U^n}-\tilde \Psi^{(\mathbb{C}_n,1)}_{U^n} \right\|\right] \leq e^{-\gamma_1n} \xrightarrow{(n)} 0, \label{auxmatchmarggenalt}
 \end{align}
thereby implying that
\begin{align}
\mathbb{E}_{\mu_n} \left[\left\| \Psi^{(\mathbb{C}_n,1)}-\tilde \Psi^{(\mathbb{C}_n,1)} \right\| \right] \leq e^{-\gamma_1n}. \label{auxmatchgenalt}
 \end{align}
Also, note that the only difference between the distributions $\tilde P^{(\mathcal{C}_n,1)}$ and $\tilde \Psi^{(\mathcal{C}_n,1)}$ is $P_{\mathsf{E}_{\mathsf{u}}}^{(\mathcal{B}_n)}$ when $U^n \notin \mathcal{T}_{[P_U]_{\delta'}}^n$. Since $Q_U=P_U$ implies $\mathbb{P}\left(U^n \notin \mathcal{T}_{[P_U]_{\delta'}}^n\big|H=1\right) \leq e^{-n \Omega(\delta')}$, it follows that 
 \begin{align}
\mathbb{E}_{\mu_n} \left[\left\| \tilde P^{(\mathbb{C}_n,1)} -\tilde \Psi^{(\mathbb{C}_n,1)} \right\| \right] \leq e^{-n \Omega(\delta')}. \label{sysauxmatchgen2alt}
 \end{align}
 Eqns. \eqref{auxmatchgenalt} and \eqref{sysauxmatchgen2alt} together imply that
  \begin{align}
\mathbb{E}_{\mu_n} \left[\left\| \tilde P^{(\mathbb{C}_n,1)} - \Psi^{(\mathbb{C}_n,1)} \right\| \right] \leq e^{-n\Omega(\delta')}+e^{-\gamma_1n} \xrightarrow{(n)}0. \label{sysauxmatchgenalt}
 \end{align}
Let $\bar {\mathbb{P}}_{\tilde P^{(\mathbb{C}_n,0)}}=\mathbb{E}_{\mu_n}\left[\mathbb{P}_{\tilde P^{(\mathbb{C}_n,0)}}\right]$ and $\bar {\mathbb{P}}_{\tilde P^{(\mathbb{C}_n,1)}}=\mathbb{E}_{\mu_n}\left[\mathbb{P}_{\tilde P^{(\mathbb{C}_n,1)}}\right]$ denote the expected probability measure (random coding measure) induced by PMF's $\tilde P^{(\mathbb{C}_n,0)}$ and  $\tilde P^{(\mathbb{C}_n,1)}$, respectively. Then, note that from \eqref{auxdist2gen}, \eqref{sysauxmatchgen}, \eqref{auxlettermatchwgen} and the weak law of large numbers,
 \begin{align}
    \bar {\mathbb{P}}_{\tilde P^{(\mathbb{C}_n,0)}}\left(\big(U^n,W^n(J)\big) \in \mathcal{T}_{[P_{UW}]_{\delta}}^n\right) \geq 1-e^{-n \Omega(\delta)} \xrightarrow{(n)} 1. \label{typicalwlln} 
 \end{align}
   \underline{\textbf{Analysis of type I and type II error probabilities}}:\\
 We analyze  type I and type II error probabilities of the coding scheme mentioned above averaged over the random ensemble $\mathbb{C}_n$.\\
    \underline{\textbf{Type I error probability}}:\\
 Note that a type I error occurs only if one of the following events occur:
\begin{flalign}
\mathcal{E}_{\mathsf{TE}}&= \left\lbrace (U^n,V^n) \notin \mathcal{T}_{[P_{UV}]_{\bar \delta}}^n\right\rbrace, \notag \\
\mathcal{E}_{\mathsf{SE}}&= \left\lbrace T \notin \mathcal{P}_n\left(\mathcal{T}_{[P_{UW}]_{\delta}}^n\right)\right\rbrace, \notag   \\
\mathcal{E}_{\mathsf{ME}}&= \left\lbrace \big(V^n,W^n(J)\big) \notin \mathcal{T}_{[P_{VW}]_{\tilde \delta}}^n \right\rbrace, \notag  \\
\mathcal{E}_{\mathsf{DE}}&=   \Bigg\{ \exists ~ l \in  \left[e^{n(I(U;W)+\eta)}\right],~l \neq J: f_{\mathbb{B}}(l)=f_{\mathbb{B}}(J),~W^n(l) \in \mathcal{T}_{[P_{W}]_{\hat\delta}}^n,~H_e\big(W^n(l)|V^n\big) \leq H_e\big(W^n(J)|V^n\big) \Bigg\}.\notag &&
\end{flalign}
Let $\mathcal{E}:= \mathcal{E}_{\mathsf{TE}} \cup \mathcal{E}_{\mathsf{SE}} \cup \mathcal{E}_{\mathsf{ME}}  \cup \mathcal{E}_{\mathsf{DE}}$. Then, the expected type I error probability over $\mathbb{C}_n$ be upper bounded as 
\begin{align}
\mathbb{E}_{\mu_n} \left[\alpha_n\left(f_n^{(\mathbb{C}_n)}, g_n^{(\mathbb{C}_n)}\right)\right] \leq      \bar {\mathbb{P}}_{\tilde P^{(\mathbb{C}_n,0)}}(\mathcal{E}).\label{expt1errprob} 
\end{align}
 Note that $    \bar {\mathbb{P}}_{\tilde P^{(\mathbb{C}_n,0)}}(\mathcal{E}_{\mathsf{TE}})$ tends to $0$ asymptotically by the weak law of large numbers. From \eqref{typicalwlln},  $    \bar {\mathbb{P}}_{\tilde P^{(\mathbb{C}_n,0)}}(\mathcal{E}_{\mathsf{SE}}) \xrightarrow{(n)} 0$.
Given $\mathcal{E}_{\mathsf{SE}}^c$ and $\mathcal{E}_{\mathsf{TE}}^c$ holds, it follows from the Markov chain relation $V-U-W$ and the Markov lemma \cite{Elgamalkim}, that $    \bar {\mathbb{P}}_{\tilde P^{(\mathbb{C}_n,0)}}(\mathcal{E}_{\mathsf{ME}}) \xrightarrow{(n)} 0$. 
Also, as in the proof of Theorem 2 in \cite{SD_2020}, it follows that
\begin{align}
  &     \bar {\mathbb{P}}_{\tilde P^{(\mathbb{C}_n,0)}}(\mathcal{E}_{\mathsf{DE}}|  ~V^n=v^n,W^n(J)=w^n,\mathcal{E}_{\mathsf{ME}}^c \cap \mathcal{E}_{\mathsf{SE}}^c \cap \mathcal{E}_{\mathsf{TE}}^c)  \leq e^{-n\left(R-I_P(U;W|V)-\delta_n^{(1)}\right)}, \label{type1bnd}
\end{align}
where $\delta_n^{(1)} \xrightarrow{(n)} \eta+ O(\delta)$. Thus, if $R> I_P(U;W|V)$, it follows by choosing $\eta=O(\delta)$ that for $\delta>0$ small enough, the R.H.S. of \eqref{type1bnd} tends to zero asymptotically. By the union bound on probability, the R.H.S. of \eqref{expt1errprob} tends to zero.\\
\underline{\textbf{Type II error probability}}:\\
Let $\delta''=|\W|\tilde \delta $. Note that a type II error occurs only if $V^n \in \mathcal{T}_{[P_{V}]_{\delta''}}^n$ and  $M \neq 0$, i.e., $U^n \in \mathcal{T}_{[P_{U}]_{\delta'}}^n$ and $T \in \mathcal{T}_{[P_{UW}]_{\delta}}^n$. Hence, we can restrict the type II error analysis to only such $(U^n,V^n)$. Denoting the event that a type II error occurs by $\mathcal{D}_0$, we have
\begin{flalign}
\mathbb{E}_{\mu_n} \left[ \beta_n\left(f_n^{(\mathbb{C}_n)}, g_n^{(\mathbb{C}_n)}\right)\right]= \sum_{u^n,v^n }     \bar {\mathbb{P}}_{\tilde P^{(\mathbb{C}_n,1)}}(U^n=u^n,V^n=v^n) ~    \bar {\mathbb{P}}_{\tilde P^{(\mathbb{C}_n,1)}}(\mathcal{D}_0|U^n=u^n,V^n=v^n). \label{firsteqspldist}&&
\end{flalign}
Let $\mathcal{E}_{\mathsf{NE}} :=  \mathcal{E}_{\mathsf{SE}}^c \cap \left\{V^n \in \mathcal{T}_{[V]_{\delta''}}^n \right\} \cap \left\{U^n \in \mathcal{T}_{[U]_{\delta'}}^n\right\} $. The last term in \eqref{firsteqspldist} can be upper bounded as follows:
 \begin{flalign}
&\bar {\mathbb{P}}_{\tilde P^{(\mathbb{C}_n,1)}}(\mathcal{D}_0|U^n=u^n,V^n=v^n) \notag \\
&= \bar {\mathbb{P}}_{\tilde P^{(\mathbb{C}_n,1)}}(\mathcal{E}_{\mathsf{NE}}|U^n=u^n,V^n=v^n) ~\bar {\mathbb{P}}_{\tilde P^{(\mathbb{C}_n,1)}}(\mathcal{D}_0|U^n=u^n,V^n=v^n,\mathcal{E}_{\mathsf{NE}}) \notag \\
& \leq\bar {\mathbb{P}}_{\tilde P^{(\mathbb{C}_n,1)}}(\mathcal{D}_0|U^n=u^n,V^n=v^n,\mathcal{E}_{\mathsf{NE}}) \notag \\
&=\sum_{j,\tilde m}\bar {\mathbb{P}}_{\tilde P^{(\mathbb{C}_n,1)}}(J=j,f_{\mathbb{B}}(J)=\tilde m|U^n=u^n,V^n=v^n,\mathcal{E}_{\mathsf{NE}})\notag \\
& \qquad \qquad \bar {\mathbb{P}}_{\tilde P^{(\mathbb{C}_n,1)}}(\mathcal{D}_0|~U^n=u^n,V^n=v^n,J=j,f_{\mathbb{B}}(J)=\tilde m,\mathcal{E}_{\mathsf{NE}})\label{symmcodexpind} \\
&=\bar {\mathbb{P}}_{\tilde P^{(\mathbb{C}_n,1)}}(\mathcal{D}_0|~U^n=u^n,V^n=v^n,J=1,f_{\mathbb{B}}(J)=1,\mathcal{E}_{\mathsf{NE}}) \label{symmcodenc} \\
&=\sum_{\substack{w^n \in \W^n }} \bar {\mathbb{P}}_{\tilde P^{(\mathbb{C}_n,1)}}(W^n(1)=w^n |U^n=u^n,V^n=v^n,J=1,f_{\mathbb{B}}(J)=1, ~\mathcal{E}_{\mathsf{NE}}) \notag \\
& \qquad \qquad \qquad \quad \bar {\mathbb{P}}_{\tilde P^{(\mathbb{C}_n,1)}}(\mathcal{D}_0|U^n=u^n,V^n=v^n,J=1,f_{\mathbb{B}}(J)=1,W^n(1)=w^n,~\mathcal{E}_{\mathsf{NE}}). \label{lsttermexpdist}&&
\end{flalign}
where \eqref{symmcodenc} follows since the term in \eqref{symmcodexpind} is independent of the indices $(j,\tilde m)$ due to the symmetry of the codebook generation, encoding and decoding procedure. 
The first term in \eqref{lsttermexpdist} can be upper bounded as
\begin{align}
 \bar {\mathbb{P}}_{\tilde P^{(\mathbb{C}_n,1)}}(W^n(1)=w^n |U^n=u^n,V^n=v^n,J=1,f_{\mathbb{B}}(J)=1, ~\mathcal{E}_{\mathsf{NE}}) \leq \frac{1}{|\mathcal{T}_{P_{\tilde W|\tilde U}}|}\leq e^{-n(H(\tilde W|\tilde U)-\frac{1}{n}|\Ucal||\W|\log(n+1))}. \label{bndcodewordprob}
\end{align}
To obtain \eqref{bndcodewordprob}, we used the fact that $P_{\mathsf{E}_{\mathsf{u}}}^{(\mathcal{B}_n)}(1|u^n)$ in \eqref{stochasticenc} is invariant to the joint type  $P_{\tilde U \tilde W}$ of $(U^n,W^n(1))=(u^n,w^n)$ (keeping all the other codewords fixed). This in turn implies that given $\mathcal{E}_{\mathsf{NE}}$, each sequence in the conditional type class $\mathcal{T}_{P_{\tilde W|\tilde U}}(u^n)$ is equally likely (in the randomness induced by $\mathbb{B}_n$ and stochastic encoding in \eqref{stochasticenc}) and its probability is upper bounded by $\frac{1}{\big|\mathcal{T}_{P_{\tilde W|\tilde U}}\big|}$. 
Defining the events
\begin{flalign}
&\mathcal{E}_{\mathsf{BE}}:=\left\lbrace \exists ~ l \in  \left[M_n'\right],~ l\neq J,~ f_{\mathbb{B}}(l)=  M,~W^n(l)) \in \mathcal{T}_{[ P_{W}]_{\hat  \delta}}^n,
  (V^n,W^n(l)) \in \mathcal{T}_{[P_{VW}]_{\tilde\delta}}^n \right\rbrace, \label{binerrevnt} \\
&\mathcal{F}:= \{U^n=u^n,V^n=v^n,J=1,f_{\mathbb{B}}(J)=1,W^n(1)=w^n,~\mathcal{E}_{\mathsf{NE}}\}, \label{ev1} \\
&\mathcal{F}_1:= \{U^n=u^n,V^n=v^n,J=1,f_{\mathbb{B}}(J)=1,W^n(1)=w^n,~\mathcal{E}_{\mathsf{NE}},~\mathcal{E}_{\mathsf{BE}}^c\}, \label{ev2} \\
\mbox{and }&\mathcal{F}_2:= \{U^n=u^n,V^n=v^n,J=1,f_{\mathbb{B}}(J)=1,W^n(1)=w^n,~\mathcal{E}_{\mathsf{NE}},~ \mathcal{E}_{\mathsf{BE}}\}, \label{ev3} &&
\end{flalign}
the last term in \eqref{lsttermexpdist} can be written as 
\begin{align}
 & \bar {\mathbb{P}}_{\tilde P^{(\mathbb{C}_n,1)}}(\mathcal{D}_0|\mathcal{F}) =  \bar {\mathbb{P}}_{\tilde P^{(\mathbb{C}_n,1)}}(\mathcal{E}_{\mathsf{BE}}^c|\mathcal{F}) ~ \bar {\mathbb{P}}_{\tilde P^{(\mathbb{C}_n,1)}}(\mathcal{D}_0|\mathcal{F}_1) +  \bar {\mathbb{P}}_{\tilde P^{(\mathbb{C}_n,1)}}(\mathcal{E}_{\mathsf{BE}}|\mathcal{F})~  \bar {\mathbb{P}}_{\tilde P^{(\mathbb{C}_n,1)}}(\mathcal{D}_0|\mathcal{F}_2).  \label{t2eecontdist}
\end{align}
The analysis of the terms in \eqref{t2eecontdist} is essentially similar to that given in the proof of Theorem 2 in \cite{SD_2020}, except for a subtle difference that we mention next. 
In order to bound the \textit{binning} error event $\mathcal{E}_{\mathsf{BE}}$, we require an upper bound similar to
\begin{align}
 \bar {\mathbb{P}}_{\tilde P^{(\mathbb{C}_n,1)}} \left(W^n(l)= \tilde w^n| \mathcal{F} \right) \leq 2~  \bar {\mathbb{P}}_{\tilde P^{(\mathbb{C}_n,1)}}(W^n(l)=\tilde w^n),~\forall ~\tilde w^n \in \W^n, \label{bndforequcdwrd}
\end{align}
that is used in
 the proof of Theorem 2 in \cite{SD_2020}. Note that the stochastic encoding scheme considered here is different from the encoding scheme in  \cite{SD_2020}. In place \eqref{bndforequcdwrd}, we will show that for $l \neq 1$,
    \begin{align}
    & \bar {\mathbb{P}}_{\tilde P^{(\mathbb{C}_n,1)}}(W^n(l)=\tilde w^n|~\mathcal{F}) \leq 3~  \bar {\mathbb{P}}_{\tilde P^{(\mathbb{C}_n,1)}}(W^n(l)=\tilde w^n), \label{bndthreeval}
\end{align}
which suffices for the proof. Note that
\begin{flalign}
 & \bar {\mathbb{P}}_{\tilde P^{(\mathbb{C}_n,1)}}(W^n(l)=\tilde w^n|\mathcal{F}) \notag \\
&= \bar {\mathbb{P}}_{\tilde P^{(\mathbb{C}_n,1)}}(W^n(l)=\tilde w^n| U^n=u^n,V^n=v^n) \frac{ \bar {\mathbb{P}}_{\tilde P^{(\mathbb{C}_n,1)}}(W^n(1)=w^n| W^n(l)=\tilde w^n, U^n=u^n,V^n=v^n)}{ \bar {\mathbb{P}}_{\tilde P^{(\mathbb{C}_n,1)}}(W^n(1)=w^n|  U^n=u^n,V^n=v^n)} \notag \\
&\qquad \frac{ \bar {\mathbb{P}}_{\tilde P^{(\mathbb{C}_n,1)}}(J=1| W^n(1)=w^n, W^n(l)=\tilde w^n, U^n=u^n,V^n=v^n)}{ \bar {\mathbb{P}}_{\tilde P^{(\mathbb{C}_n,1)}}(J=1| W^n(1)=w^n, U^n=u^n,V^n=v^n)}  \label{boundcdwrd} \\
&\qquad \frac{ \bar {\mathbb{P}}_{\tilde P^{(\mathbb{C}_n,1)}}(f_{\mathbb{B}}(J)=1| J=1, W^n(1)=w^n, W^n(l)=\tilde w^n, U^n=u^n,V^n=v^n)}{ \bar {\mathbb{P}}_{\tilde P^{(\mathbb{C}_n,1)}}(f_{\mathbb{B}}(J)=1| J=1, W^n(1)=w^n, U^n=u^n,V^n=v^n)} \notag \\
&\qquad \frac{ \bar {\mathbb{P}}_{\tilde P^{(\mathbb{C}_n,1)}}(\mathcal{E}_{\mathsf{NE}}|f_{\mathbb{B}}(J)=1, J=1, W^n(1)=w^n, W^n(l)=\tilde w^n, U^n=u^n,V^n=v^n)}{ \bar {\mathbb{P}}_{\tilde P^{(\mathbb{C}_n,1)}}(\mathcal{E}_{\mathsf{NE}}|f_{\mathbb{B}}(J)=1, J=1, W^n(1)=w^n, U^n=u^n,V^n=v^n)} \label{equtmp0} &&
\end{flalign}
Since the codewords are generated independently of each other and the binning operation is done independent of the codebook generation, we have 
\begin{align}
& \bar {\mathbb{P}}_{\tilde P^{(\mathbb{C}_n,1)}}(W^n(1)=w^n| W^n(l)=\tilde w^n, U^n=u^n,V^n=v^n)= \bar {\mathbb{P}}_{\tilde P^{(\mathbb{C}_n,1)}}(W^n(1)=w^n| U^n=u^n,V^n=v^n), \label{equtmp1}
\end{align}
and 
\begin{align}
& \bar {\mathbb{P}}_{\tilde P^{(\mathbb{C}_n,1)}}(f_{\mathbb{B}}(J)=1| J=1, W^n(1)=w^n, W^n(l)=\tilde w^n, U^n=u^n,V^n=v^n) \notag \\
&= \bar {\mathbb{P}}_{\tilde P^{(\mathbb{C}_n,1)}}(f_{\mathbb{B}}(J)=1| J=1, W^n(1)=w^n, U^n=u^n,V^n=v^n). \label{equtmp2}
\end{align}
Also, note that 
\begin{align}
    & \bar {\mathbb{P}}_{\tilde P^{(\mathbb{C}_n,1)}}(\mathcal{E}_{\mathsf{NE}}|f_{\mathbb{B}}(J)=1, J=1, W^n(1)=w^n, W^n(l)=\tilde w^n, U^n=u^n,V^n=v^n) \notag \\
    &= \bar {\mathbb{P}}_{\tilde P^{(\mathbb{C}_n,1)}}(\mathcal{E}_{\mathsf{NE}}|f_{\mathbb{B}}(J)=1, J=1, W^n(1)=w^n, U^n=u^n,V^n=v^n).\label{equtmp3}
\end{align}
Next, consider the term in \eqref{boundcdwrd}. Let
\begin{align}
 &\mathcal{F}':=\{W^n(1)=w^n, U^n=u^n,V^n=v^n\}, \notag \\
 &\mathcal{F}'':=\{ W^n(1)=w^n, W^n(l)=\tilde w^n, U^n=u^n,V^n=v^n\}. \notag 
\end{align}
 Then, the numerator and denominator of \eqref{boundcdwrd} can be written as
\begin{flalign}
    & \bar {\mathbb{P}}_{\tilde P^{(\mathbb{C}_n,1)}}(J=1|\mathcal{F}'') \notag\\
    &= \mathbb{E}_{\mu_n}\left[\frac{\prod_{i=1}^n  P_{U|W}(u_i|w_i)}{\prod_{i=1}^n  P_{U|W}(u_i|w_i)+\prod_{i=1}^n  P_{U|W}(u_i|\tilde w_i)+\sum_{j \neq 1,l}\prod_{i=1}^n  P_{U|W}(u_i|W_{i}(j)) }\right] \notag \\
    & \leq \mathbb{E}_{\mu_n}\left[\frac{\prod_{i=1}^n  P_{U|W}(u_i|w_i)}{\prod_{i=1}^n  P_{U|W}(u_i|w_i)+\sum_{j \neq 1,l}\prod_{i=1}^n  P_{U|W}(u_i|W_{i}(j)) }\right], \label{oneandlnotpres} &&
\end{flalign}
and  
\begin{flalign}
   & \bar {\mathbb{P}}_{\tilde P^{(\mathbb{C}_n,1)}}(J=1|\mathcal{F}')= \mathbb{E}_{\mu_n}\left[\frac{\prod_{i=1}^n  P_{U|W}(u_i|w_i)}{\prod_{i=1}^n  P_{U|W}(u_i|w_i)+\sum_{j \neq 1}\prod_{i=1}^n  P_{U|W}(u_i|W_{i}(j)) }\right], \label{onenotpres}&&
\end{flalign}
respectively. 
The R.H.S. of \eqref{oneandlnotpres}  (resp. \eqref{onenotpres}) denote the average probability that $J=1$ is chosen by  $P_{\mathsf{E}_{\mathsf{u}}}^{(\mathbb{B}_n)}$ given $W^n(1)=w^n$, $U^n=u^n$ and  $ M_n' -2$ (resp. $ M_n'-1$) other independent codewords in $\mathbb{B}_n$. 
Let
\begin{flalign}
    \mathcal{E}_{l}:=\left \lbrace \prod_{i=1}^n  P_{U|W}(u_i|W_{i}(l)) \geq  \max \left( \left\lbrace\prod_{i=1}^n  P_{U|W}(u_i|W_{i}(j)),~j \in \lceil M_n'\rceil \backslash\{1\} \right \rbrace  \cup \left\lbrace\prod_{i=1}^n  P_{U|W}(u_i|w_i)\right\rbrace\right) \right \rbrace. &&
\end{flalign} 
\revised{Note that
\begin{flalign}
 &   \mathbb{E}_{\mu_n|\mathcal{E}_{l}^c}\left[\frac{\prod_{i=1}^n  P_{U|W}(u_i|w_i)}{\prod_{i=1}^n  P_{U|W}(u_i|w_i)+\sum_{j \neq 1}\prod_{i=1}^n  P_{U|W}(u_i|W_{i}(j)) }\right] \notag \\
 &\geq \frac 12 \mathbb{E}_{\mu_n|\mathcal{E}_{l}^c}\left[\frac{\prod_{i=1}^n  P_{U|W}(u_i|w_i)}{\prod_{i=1}^n  P_{U|W}(u_i|w_i)+\sum_{j \neq 1,l}\prod_{i=1}^n  P_{U|W}(u_i|W_{i}(j)) }\right]. \label{bndtwiceprobstoch} &&
\end{flalign}
Hence, denoting by $\bar \mu_n$ the probability measure induced by $\mu_n$, we have
\begin{flalign}
 &\frac{ \bar {\mathbb{P}}_{\tilde P^{(\mathbb{C}_n,1)}}(J=1|\mathcal{F}'')}{ \bar {\mathbb{P}}_{\tilde P^{(\mathbb{C}_n,1)}}(J=1|\mathcal{F}')} \notag \\
 &\leq \frac{\mathbb{E}_{\mu_n}\left[\frac{\prod_{i=1}^n  P_{U|W}(u_i|w_i)}{\prod_{i=1}^n  P_{U|W}(u_i|w_i)+\sum_{j \neq 1,l}\prod_{i=1}^n  P_{U|W}(u_i|W_{i}(j)) }\right]}{\mathbb{E}_{\mu_n}\left[\frac{\prod_{i=1}^n  P_{U|W}(u_i|w_i)}{\prod_{i=1}^n  P_{U|W}(u_i|w_i)+\sum_{j \neq 1}\prod_{i=1}^n  P_{U|W}(u_i|W_{i}(j)) }\right]} \notag \\
&\leq  \frac{\mathbb{E}_{\mu_n}\left[\frac{\prod_{i=1}^n  P_{U|W}(u_i|w_i)}{\prod_{i=1}^n  P_{U|W}(u_i|w_i)+\sum_{j \neq 1,l}\prod_{i=1}^n  P_{U|W}(u_i|W_{i}(j)) }\right]}{\bar{\mu}_n(\mathcal{E}_{l}^c)\mathbb{E}_{\mu_n|\mathcal{E}_{l}^c}\left[\frac{\prod_{i=1}^n  P_{U|W}(u_i|w_i)}{\prod_{i=1}^n  P_{U|W}(u_i|w_i)+\sum_{j \neq 1}\prod_{i=1}^n  P_{U|W}(u_i|W_{i}(j)) }\right]} \notag &&      \end{flalign}
      \begin{flalign}
 &\leq \frac{\mathbb{E}_{\mu_n}\left[\frac{\prod_{i=1}^n  P_{U|W}(u_i|w_i)}{\prod_{i=1}^n  P_{U|W}(u_i|w_i)+\sum_{j \neq 1,l}\prod_{i=1}^n  P_{U|W}(u_i|W_{i}(j)) }\right]}{\frac 12 \bar{\mu}_n(\mathcal{E}_{l}^c) \mathbb{E}_{\mu_n|\mathcal{E}_{l}^c}\left[\frac{\prod_{i=1}^n  P_{U|W}(u_i|w_i)}{\prod_{i=1}^n  P_{U|W}(u_i|w_i)+\sum_{j \neq 1,l}\prod_{i=1}^n  P_{U|W}(u_i|W_{i}(j)) }\right]} \label{bndtwiceprobstochapp}\\
  &=\frac{\mathbb{E}_{\mu_n}\left[\frac{\prod_{i=1}^n  P_{U|W}(u_i|w_i)}{\prod_{i=1}^n  P_{U|W}(u_i|w_i)+\sum_{j \neq 1,l}\prod_{i=1}^n  P_{U|W}(u_i|W_{i}(j)) }\right]}{\frac 12  \mathbb{E}_{\mu_n}\left[\frac{\prod_{i=1}^n  P_{U|W}(u_i|w_i)}{\prod_{i=1}^n  P_{U|W}(u_i|w_i)+\sum_{j \neq 1,l}\prod_{i=1}^n  P_{U|W}(u_i|W_{i}(j)) }\right]-\frac 12  \bar{\mu}_n(\mathcal{E}_{l}) \mathbb{E}_{\mu_n|\mathcal{E}_{l}}\left[\frac{\prod_{i=1}^n  P_{U|W}(u_i|w_i)}{\prod_{i=1}^n  P_{U|W}(u_i|w_i)+\sum_{j \neq 1,l}\prod_{i=1}^n  P_{U|W}(u_i|W_{i}(j)) }\right]} \label{bndeqterm} \\
  &\leq \frac{\mathbb{E}_{\mu_n}\left[\frac{\prod_{i=1}^n  P_{U|W}(u_i|w_i)}{\prod_{i=1}^n  P_{U|W}(u_i|w_i)+\sum_{j \neq 1,l}\prod_{i=1}^n  P_{U|W}(u_i|W_{i}(j)) }\right]}{\frac 12  \mathbb{E}_{\mu_n}\left[\frac{\prod_{i=1}^n  P_{U|W}(u_i|w_i)}{\prod_{i=1}^n  P_{U|W}(u_i|w_i)+\sum_{j \neq 1,l}\prod_{i=1}^n  P_{U|W}(u_i|W_{i}(j)) }\right]-\frac 12  \bar{\mu}_n(\mathcal{E}_{l})} \label{bndtotvarprob} \\
 &\leq \frac{\mathbb{E}_{\mu_n}\left[\frac{\prod_{i=1}^n  P_{U|W}(u_i|w_i)}{\prod_{i=1}^n  P_{U|W}(u_i|w_i)+\sum_{j \neq 1,l}\prod_{i=1}^n  P_{U|W}(u_i|W_{i}(j)) }\right]}{\frac 12  \mathbb{E}_{\mu_n}\left[\frac{\prod_{i=1}^n  P_{U|W}(u_i|w_i)}{\prod_{i=1}^n  P_{U|W}(u_i|w_i)+\sum_{j \neq 1,l}\prod_{i=1}^n  P_{U|W}(u_i|W_{i}(j)) }\right]-e^{-e^{n(I_P(U;W)+\eta')}}} \label{doubexpapply} \\
 &\leq 2+o(1) \leq 3, \label{finineqbndratio} &&
\end{flalign}
where \eqref{bndtwiceprobstochapp} is due to \eqref{bndtwiceprobstoch}; \eqref{bndtotvarprob} is since the term within $\mathbb{E}_{\mu_n|\mathcal{E}_{l}}[\cdot]$  in \eqref{bndeqterm} is upper bounded by one; \eqref{doubexpapply} is since $ \bar{\mu}_n(\mathcal{E}_{l}) \leq e^{-e^{n(I_P(U;W)+\eta')}}$ for some $\eta'>0$ which follows similar to \cite[Section 3.6.3]{Elgamalkim},
and \eqref{finineqbndratio} follows since the term within the expectation which is exponential in order dominates the double exponential term.} From \eqref{equtmp0}, \eqref{equtmp1}, \eqref{equtmp2}, \eqref{equtmp3} and  \eqref{finineqbndratio}, \eqref{bndthreeval} follows.
The analysis of the other terms in \eqref{t2eecontdist} is the same as in the SHA scheme in \cite{Shimokawa}, and results in the error-exponent (within a additive $O(\delta)$ term) claimed in the Theorem. We refer the reader to \cite[Theorem 2]{SD_2020} for a detailed proof \footnote{In \cite{SD_2020}, the  communication channel between the observer and the detector is a DMC. However, since the coding scheme  used in the achievability part of Theorem 2 in \cite{SD_2020} is a separation based scheme, the error-exponent when the channel is noiseless can be recovered by setting  $E_3(\cdot)$ and $E_4(\cdot)$ in Theorem 2 to $\infty$.}.
By the random coding argument followed by the standard expurgation technique \cite{Gallager-codthm} (see \cite[Proof of Theorem 2]{SD_2020}), there exists a deterministic codebook and binning function pair  $\mathcal{C}_n=(\mathcal{B}_n,f_{\mathsf{B}})$ such that the type I  and type II error probabilities  are within a constant multiplicative factor of their average values over the random ensemble $\mathbb{C}_n$, and 
\begin{flalign}
&S_i-(w_i(J),V_i)-(M,w^n(J),V^n,S^{i-1}),~i \in [n],  \label{detcdcond0}\\
&\big\| \tilde P^{(\mathcal{C}_n,0)} -  \Psi^{(\mathcal{C}_n,0)} \big\|  \leq e^{-\gamma_{4} n}, \label{detcdcond1}\\
&\big\| \tilde P^{(\mathcal{C}_n,1)} - \Psi^{(\mathcal{C}_n,1)} \big\|  \leq e^{-\gamma_{4} n},~~ \mbox{if } Q_U=P_U, \label{detcdcond2}\\
\mbox{and }&\sum_{i=1}^n \big\| P_{W}-  \Psi^{(\mathcal{C}_n,l)}_{w_{i}(J)} \big\|  \leq e^{-\gamma_{5} n},~ l=0,1,  \label{detcdcond3}
\end{flalign}
where $\gamma_{4}$ and $\gamma_{5}$ are some positive numbers.
Since the average type I error probability for our scheme tends to zero asymptotically, and the error-exponent is unaffected by a constant multiplicative scaling of the type II error probability, this codebook achieves the same type I error probability and error-exponent as the  average over the random ensemble.  
 Using this deterministic codebook for encoding and decoding, we first lower bound the equivocation and average distortion of $S^n$ at the detector as follows: 
 
First consider the equivocation of $S^n$ under the null hypothesis.
\begin{flalign}
    H_{\tilde P^{(\mathcal{C}_n,0)}}(S^n|M,V^n) &\geq \mathbb{P}_{\tilde P^{(\mathcal{C}_n,0)}}(M \neq 0) H(S^n|M \neq 0,V^n) \notag &&      \end{flalign}
      \begin{flalign}
    & \geq (1-e^{-n \Omega(\delta)})~H_{\tilde P^{(\mathcal{C}_n,0)}}(S^n|M \neq 0,V^n) \label{applymsgn0}\\
    &\geq (1-e^{-n \Omega(\delta)})~H_{\tilde P^{(\mathcal{C}_n,0)}}(S^n|w^n(J),V^n) \label{applymarkchnmsgcd}\\
    &= (1-e^{-n \Omega(\delta)}) ~H_{\tilde P^{(\mathcal{C}_n,0)}}(S^n|w^n(J),V^n) \\
    &\geq (1-e^{-n \Omega(\delta)}) ~H_{ \Psi^{(\mathcal{C}_n,0)}}(S^n|w^n(J),V^n)-2 e^{-\gamma_{4} n} \log\left(\frac{|\mathcal{S}|^n|\V|^n}{e^{-\gamma_{4} n} }\right) \label{bndtotvarstr}\\
    & = \sum_{i=1}^n H_{ \Psi^{(\mathcal{C}_n,0)}}(S_i|w_i(J),V_i)- e^{-n \Omega(\delta)} \sum_{i=1}^n H_{ \Psi^{(\mathcal{C}_n,0)}}(S_i|w_i(J),V_i)-o(1) \label{memauxdist}\\
    & \geq  \sum_{i=1}^n H_{ \Psi^{(\mathcal{C}_n,0)}}(S_i|w_i(J),V_i)-n e^{-n \Omega(\delta)} H_P(S|V)-o(1) \\
    &=\left[\sum_{i=1}^n H_{ \Psi^{(\mathcal{C}_n,0)}}(S_i|w_i(J),V_i)\right]-o(1) \\
    &=n H_P(S|W,V)-o(1). \label{finstrentbnd} &&
\end{flalign}
Here, \eqref{applymsgn0} follows from \eqref{msgzeroprob}; \eqref{applymarkchnmsgcd} follows since $M$ is a function of $w^n(J)$ for a deterministic codebook; \eqref{bndtotvarstr} follows from \eqref{detcdcond1} and Lemma \ref{contentrtv}; \eqref{memauxdist} follows from \eqref{auxdist2gen}; and \eqref{finstrentbnd} follows from \eqref{detcdcond3} and $\Psi_{S_iV_i|w_{i}}^{(0)}=P_{SV|W}^{(0)},~ i \in [n]$.

If $Q_U=P_U$, it follows similarly to above that
\begin{flalign}
   H_{\tilde P^{(\mathcal{C}_n,1)}}(S^n|M,V^n)  &\geq \left(1-e^{-n \Omega(\delta)}\right) ~H_{\Psi^{(\mathcal{C}_n,1)}}(S^n|w^n(J),V^n)-2 e^{-\gamma_{4} n} \log\left(\frac{|\mathcal{S}|^n|\V|^n}{e^{-\gamma_{4} n}}\right) \\
   &=\sum_{i=1}^n H_{\Psi^{(\mathcal{C}_n,1)}}(S_i|w_i(J),V_i)- e^{-n \Omega(\delta)} \sum_{i=1}^n H_{\Psi^{(\mathcal{C}_n,1)}}(S_i|w_i(J),V_i)-o(1) \\
    & \geq \sum_{i=1}^n H_{\Psi^{(\mathcal{C}_n,1)}}(S_i|w_i(J),V_i)-n e^{-n \Omega(\delta)} H_Q(S|V)-o(1) \\
    &=\left[\sum_{i=1}^n H_{\Psi^{(\mathcal{C}_n,1)}}(S_i|w_i(J),V_i)\right]-o(1) \\
    &=n H_Q(S|W,V)-o(1). &&
\end{flalign}
Finally, consider the case $H=1$ and $Q_U \neq P_U$. We have for $\delta'$ small enough that,
\begin{align}
    \mathbb{P}_{\tilde P^{(\mathcal{C}_n,1)}}\left(M=0\right)= \mathbb{P}_{\tilde P^{(\mathcal{C}_n,1)}}\left(U^n \notin \mathcal{T}_{[P_U]_{\delta'}}^n\right) \geq  1-e^{-n(D(P_U||Q_U)-O(\delta'))} \xrightarrow{(n)} 1. \label{probtendexp1}
\end{align} 
Hence, for $\delta'$ small enough, we can write 
\begin{flalign}
& H_{\tilde P^{(\mathcal{C}_n,1)}}(S^n|M, V^n) \geq H_{\tilde P^{(\mathcal{C}_n,1)}}(S^n|M, V^n,\Pi(U^n,\delta',P_U))\notag\\ 
&\geq \left( 1-e^{-n(D(P_U||Q_U)-O(\delta'))}\right)  ~H_{\tilde P^{(\mathcal{C}_n,1)}}(S^n|M, V^n,\Pi(U^n,\delta',P_U)=1) \label{applyprbt1}\\
&= \left( 1-e^{-n(D(P_U||Q_U)-O(\delta'))}\right)~H_{\tilde P^{(\mathcal{C}_n,1)}}(S^n| V^n,\Pi(U^n,\delta',P_U)=1) \label{msgeqzerocond} \\
& \geq \left( 1-e^{-n(D(P_U||Q_U)-O(\delta'))}\right) ~\big(H_{\tilde P^{(\mathcal{C}_n,1)}}(S^n|V^n)-o(1)\big) \label{aplylemtotvar} \\
& = n H_{Q}(S|V)-ne^{-n(D(P_U||Q_U)-O(\delta'))}H_{Q}(S|V)- o(1)=n H_{Q}(S|V)- o(1).&&
\end{flalign}
Here, \eqref{applyprbt1} follows from \eqref{probtendexp1}; \eqref{msgeqzerocond} follows since $\Pi(U^n,\delta',P_U)=1$ implies $M=0$; \eqref{aplylemtotvar} follows from Lemma \ref{contentrtv} and \eqref{althypapproxdist}. Thus,  since $\delta>0$ is arbitrary, we have shown that for $\epsilon \in (0,1)$, $(R, \kappa, \Lambda_0, \Lambda_1) \in \mathcal{R}_e(\epsilon)$ if  \eqref{rtconsteqgen}-\eqref{distconsteq1gen} holds.

On the other hand, average distortion of $S^n$ at the detector can be lower bounded under $H=0$ as follows:
\revised{\begin{flalign}
&\min_{g^{(r)}_{i,n}}\mathbb{E}\left[d \left(S^n, \hat S^n\right) |H=0 \right] \notag \\
 & = \min_{\big\{\bar{\phi}_{i,n}(m,v^n,s^{i-1})\big\}_{i=1}^n} \mathbb{E}_{\tilde P^{(\mathcal{C}_n,0)}} \left[\sum_{i=1}^n  d \left(S_i, \bar \phi_i(m,v^n,s^{i-1}) \right) \right] \label{determadvapp1} \\
 &\geq \min_{\big\{\bar \phi_i(m,v^n,s^{i-1})\big\}_{i=1}^n}\mathbb{E}_{ \Psi^{(\mathcal{C}_n,0)}}  \left[\sum_{i=1}^n d(S_i,\bar \phi_i(m,v^n,s^{i-1})) \right]  -ne^{-n\gamma_4} D_m  \label{changedistgen} \\
  &\geq \min_{\big\{\bar \phi_i( \cdot,\cdot)\big\}_{i=1}^n}\mathbb{E}_{ \Psi^{(\mathcal{C}_n,0)}}  \left[\sum_{i=1}^n d(S_i,\bar \phi_i(w_i(J),V_i)) \right]  -ne^{-n\gamma_4} D_m  \label{Markchainauxdist}\\
 & \geq n \min_{\big\{\phi( \cdot,\cdot)\big\}}\mathbb{E}_{P}   \left[  d(S,\phi(W,V)) \right]  - n\left(e^{-n\gamma_4}+e^{-n\gamma_5}\right)D_m \label{auxapplydist} \\
 & =n\min_{\big\{\phi( \cdot,\cdot)\big\}_{i=1}^n}\mathbb{E}_{P}   \left[  d(S,\phi(W,V)) \right] -o(1).&&
  \end{flalign}}
 Here, \eqref{determadvapp1} follows from Lemma \ref{determfnatadv}; \eqref{changedistgen} follows from \cite[Property 2(b)]{Schieler-Cuff-2014} due to \eqref{detcdcond1} and boundedness of distortion measure; \eqref{Markchainauxdist} follows from the Markov chain in  \eqref{detcdcond0}; \eqref{auxapplydist} follows from \eqref{detcdcond3} and the fact that $\Psi_{S_iV_i|w_{i}(J)}^{(0)}=P_{SV|W}^{(0)},~ i \in [n]$.
 
Next, consider the case $H=1$ and $Q_U=P_U$. Then, similarly to above, we can write
\revised{\begin{flalign}
&\min_{g^{(r)}_{i,n}}\mathbb{E}\left[d \left(S^n, \hat S^n\right) |H=1 \right] \notag \\
&=   \min_{\big\{\bar \phi_i(m,v^n,s^{i-1})\big\}_{i=1}^n} \mathbb{E}_{\tilde P^{(\mathcal{C}_n,1)}} \left[\sum_{i=1}^n  d \left(S_i, \phi_i(M,V^n,S^{i-1}) \right) \right] \notag \\
 &\geq \min_{\big\{\bar \phi_i(m,v^n,s^{i-1})\big\}_{i=1}^n}\mathbb{E}_{\Psi^{(\mathcal{C}_n,1)}}  \left[\sum_{i=1}^n d(S_i,\phi_i(M,V^n,S^{i-1})) \right]  - ne^{-n\gamma_4}D_m  \label{changedistgenalt2} \\
  &\geq \min_{\big\{\phi_i( \cdot,\cdot)\big\}_{i=1}^n}\mathbb{E}_{\Psi^{(\mathcal{C}_n,1)}}  \left[\sum_{i=1}^n d(S_i,\phi_i(w_i,V_i)) \right]  - ne^{-n\gamma_4} D_m \label{Markchainauxdistalt2} \\
 & \geq  n\min_{\big\{\phi( \cdot,\cdot)\big\}_{i=1}^n}\mathbb{E}_{Q}   \left[  d(S,\phi(W,V)) \right] - n(e^{-n\gamma_4}+e^{-n\gamma_5})D_m . \label{auxapplydistalt2} \\
 & =n\min_{\big\{\phi( \cdot,\cdot)\big\}_{i=1}^n}\mathbb{E}_{Q}   \left[  d(S,\phi(W,V)) \right] -o(1).&&
 \end{flalign}}
If $H=1$ and $Q_U \neq P_U$, we have
\revised{
\begin{flalign}
\min_{g^{(r)}_{i,n}}\mathbb{E}\left[d \left(S^n, \hat S^n\right) |H=1 \right] &\geq  \mathbb{P}_{\tilde P^{(\mathcal{C}_n,1)}}\left(M=0 |H=1\right) \min_{\big\{\bar \phi_i(m,v^n,s^{i-1})\big\}_{i=1}^n} \sum_{i=1}^n \mathbb{E}_{\tilde P^{(\mathcal{C}_n,1)}} \left[ d \left(S_i, \phi_i(0,V^n,S^{i-1}) \right) \right] \notag   \\
 &\geq \mathbb{P}_{\tilde P^{(\mathcal{C}_n,1)}}\left(M=0 |H=1\right) \left[\min_{\big\{\phi_i'( v)\big\}_{i=1}^n}\mathbb{E}_{Q}  \left[\sum_{i=1}^n d(S_i,\phi_i'(V_i)) \right]  - D_m o(1)\right]  \label{applylempneqq}\\
 & = n\min_{\big\{\phi'( \cdot)\big\}}\mathbb{E}_{Q}   \left[  d(S,\phi'(V)) \right]  - o(1). \label{finbndqneqp} &&
\end{flalign} }
Here, \eqref{applylempneqq} follows from \eqref{althypapproxdist} in Lemma \ref{approxdistmequzero} and \eqref{finbndqneqp} follows from \eqref{probtendexp1}. Thus, since $\delta>0$ is arbitrary, we have shown that $(R, \kappa, \Delta_0, \Delta_1) \in \mathcal{R}_d(\epsilon)$, $\epsilon \in (0,1)$, provided that  \eqref{rtconsteqgen}, \eqref{expconsteqgen}, \eqref{distconst0} and \eqref{distconst1} are satisfied. This completes the proof of the theorem.
\section{Proof of Lemma \ref{cardauxrv}}\label{cardbndauxproof}
Consider the $|\Ucal|+2$ functions 
 of $P_{U|W}$,
\begin{align}
  &P_{U}(u_i)= \sum_{w \in \W} P_{W}(w) P_{U|W}(u_i|w), i=1,2,\ldots, |\Ucal|-1, \label{preserv1} \\
   & H_P(U|W,Z)= \sum_{w} P_{W}(w) g_1(P_{U|W},w),\label{preserv2} \\
   &  H_P(Y|W,Z)= \sum_{w} P_{W}(w) g_2(P_{U|W},w), \label{preserv3} \\
    & H_{P}(S|W,Y,Z)=\sum_{w} P_{W}(w) g_3(P_{U|W},w), \label{preserv4}
\end{align}
where
\begin{align}
  g_1(P_{U|W},w)&=   -\sum_{u,z}P_{U|W}(u|w)P_{Z|U}(z|u) \log\left(\frac{ P_{U|W}(u|w)P_{Z|U}(z|u)}{\sum_{u}P_{U|W}(u|w)P_{Z|U}(z|u)}\right), \notag \\
    g_2(P_{U|W},w)&=   -\sum_{y,z,u}P_{U|W}(u|w)P_{YZ|U}(y,z|u) \log\left(\frac{ \sum_{u}P_{U|W}(u|w)P_{YZ|U}(y,z|u)}{\sum_{u}P_{U|W}(u|w)P_{Z|U}(z|u)}\right),\notag \\
    g_3(P_{U|W},w)&= -\sum_{s,y,z,u}P_{U|W}(u|w)P_{SYZ|U}(s,y,z|u) \log\left(\frac{ \sum_{u}P_{U|W}(u|w)P_{SYZ|U}(s,y,z|u)}{\sum_{u}P_{U|W}(u|w)P_{YZ|U}(y,z|u)}\right).\notag
\end{align}
 Thus, by the Fenchel-Eggleston-Carath\'eodory's theorem \cite{Elgamalkim}, it is sufficient to have at most $|\Ucal|-1$ points in the support of $W$ to preserve $P_U$ and three more to preserve $H_P(U|W,Z)$, $H_P(Y|W,Z)$ and $ H_{P}(S|W,Z,Y)$. Noting that $H_P(Y|Z)$ and $H_P(U|Z)$ are automatically preserved since $P_U$ is preserved (and $(Y,Z,S)-U-W$ holds), $|\W|=|\Ucal|+2$ points are sufficient to preserve the R.H.S. of equations \eqref{expconsteq}-\eqref{distconsteq0}. This completes the proof for the case of $\mathcal{R}_e$.
 Similarly, considering the $|\Ucal|+1$ functions of $P_{W|U}$ given in \eqref{preserv1}-\eqref{preserv3} and 
\begin{align}
\mathbb{E}_P\left[d \left(S, \phi(W,Y,Z)\right)\right]&= \sum_{w} P_W(w) g_4(w,P_{W|U}), \notag 
\end{align}
where
\begin{align}
g_4(w,P_{W|U})&= \sum_{s,u,y,z}P_{U|W}(u|w)P_{YZS|U}(y,z,s|u) ~d(s,\phi(w,y,z)),\notag
\end{align}
similar result holds also for the case of $\mathcal{R}_d$.

\end{appendices}

\bibliographystyle{IEEEtran}
\bibliography{refarxiv}

\begin{thebibliography}{10}
\providecommand{\url}[1]{#1}
\csname url@samestyle\endcsname
\providecommand{\newblock}{\relax}
\providecommand{\bibinfo}[2]{#2}
\providecommand{\BIBentrySTDinterwordspacing}{\spaceskip=0pt\relax}
\providecommand{\BIBentryALTinterwordstretchfactor}{4}
\providecommand{\BIBentryALTinterwordspacing}{\spaceskip=\fontdimen2\font plus
\BIBentryALTinterwordstretchfactor\fontdimen3\font minus
  \fontdimen4\font\relax}
\providecommand{\BIBforeignlanguage}[2]{{%
\expandafter\ifx\csname l@#1\endcsname\relax
\typeout{** WARNING: IEEEtran.bst: No hyphenation pattern has been}%
\typeout{** loaded for the language `#1'. Using the pattern for}%
\typeout{** the default language instead.}%
\else
\language=\csname l@#1\endcsname
\fi
#2}}
\providecommand{\BIBdecl}{\relax}
\BIBdecl

\bibitem{ITW-18}
S.~Sreekumar, D.~G\"und\"uz, and A.~Cohen, ``Distributed hypothesis testing
  under privacy constraints,'' in \emph{IEEE Inf. Theory Workshop}, Guangzhou,
  China, Nov. 2018.

\bibitem{AJ-2010}
A.~Appari and E.~Johnson, ``Information security and privacy in healthcare:
  current state of research,'' \emph{Int. Journ. Internet and Enterprise
  Management}, vol.~6, no.~4, pp. 279--314, 2010.

\bibitem{GA-2005}
R.~Gross and A.~Acquisti, ``Information revelation and privacy in online social
  networks,'' in \emph{ACM workshop on Privacy in Electronic Society},
  Alexandria, VA, USA, Nov. 2005.

\bibitem{MF-2010}
A.~Miyazaki and A.~Fernandez, ``Consumer perceptions of privacy and security
  risks for online shopping,'' \emph{Journ. of Consumer Affairs}, vol.~35,
  no.~1, pp. 27--44, 2001.

\bibitem{GD-2016}
G.~Giaconi, D.~G\"{u}nd\"{u}z, and H.~V. Poor, ``Privacy-aware smart metering:
  Progress and challenges,'' \emph{IEEE Signal Processing Magazine}, vol.~35,
  no.~6, pp. 59--78, Nov. 2018.

\bibitem{Ahlswede-Csiszar}
R.~Ahlswede and I.~Csisz\'{a}r, ``Hypothesis testing with communication
  constraints,'' \emph{{IEEE} Trans. Inf. Theory}, vol.~32, no.~4, pp.
  533--542, Jul. 1986.

\bibitem{Han}
T.~S. Han, ``Hypothesis testing with multiterminal data compression,''
  \emph{{IEEE} Trans. Inf. Theory}, vol.~33, no.~6, pp. 759--772, Nov. 1987.

\bibitem{Shimokawa}
H.~Shimokawa, T.~S. Han, and S.~Amari, ``Error bound of hypothesis testing with
  data compression,'' in \emph{IEEE Int. Symp. Inf. Theory}, Trondheim, Norway,
  1994.

\bibitem{Shalaby-pap}
H.~M.~H. Shalaby and A.~Papamarcou, ``Multiterminal detection with zero-rate
  data compression,'' \emph{{IEEE} Trans. Inf. Theory}, vol.~38, no.~2, pp.
  254--267, Mar. 1992.

\bibitem{Zhao-Lai}
W.~Zhao and L.~Lai, ``Distributed testing against independence with multiple
  terminals,'' in \emph{52nd Annual Allerton Conf.}, IL, USA, Oct. 2014.

\bibitem{Katz-estdetjourn}
G.~Katz, P.~Piantanida, and M.~Debbah, ``Distributed binary detection with
  lossy data compression,'' \emph{{IEEE} Trans. Inf. Theory}, vol.~63, no.~8,
  pp. 5207--5227, Aug. 2017.

\bibitem{Rahman-Wagner}
M.~S. Rahman and A.~B. Wagner, ``On the optimality of binning for distributed
  hypothesis testing,'' \emph{{IEEE} Trans. Inf. Theory}, vol.~58, no.~10, pp.
  6282--6303, Oct. 2012.

\bibitem{SG_isit17}
S.~Sreekumar and D.~G\"und\"uz, ``Distributed hypothesis testing over noisy
  channels,'' in \emph{IEEE Int. Symp. Inf. Theory}, Aachen, Germany, Jun.
  2017.

\bibitem{SD_2020}
------, ``Distributed hypothesis testing over discrete memoryless channels,''
  \emph{{IEEE} Trans. Inf. Theory}, vol.~66, no.~4, Apr. 2020.

\bibitem{Sadaf_Wigger_Timo}
S.~Salehkalaibar, M.~Wigger, and R.~Timo, ``On hypothesis testing against
  conditional independence with multiple decision centers,'' \emph{{IEEE}
  Trans. Commun.}, Jan 2018.

\bibitem{Sadaf-Wigger-HTN}
S.~Salehkalaibar and M.~Wigger, ``Distributed hypothesis testing based on
  unequal-error protection codes,'' \emph{arXiv:1806.05533}.

\bibitem{HK-1989}
T.~S. Han and K.~Kobayashi, ``Exponential-type error probabilities for
  multiterminal hypothesis testing,'' \emph{{IEEE} Trans. Inf. Theory},
  vol.~35, no.~1, pp. 2--14, Jan. 1989.

\bibitem{HK-2017}
E.~Haim and Y.~Kochman, ``On binary distributed hypothesis testing,''
  \emph{arXiv:1801.00310}.

\bibitem{WK-2017}
N.~Weinberger and Y.~Kochman, ``On the reliability function of distributed
  hypothesis testing under optimal detection,'' \emph{{IEEE} Trans. Inf.
  Theory}, vol.~65, no.~8, pp. 4940--4965, Apr. 2019.

\bibitem{Bayardo-Agrawal}
R.~Bayardo and R.~Agrawal, ``Data privacy through optimal k-anonymization,'' in
  \emph{Int. Conf. on Data Engineering}, Tokyo, Japan, Apr. 2005.

\bibitem{Rakesh-Srikant}
R.~Agrawal and R.~Srikant, ``Privacy-preserving data mining,'' in \emph{ACM
  SIGMOD Int. Conf. on Management of data}, Dallas, USA, May. 2000.

\bibitem{Bert15}
E.~Bertino, ``Big data-security and privacy,'' in \emph{IEEE Int. Congress on
  BigData}, New York, USA, 2015.

\bibitem{Yael2000}
Y.~Gertner, Y.~Ishai, E.~Kushilevitz, and T.~Malkin, ``Protecting data privacy
  in private information retrieval schemes,'' \emph{Journ. of Computer and
  System Sciences}, vol.~60, no.~3, pp. 592--629, Jun 2000.

\bibitem{Hay08}
M.~Hay, G.~Miklau, D.~Jensen, D.~Towsley, and P.~Weis, ``Resisting structural
  re-identification in anonymized social networks,'' \emph{Journ. Proc. of the
  VLDB Endowment}, vol.~1, no.~1, pp. 102--114, Aug. 2008.

\bibitem{Narayanan-Shmatikov}
A.~Narayanan and V.~Shmatikov, ``De-anonymizing social networks,'' in
  \emph{IEEE Symp. on Security and Privacy}, Berkeley, USA, 2009.

\bibitem{Liao-Lal-Tan-Cal}
J.~Liao, L.~Sankar, V.~Tan, and F.~Calmon, ``Hypothesis testing under mutual
  information privacy constraints in the high privacy regime,'' \emph{{IEEE}
  Trans. Inf. Forensics and Security}, vol.~13, no.~4, pp. 1058--1071, Apr.
  2018.

\bibitem{Jiao-Lal-Cal}
J.~Liao, L.~Sankar, F.~Calmon, and V.~Tan, ``Hypothesis testing under maximal
  leakage privacy constraints,'' in \emph{IEEE Int. Symp. Inf. Theory}, Aachen,
  Germany, Jun. 2017.

\bibitem{GSSV-2018}
A.~Gilani, S.~B. Amor, S.~Salehkalaibar, and V.~Tan, ``Distributed hypothesis
  testing with privacy constraints,'' \emph{Entropy}, vol.~21, no. 478, pp.
  1--27, May 2019.

\bibitem{GEP-08}
D.~G\"und\"uz, E.~Erkip, and H.~V. Poor, ``Secure lossless compression with
  side information,'' in \emph{IEEE Inf. Theory Workshop}, Porto, Portugal,
  May. 2008.

\bibitem{GEP-08-ISIT}
------, ``Lossless compression with security constraints,'' in \emph{IEEE Int.
  Symp. Inf. Theory}, Jul. 2008.

\bibitem{MP-2015}
M.~Mhanna and P.~Piantanida, ``On secure distributed hypothesis testing,'' in
  \emph{Proc. IEEE Int. Symp. Inf. Theory}, Hong Kong, China, 2015.

\bibitem{Sweeney-2002}
L.~Sweeney, ``K-anonymity: a model for protecting privacy,'' \emph{Int. Journ.
  on Uncertainty, Fuzziness and Knowledge based Systems}, 2002.

\bibitem{Dwork-2006}
C.~Dwork, ``Differential privacy,'' \emph{Automata, Languages and Programming.
  Springer}, vol. 4052, pp. 1--12, 2006.

\bibitem{Calmon-Fawaz}
F.~Calmon and N.~Fawaz, ``Privacy against statistical inference,'' in
  \emph{50th Annual Allerton Conf.}, IL, USA, Oct.2012.

\bibitem{PF-2014}
A.~Makhdoumi, S.~Salamatian, N.~Fawaz, and M.~Medard, ``From the information
  bottleneck to the privacy funnel,'' in \emph{IEEE Inf. Theory Workshop},
  Hobart, Australia, Nov. 2014.

\bibitem{Cal-Ali-Muriel}
F.~Calmon, A.~Makhdoumi, and M.~Medard, ``Fundamental limits of perfect
  privacy,'' in \emph{IEEE Int. Symp. Inf. Theory}, Hong Kong, China, Jun.
  2015.

\bibitem{Issa-Kamath-Wagner-2016}
I.~Issa, S.~Kamath, and A.~B. Wagner, ``An operational measure of information
  leakage,'' in \emph{Annual Conf. on Inf. Science and Systems}, Princeton,
  USA, Mar. 2016.

\bibitem{Rassouli-Gunduz-TIFS19}
B.~Rassouli and D.~G\"und\"uz, ``Optimal utility-privacy trade-off with total
  variation distance as a privacy measure,'' \emph{{IEEE} Trans. Inf. Forensics
  and Security}, to appear, 2019.

\bibitem{Wagner-Eckhoff}
I.~Wagner and D.~Eckhoff, ``Technical privacy metrics: a systematic survey,''
  \emph{{arXiv:1512.00327v1 [cs.CR]}}.

\bibitem{GM-1984}
S.~Goldwasser and S.~Micali, ``Probabilistic encryption,'' \emph{Journ. of
  Computer and System Sciences}, vol.~28, no.~2, pp. 270--299, Apr. 1984.

\bibitem{BTV-2012}
M.~Bellare, S.~Tessaro, and A.~Vardy, ``Semantic security for the wiretap
  channel,'' in \emph{Advances in Cryptology-CRYPTO 2012}, Heidelberg, Germany,
  2012.

\bibitem{Yamamoto-1988}
H.~Yamamoto, ``A rate-distortion problem for a communication system with a
  secondary decoder to be hindered,'' \emph{{IEEE} Trans. Inf. Theory},
  vol.~34, no.~4, pp. 835--842, Jul. 1988.

\bibitem{TSP-2013}
R.~Tandon, L.~Sankar, and H.~V. Poor, ``Discriminatory lossy source coding:
  Side information privacy,'' \emph{{IEEE} Trans. Inf. Theory}, vol.~59, no.~9,
  pp. 5665--5677, Sep. 2013.

\bibitem{Schieler-Cuff-2014}
C.~Schieler and P.~Cuff, ``Rate-distortion theory for secrecy systems,''
  \emph{{IEEE} Trans. Inf. Theory}, vol.~60, no.~12, pp. 7584--7605, Dec. 2014.

\bibitem{Gaurav-thesis}
G.~K. Agarwal, ``On information theoretic and distortion-based security,''
  \emph{PhD Thesis-UCLA [Online]. Available:
  https://escholarship.org/uc/item/\\7qs7z91g}, 2019.

\bibitem{LOG-2019}
Z.~Li, T.~Oechtering, and D.~G\"und\"uz, ``Privacy against a hypothesis testing
  adversary,'' \emph{{IEEE} Trans. Inf. Forensics and Security}, vol.~14,
  no.~6, pp. 1567--1581, Jun. 2019.

\bibitem{Csiszar-Korner}
I.~Csisz\'{a}r and J.~K\"{o}rner, \emph{Information Theory: Coding Theorems for
  Discrete Memoryless Systems}.\hskip 1em plus 0.5em minus 0.4em\relax
  Cambridge University Press, 2011.

\bibitem{WBI-2017}
Y.~Wang, Y.~O. Basciftci, and P.~Ishwar, ``Privacy-utility tradeoffs under
  constrained data release mechanisms,'' \emph{arXiv:1710.09295}.

\bibitem{Cuff-2013}
P.~Cuff, ``Distributed channel synthesis,'' \emph{{IEEE} Trans. Inf. Theory},
  vol.~59, no.~11, pp. 7071--7096, Nov. 2013.

\bibitem{Song-Cuff-Poor-2016}
E.~C. Song, P.~Cuff, and H.~V. Poor, ``The likelihood encoder for lossy
  compression,'' \emph{{IEEE} Trans. Inf. Theory}, vol.~62, no.~4, pp.
  1836--1849, Apr. 2016.

\bibitem{Wyner-1975-comminfo}
A.~D. Wyner, ``The common information of two dependent random variables,''
  \emph{{IEEE} Trans. Inf. Theory}, vol.~21, no.~2, pp. 163--179, Mar. 1975.

\bibitem{Han-Verdu-1993}
T.~S. Han and S.~Verd\'{u}, ``Approximation theory of output statistics,''
  \emph{{IEEE} Trans. Inf. Theory}, vol.~39, no.~3, pp. 752--772, May. 1993.

\bibitem{Elgamalkim}
A.~E. Gamal and Y.-H. Kim, \emph{Network Information theory}.\hskip 1em plus
  0.5em minus 0.4em\relax Cambridge University Press, 2011.

\bibitem{YP-Thesis}
Y.~Polyanskiy, \emph{Channel coding: non-asymptotic fundamental limits}.\hskip
  1em plus 0.5em minus 0.4em\relax PhD Thesis, Princeton University, 2010.

\bibitem{YCDP-2015}
W.~Yang, G.~Caire, G.~Durisi, and Y.~Polyanskiy, ``Optimum power control at
  finite blocklength,'' \emph{{IEEE} Trans. Inf. Theory}, vol.~61, no.~9, pp.
  4598--4615, Sep. 2015.

\bibitem{Villard-Pablo-journal}
J.~Villard and P.~Piantanida, ``Secure multiterminal source coding with side
  information at the eavesdropper,'' \emph{{IEEE} Trans. Inf. Theory}, vol.~59,
  no.~6, pp. 3668--3692, Jun. 2013.

\bibitem{Gallager-codthm}
R.~G. Gallager, ``A simple derivation of the coding theorem and some
  applications,'' \emph{{IEEE} Trans. Inf. Theory}, vol.~11, pp. 3--18, Jan
  1965.

\end{thebibliography}
    
\end{document}